\documentclass[a4paper,11pt]{article}
\usepackage{mathtools,amsfonts,amssymb,mathrsfs,amsthm,tabularx,multicol,multirow,calc,graphicx,color,rotating,enumitem,bbm,geometry}
\usepackage[OT1]{fontenc}    %
\usepackage{arydshln}        
\allowdisplaybreaks
\usepackage[round,longnamesfirst]{natbib}
\definecolor{dblue}{rgb}{0,0,0.5}
\usepackage[linktoc=all,hypertexnames=false]{hyperref} 
\hypersetup{
  colorlinks=true,
  linkcolor=dblue,
  anchorcolor=dblue,
  citecolor=dblue,
  filecolor=dblue,
  menucolor=dblue,
  urlcolor=dblue,
  pdftitle={Improved Density and CDF Estimation},
  pdfauthor={V. Oryshchenko and R. J. Smith},
  pdfcreator={\LaTeX\ with package \flqq hyperref\frqq}, 
  final=true,
  breaklinks=true
  }
\usepackage{titling}  
\newcommand{\I}{\mathbbm{1}}                           
\newcommand{\abs}[1]{\lvert#1\rvert}                   %
\newcommand{\norm}[1]{\lVert#1\rVert}                  %
\newcommand{\transp}{\top}                             
\newcommand{\trace}{\mathop{\mathrm{tr}}}              
\newcommand{\littleO}[1][]{\mathit{o}_{#1}}            
\newcommand{\bigO}[1][]{\mathit{O}_{#1}}               
\newcommand{\R}{\mathbbm{R}}
\newcommand{\interior}{\mathop{\mathrm{int}}}
\DeclareMathOperator{\E}{E}

\DeclareMathOperator{\Var}{Var}
\DeclareMathOperator{\Cov}{Cov}
\DeclareMathOperator{\MISE}{MISE}
\DeclareMathOperator{\MSE}{MSE}

\DeclareMathOperator{\ISB}{ISB}
\DeclareMathOperator{\IVar}{IVar}
\DeclareMathOperator{\arsinh}{arsinh}

\DeclareMathOperator*{\argmin}{argmin}
\DeclareMathOperator*{\argmax}{argmax}
\newcommand{\M}[4][]{\prescript{#1}{#2}{\mathrm{M}}^{#4}_{#3}}
\def\tint{{\begingroup\textstyle\int\endgroup}}  
\def\tsum{{\begingroup\textstyle\sum\endgroup}}
\newcommand{\bv}{e}  %
\usepackage[sc]{titlesec}  
\newtheoremstyle{examplestyle}
  {\topsep}
  {\topsep}
  {\upshape}
  {0pt}
  {\Large\scshape}
  { }
  { }
  {\thmname{#1} \thmnumber{#2} \thmnote{(#3)}\\}
\theoremstyle{examplestyle} \newtheorem{example}{Example}
\newtheoremstyle{scrmstyle}{\topsep}{\topsep}{\upshape}{0pt}{\scshape}{}{ }{\thmname{#1} \thmnumber{#2}.\thmnote{(#3)}}
\theoremstyle{scrmstyle} \newtheorem{remark}{Remark}
\newtheoremstyle{scitstyle}{\topsep}{\topsep}{\itshape}{0pt}{\scshape}{}{ }{\thmname{#1} \thmnumber{#2}.\thmnote{(#3)}}
\theoremstyle{scitstyle} \newtheorem{theorem}{Theorem}
\theoremstyle{scitstyle} \newtheorem{lemma}{Lemma}
\theoremstyle{scitstyle} \newtheorem{corollary}{Corollary}
\theoremstyle{scitstyle} \newtheorem{assumption}{Assumption}

\makeatletter
\newcommand{\customlabel}[2]{%
   \protected@write \@auxout {}{\string \newlabel {#1}{{#2}{\thepage}{#2}{#1}{}} }%
   \hypertarget{#1}{#2}
}
\makeatother

\begin{document}

\title{\scshape Improved Density and Distribution Function Estimation} 
\author{\protect\begin{tabular}{cp{10mm}c}
Vitaliy Oryshchenko
\thanks{Address for correspondence: 2.068 Arthur Lewis Building, Department of Economics, School of Social Sciences, University of Manchester, Oxford Road, Manchester M13 9PL, United Kingdom. E-mail: \href{mailto:v.oryshchenko@cantab.net}{v.oryshchenko@cantab.net}.} 
&  & Richard J. Smith \\ 
Department of Economics  & & c{\it e}mmap, U.C.L and I.F.S.\\
University of Manchester & & Faculty of Economics, University of Cambridge \\
 & & Department of Economics, University of Melbourne \\
 & & ONS Economic Statistics Centre of Excellence
 \protect\end{tabular}}
\date{\today}
\maketitle

\begin{abstract}
Given additional distributional information in the form of moment restrictions, kernel density and distribution function estimators with implied generalised empirical likelihood probabilities as weights achieve a reduction in variance due to the systematic use of this extra information. The particular interest here is the estimation of densities or distributions of (generalised) residuals in semi-parametric models defined by a finite number of moment restrictions. Such estimates are of great practical interest, being potentially of use for diagnostic purposes, including tests of parametric assumptions on an error distribution, goodness-of-fit tests or tests of overidentifying moment restrictions. The paper gives conditions for the consistency and describes the asymptotic mean squared error properties of the kernel density and distribution estimators proposed in the paper. A simulation study evaluates the small sample performance of these estimators. Supplements provide analytic examples to illustrate situations where kernel weighting provides a reduction in variance together with proofs of the
results in the paper.

\vspace*{0.3\baselineskip}

\vspace*{0.3\baselineskip}

\end{abstract}


\numberwithin{remark}{section}
\numberwithin{equation}{section}
\numberwithin{assumption}{section}
\numberwithin{theorem}{section}

\section{Introduction}
In many statistical and economic applications, additional distributional information about the data observation $d_{z}$-vector $z$ may be available in the form of moment restrictions on its distribution. These constraints may arise from a particular economic or physical law, e.g., \citet[Section 5]{chen1997}, be implied by estimating equations, \citet[Example 1]{qin1994}, or correspond to known population moments of another observable random vector correlated with $z$, e.g., in survey samples with  auxiliary population information available from census data, e.g., \citet{chen1993} and \citet[Example 2]{qin1994}.
The primary purpose of the paper is to explore the advantages of this additional information for the estimation of the density and distribution function of a scalar residual-like function of $z$ which may depend on unknown parameters.

To this end, let $g(z,\beta)$ denote a $d_{g}$-vector of known functions of the data observation $d_{z}$-vector $z\in\mathcal{Z}$ and the $d_{\beta}$-vector $\beta\in\mathcal{B}$ of parameters where the sample space $\mathcal{Z}\subseteq\R^{d_{z}}$ and parameter space $\mathcal{B}\subset\R^{d_{\beta}}$ with $d_{\beta}\leq d_{g}$. The moment indicator vector $g(z,\beta)$ will form the basis for inference in the following discussion and analysis. In particular, it is assumed that the true value $\beta_{0}$ taken by $\beta$ uniquely satisfies the population unconditional moment equality condition 
\begin{equation}\label{Eq:GEL.constraints}
\E[g(z,\beta_{0})]=0,
\end{equation}
where $\E[\cdot]$ denotes expectation taken with respect to the true population probability law of $z$. The true parameter value $\beta_{0}$ is generally unknown, but can also be fully or partially known in particular applications.

Models specified in the form of unconditional moment restrictions \eqref{Eq:GEL.constraints} convey partial information about the distribution $F^{z}$ of $z$ and are ubiquitous in economics; see, e.g., the monographs \citet{hall2005} and \citet{matyas1999}. Many other commonly used models lead to estimators that can be reformulated as solutions to a set of moment restrictions. Clearly, models given by conditional moment restrictions imply \eqref{Eq:GEL.constraints}. Traditionally, such models are estimated by the generalised method of moments (GMM). However, the performance of GMM estimators and associated test statistics is
often poor in finite samples, which has lead to the development of a number of (information-theoretic) alternatives to GMM.

This paper focuses on the class of generalised (G) empirical likelihood (EL) estimators, which has attractive large sample properties; see, e.g., \citet{newey2004}, \citet{smith1997,smith2011}, and \citet{parente2014} for a recent review. Special cases of GEL include EL, \citep{owen1988,owen1990}, \citet{qin1994}, exponentially tilting (ET), \citet{corcoran1998}, \citet{kitamura1997}, \citet{imbens1998}, and continuous-updating (GMM) estimators (CUE), \citet{hansen1996}; see also Euclidean EL, \citet{antoine2007}. Of these estimators, EL has the attractive property of being Bartlett-correctable; see \citet{chen2007}.

When the parameter vector $\beta_{0}$ is overidentified by the moment restriction \eqref{Eq:GEL.constraints}, i.e., $d_{\beta}<d_{g}$, these constraints generally carry useful additional information about $F^{z}$. Given a random sample $z_{i}$, $i=1,\ldots,n$, of observations on $z$, such information is captured by the associated (G)EL implied probabilities $\pi_{i}$, $i=1,\ldots,n$, which enable a nonparametric description of $F^{z}$ satisfying the moment condition \eqref{Eq:GEL.constraints} given by the estimator $F_{\pi}^{z}(z)=\sum_{i=1}^{n}\pi_{i}\I\{z_{i}\leq z\}$, where $\I\{\cdot\}$ denotes the indicator function, \citet{back1993}, \citet{qin1994}. In the absence of the moment information \eqref{Eq:GEL.constraints} or when $\beta_{0}$ is just identified, $d_{\beta}=d_{g}$, $F_{\pi}^{z}(z)$ reduces to the empirical distribution function (EDF) $F_{n}^{z}(z)=n^{-1}\sum_{i=1}^{n}\I\{z_{i}\leq z\}$. In general, if $d_{\beta}<d_{g}$, $F_{\pi}^{z}(z)$ is a more efficient estimator of $F^{z}$ than the EDF $F_{n}^{z}(z)$ reflecting the value of the overidentifying information in \eqref{Eq:GEL.constraints}. This observation suggests therefore that estimation of the functionals of $F^{z}$, $T(F^{z})$, by $T(F_{\pi}^{z})$ rather than $T(F_{n}^{z})$ will be similarly more efficient. Indeed this is the case when estimating expectations of certain known functions of $z$, see \citet{brown1998}. A similar advantage is apparent for EL estimation of quantile functions with known $\beta_{0}$, e.g., \citet{chen1993} and \citet{zhang1995}, general EL-based quantile estimation, \citet{yuan2014}, and EL-based kernel estimation of a univariate density function, e.g., \citet{chen1997} and \citet{zhang1998}. 

The concern of this paper is with efficient kernel estimation of the probability density (p.d.f.) and distribution (c.d.f.)  functions of a scalar-valued function $u(z,\beta_{0})$ of the data observation $z$ with either known or unknown parameter vector $\beta_{0}$. The former case, when  $\beta_{0}$ is known, is the classical situation briefly mentioned above. The central case of interest, when $\beta_{0}$ is unknown, is estimation of the p.d.f and c.d.f. of an error term based on the estimated residuals. Such estimates are routinely computed by practitioners and are used for both visual diagnostics, e.g., potentially revealing omitted structure such as multimodality or other features of interest, and formal diagnostic tests, e.g., goodness-of-fit and tests of parametric assumptions on the error distribution. The importance of obtaining residual density estimates with
good (higher order) properties can hardly be understated. Yet, as discussed below, simply applying standard kernel estimators with default bandwidths to estimated residuals may result in an inconsistent p.d.f. or c.d.f. estimators as further conditions on the kernel function and bandwidth are generally required. Similar conclusions have been reached elsewhere in related literature on residual density estimation in nonparametric regression and other settings; see, e.g., 
\citet{ahmad1992}, \citet{cheng2004}, \citet{kiwitt2008}, \citet{gyorfi2012} and the discussion and references in \citet{bott2013}. 

When $\beta_{0}$ is known, kernel density and distribution function estimators exploiting the (G)EL implied probabilities instead of the uniform EDF $n^{-1}$ weights achieve a reduction of higher order variance due to the systematic use of the extra moment information in \eqref{Eq:GEL.constraints}. The efficiency gains are first order asymptotically in the c.d.f. case and
second order for p.d.f. estimation. In contradistinction, for residual p.d.f. and c.d.f. estimation, such gains will not always  be realised. One can, however, expect efficiency gains from the knowledge that the mean of residuals is zero.

The outline of the paper is as follows. Section \ref{Sec:GEL.overview} briefly describes (G)EL estimation and the associated (G)EL implied probabilities. The main results concerning p.d.f. and c.d.f. estimators are given in Sections \ref{Sec:GEL.based.KDE} and \ref{Sec:GEL.based.CDF.KE} for both known and unknown $\beta_{0}$ cases. The finite sample performance of the proposed estimators is evaluated via a simulation study reported in Section \ref{Sec:Monte.Carlo}. Section \ref{sec:conclusions} concludes. 
Supplement Supplement \ref{Supp:Proofs}: Proofs and \ref{Supp:Examples}: Examples in the Supplementary Information respectively 
details some additional assumptions for and the proofs of the results in the main text and 
analyses a number of examples to illustrate the the properties of the estimators developed in the paper.

\section{Generalised Empirical Likelihood}\label{Sec:GEL.overview}
The GEL class of estimators for $\beta_{0}$ is defined in terms of a real valued scalar carrier function $\rho:\mathcal{V}\mapsto\R$ that is concave on an open interval $\mathcal{V}$ containing zero with derivatives $\rho^{(j)}(v)=\mathrm{d}^{j}\rho(v)/\mathrm{d}v^{j}$ and $\rho_{j}=\rho^{(j)}(0)$, $j=1,2,\ldots$, normalized without loss of generality such that $\rho_{1}=\rho_{2}=-1$. The special cases $\rho(v)=\ln(1-v)$ for $\mathcal{V}=(-\infty, 1)$, $\rho(v)=-\exp(v)$ and $\rho(v)=-v^{2}/2-v$ correspond to EL, ET and CUE respectively and are all members of the \citet{cressie1984} family where $\rho(v)=-(1+\gamma v)^{(\gamma +1)/\gamma}/(\gamma+1)$.

Given a random sample $z_{i}$, $i=1,\ldots,n$, of size $n$ of observations on the $d_{z}$-dimensional vector $z$, let $g_{i}(\beta)=g(z_{i},\beta)$, $g_{i}=g_{i}(\beta_{0})$, and $G_{i}(\beta)=\partial g(z_{i},\beta)/\partial\beta^{\transp}$, $G_{i}=G_{i}(\beta_{0})$, $i=1,\ldots,n$. Also let $\Lambda _{n}(\beta)=\{\lambda : \lambda^{\transp}g_{i}(\beta)\in\mathcal{V}, i=1,\ldots,n\}$. The GEL criterion $P_{n}(\beta,\lambda)$ is defined by $P_{n}(\beta,\lambda)=n^{-1}\sum_{i=1}^{n}\rho(\lambda^{\transp}g_{i}(\beta))-\rho(0)$, with $\lambda$ a $d_{g}$-vector of auxiliary parameters, each element of which corresponding to an
element of the moment function vector $g(z,\beta)$; for members of the \citet{cressie1984} family of power divergence criteria $\lambda$ is the Lagrange multiplier vector associated with imposition of the moment restriction \eqref{Eq:GEL.constraints}. The GEL estimator $\hat{\beta}$ is the solution to the saddle point problem 
\begin{equation}\label{Eq:GEL.saddle.point.problem}
\hat{\beta}=\argmin_{\beta\in\mathcal{B}}\sup_{\lambda\in\Lambda_{n}(\beta)}P_{n}(\beta,\lambda).
\end{equation}

If Supplement \ref{Supp:Proofs}: Assumptions \ref{Assumption.GEL.1} and \ref{Assumption.GEL.2} are satisfied, in particular, the
population Jacobian $G=\E[\partial g(z,\beta_{0})/\partial\beta^{\transp}]$ and variance $\Omega =\E[g(z,\beta_{0})g(z,\beta_{0})^{\transp}]$ matrices are full column rank and positive definite respectively, then all GEL estimators share the same first order large sample properties, see, e.g., \citet[Theorems 3.1 and 3.2]{newey2004}, i.e., $n^{1/2}(\hat{\beta}
-\beta _{0})\xrightarrow{d}N(0,\Sigma)$, achieving the semiparametric efficiency lower bound $\Sigma =(G^{\transp}\Omega^{-1}G)^{-1}$, \citet[Theorem 2]{chamberlain1987}. Furthermore, if the additional Supplement \ref{Supp:Proofs}: Assumption  \ref{Assumption.GEL.3} is imposed, defining $H=\Sigma G^{\transp}\Omega^{-1}$ and $P=\Omega^{-1}-\Omega^{-1}G(G^{\transp}\Omega^{-1}G)^{-1}G^{\transp}\Omega^{-1}$, the second order bias of $\hat{\beta}$ is 
$\E[\hat{\beta}]-\beta _{0}=n^{-1}H\zeta_{\lambda}+\bigO(n^{-2})$,
where 
\begin{equation}\label{Eq:GEL.stochastic.expansion.2nd.order.bias.term}
\zeta_{\lambda} = -a+\E[G_{i}Hg_{i}]+c_{\rho}\E[g_{i}g_{i}^{\transp}Pg_{i}],
\end{equation}
with $c_{\rho}=1+\rho_{3}/2$ and $a$ a $d_{g}$-vector with elements $a^{j}=\trace(\Sigma\E[\partial^{2}g^{j}(z,\beta_{0})/\partial\beta\partial\beta^{\transp}])$, $j=1,\ldots,d_{g}$; see \citet[Theorem 4.2]{newey2004}.

\begin{remark}
The validity of the higher order bias and variance calculations, and hence the validity of the results reported below can be
formally justified by that of an Edgeworth expansion of order $\littleO(n^{-1})$ for the distribution of GEL parameter estimators. If $z$ is continuously distributed, appropriate conditions may be found in \citet{bhattacharya1978} for general smooth functions of sample moments and \citet{kundhi2012} for Edgeworth expansions for (G)EL estimators. If some of
the elements of $z$ are discretely distributed, \citet{jensen1989} provides appropriate conditions.
\end{remark}

For given $\beta$, the auxiliary parameter estimator is defined by $\lambda(\beta)=\argmax_{\lambda\in\Lambda_{n}(\beta)}P_{n}(\beta,\lambda)$. Whenever the constraint in $\lambda\in\Lambda_{n}(\beta)$ is not binding, $\lambda(\beta)$ solves the first-order conditions \\ $n^{-1}\sum_{i=1}^{n}\rho^{(1)}(\lambda(\beta)^{\transp}g_{i}(\beta))g_{i}(\beta)=0$.
The GEL implied probabilities are then 
\begin{equation*}
\pi_{i}(\beta)=\frac{\rho^{(1)}(\lambda(\beta)^{\transp}g_{i}(\beta))}{n^{-1}\tsum_{j=1}^{n}\rho^{(1)}(\lambda(\beta)^{\transp}g_{j}(\beta))}, \quad i=1,\ldots,n.
\end{equation*}
The sample moment constraint $\sum_{i=1}^{n}\pi_{i}(\beta)g_{i}(\beta)=0$ holds whenever the first order conditions for $\lambda
(\beta)$ hold. In what follows, $\hat{\pi}_{i}=\pi_{i}(\hat{\beta})$, $i=1,\ldots,n$, corresponds to the solution $\hat{\lambda}=\lambda(\hat{\beta})$, and, if $\beta_{0}$ is known, $\tilde{\pi}_{i}=\pi_{i}(\beta_{0})$, $i=1,\ldots,n$, with auxiliary parameter estimator $\tilde{\lambda}=\lambda(\beta_{0})$. The generic notation $\pi_{i}$, $i=1,\ldots,n$, is used whenever the distinction is unnecessary. 

\begin{remark}
Properties of the GEL implied probabilities relevant to the subsequent developments are summarized in Supplement \ref{Supp:Proofs}: Lemmas \ref{Lemma.Implied.probabilities} and \ref{Lemma.Implied.probabilities.moments}. Although $\pi_{i}(\beta)$, $i=1,\ldots,n$, sum to unity and are positive if $\pi_{i}(\beta)g_{i}(\beta)$ is small uniformly in $i$, they are not guaranteed to be non-negative. The shrinkage estimator $\pi_{i}^{\varepsilon}=(\pi_{i}+\varepsilon_{n})/\sum_{j=1}^{n}(\pi_{j}+\varepsilon_{n})$, $i=1,\ldots,n$, where $\varepsilon_{n}=-\min[\min_{1\leq i\leq n}\pi_{i},0]$, see \citet{antoine2007}, \citet{smith2011},
ensures non-negativity $\pi_{i}^{\varepsilon}\geq0$, $i=1,\ldots,n$, and $\sum_{i=1}^{n}\pi_{i}^{\varepsilon}=1$. Alternative solutions relevant to probability density and distribution function estimation respectively are discussed in Sections \ref{Sec:GEL.based.KDE} and \ref{Sec:GEL.based.CDF.KE}. 
\end{remark}

\begin{remark}
The implied probabilities  were given for EL by \citet{owen1988}, for ET by \citet{kitamura1997}, for quadratic $\rho(\cdot)$ by \citet{back1993}, and for the general case in the 1992 working paper version of \citet{brown2002}; 
see also \citet{smith1997}. For any function $a(z,\beta)$ and GEL estimator $\hat{\beta}$ the implied probabilities can be used to form a semiparametrically efficient estimator $\sum_{i=1}^{n}\hat{\pi}_{i}a(z_{i},\hat{\beta})$ of $\E[a(z,\beta_{0})]$ as in \citet{brown1998}.
\end{remark}

\section{GEL-Based Density Estimation}\label{Sec:GEL.based.KDE}

Suppose the p.d.f. $f(\cdot)$ of the scalar random variable $u=u(z,\beta _{0})$ is of interest,  where the scalar function $u:\mathcal{Z}\times\mathcal{B}\mapsto\mathcal{U}\subseteq\R$  is known up to the parameter vector $\beta_{0}$. 

Let $\mathcal{N}$ denote an open neighbourhood of $\beta_{0}$.

\begin{assumption}\label{Assumption.reformulation.bijection}
For all $\beta\in\mathcal{N}$ there exists a function $v:\mathcal{Z}\times\mathcal{B}\mapsto\mathcal{V}\subseteq\R^{d_{z}-1}$ such that the vector of functions $(u(z,\beta),\: v(z,\beta)^{\transp})^{\transp}$ is a bijection between $\mathcal{Z}$ and $\mathcal{U}\times \mathcal{V}$.
\end{assumption}

\begin{remark}
Equivalently Assumption \ref{Assumption.reformulation.bijection} may be restated as requiring that for every $\beta\in \mathcal{N}$ there exists a bijection between $z$ and some $d_{z}$-vector $w=w(z,\beta)$ such that, given $\{w^{j}(z,\beta)\}_{j=2}^{d_{z}}$, $u(z,\beta)$ and $w^{1}(z,\beta)$ are bijective. That is to say, $z$ may be solved for uniquely given values for $u$, $v$ and $\beta$.
\end{remark}

\begin{remark}
A function $u(z,\beta)$ satisfying Assumption  \ref{Assumption.reformulation.bijection} may be thought of as defining a generalised residual in the sense of \citet{cox1968} and \citet{loynes1969}, with $\hat{u}_{i}=u(z_{i},\hat{\beta})$, $i=1,\ldots,n$, the estimated residuals. Of course, other possibilities of interest are included, e.g., estimating the density of an element of $z$ subject to the extra information available in the moment condition \eqref{Eq:GEL.constraints}.
\end{remark}

\subsection{Known \texorpdfstring{$\beta_{0}$}{beta}}
Suppose that $u_{i}=u(z_{i},\beta _{0})$, $i=1,\ldots,n$, are observed. Then the classical kernel density estimator for the p.d.f. $f$ of $u=u(z,\beta_{0})$ can be employed; \textit{viz}. 
\begin{equation} \label{Eq:gelkde.known.beta.unweighted}
\tilde{f}(u)  = n^{-1}\tsum_{i=1}^{n}k_{b}(u-u_{i}),
\end{equation}
where $k_{b}(x)=k(x/b)/b$, $k(\cdot)$ is a kernel function and $b=b_{n}>0$ is a bandwidth sequence; see \citet{rosenblatt1956} and \citet{parzen1962}. The estimator $\tilde{f}$ \eqref{Eq:gelkde.known.beta.unweighted} will serve as a benchmark for later comparisons.

The properties of $\tilde{f}$ are well known and can be formally established under different combinations of smoothness and integrability conditions on the kernel $k$ and density $f$; see, e.g., \citet[Section 2.1]{rao1983}. A standard set of such conditions is given in Assumption \ref{Assumption.KDE.default} below. If $k$ is square integrable, but not absolutely integrable, as is the case for the sinc kernel, conditions such as those in \citet[Theorem 1.5]{tsybakov2009} can be imposed.

Let $R(k)=\int_{-\infty}^{\infty}k(x)^{2}\mathrm{d}x$  for any square integrable function $k$; the limits of integration are omitted whenever there is little scope for confusion. Also let $f^{(j)}(u)=\mathrm{d}^{j}f(u)/\mathrm{d}u^{j}$ for any $j$th order differentiable function $f$. 

\begin{assumption}\label{Assumption.KDE.default} 
\hspace*{-0.5em}(a)(i) $\sup_{-\infty<x<\infty}\abs{k(x)}<\infty$, $\int\abs{k(x)}\mathrm{d}x<\infty$, 
$\int k(x)\mathrm{d}x=1$, and $\lim_{\abs{x}\to\infty}\abs{xk(x)} =0$; 
(ii) $k$ is a $(2r)$th order kernel, i.e., an even function such that, for some $r\geq1$, $\mu_{0}(k)=1$, $\mu_{j}(k)=0$, $j=1,\ldots,2r-1$, and $\mu_{2r}(k)<\infty$,  where $\mu_{j}(k)=\int x^{j}k(x)\mathrm{d}x$; (iii)  $R(k)<\infty$;
(b) $f(\cdot)$ is $s$ times continuously differentiable and $R(f^{(j)})<\infty$, $j=0,1,\ldots,s$. 
(c) as $n\to\infty$, $b\to0$ and $nb\to\infty$.
\end{assumption}

\begin{remark}
If Assumption \ref{Assumption.KDE.default}(a)(i) holds, then by Supplement \ref{Supp:Proofs}: Lemma \ref{Lemma:convergence}, $\E[\tilde{f}(u)]\to f(u)$ as $b\to0$ at all points  $u$ of continuity of $f$ and if, in addition, Assumption \ref{Assumption.KDE.default}(c) holds, then the mean squared error (MSE), $\MSE[\tilde{f}(u)]=\E[(\tilde{f}(u)-f(u))^{2}]\to 0$ as $n\to\infty$; see, e.g., \citet{parzen1962}.
\end{remark}

\begin{remark}
Higher order approximations to $\MSE[\tilde{f}(u)]$ can be obtained if $f$ is sufficiently smooth. See, e.g.  \citet[Theorem 2.1.5]{rao1983}, \citet[Section 2.8]{wand1995} or \citet[Section 2.4.3]{pagan1999}. The idea of using higher order kernels as a bias reduction technique originates at least as far back as \citet{bartlett1963}.
\end{remark}

Let $1\leq r<\infty$. Suppose that Assumptions \ref{Assumption.KDE.default}(a)(ii), \ref{Assumption.KDE.default}(b) with $s=2r+2$, \ref{Assumption.KDE.default}(c) together with $\mu_{2r+2}(k)<\infty$ and $\int x^{2}k(x)^{2}\mathrm{d}x<\infty$ hold. Then 
\begin{align*}
\E[\tilde{f}(u)]   &=f(u)+(2r)!^{-1}\mu_{2r}(k)f^{(2r)}(u)b^{2r}+\bigO(b^{2r+2}),\\
\Var[\tilde{f}(u)] &=(nb)^{-1}R(k)f(u)-n^{-1}f(u)^{2}+\bigO(n^{-1}b).
\end{align*}
Hence, 
\begin{equation} \label{KDE.observed.MSE.asy.approximation}
\MSE[\tilde{f}(u)] = (nb)^{-1}R(k)f(u)+(2r)!^{-2}\mu_{2r}(k)^{2}f^{(2r)}(u)^{2}b^{4r}-n^{-1}f(u)^{2}+\bigO(b^{4r+2}\vee n^{-1}b).
\end{equation}

\begin{remark}
If  $k$ is a $(2r)$th order kernel and Assumption \ref{Assumption.KDE.default}(b) holds with $s=2r$, the remainder term in 
$\E[\hat{f}(u)]$ is $\littleO(b^{2r})$. The $\sim n^{-1}$ term is kept explicit with $\bigO$ remainder for reasons that will become apparent below.
\end{remark}

The mean integrated squared error (MISE), $\MISE[\tilde{f}]=\E[\int(\tilde{f}(u)-f(u))^{2}\mathrm{d}u]$, is a commonly used global measure of performance. The optimal bandwidth is then defined as that value of $b>0$ minimising MISE, or an approximation thereof. In particular, the asymptotically optimal bandwidth is defined as the value $b^{\ast}$ minimising the two leading terms in the expansion 
\begin{equation} \label{KDE.observed.MISE.asy.approximation}
\MISE[\tilde{f}(\cdot;b)] = (nb)^{-1}R(k)+(2r)!^{-2}\mu_{2r}(k)^{2}R(f^{(2r)})b^{4r}-n^{-1}R(f)+\bigO(b^{4r+2}\vee n^{-1}b), 
\end{equation}
i.e., $b^{\ast}=cn^{-1/(4r+1)}$ where  $c=[(2r)!^{2}R(k)/(4r\mu_{2r}(k)^{2}R(f^{(2r)})]^{1/(4r+1)}$. The asymptotically optimal MISE is thereby 
\begin{equation*} 
\MISE[\tilde{f}(\cdot;b^{\ast})] = 
n^{-4r/(4r+1)}c^{-1}R(k)\left[1 +(4r)^{-1}\right] -n^{-1}R(f)+\bigO(n^{-1-1/(4r+1)}).
\end{equation*}

\begin{remark}
If $k$ is of order greater than two, it necessarily takes negative values. Hence $\tilde{f}$ \eqref{Eq:gelkde.known.beta.unweighted} itself need not be a density function. Note, however, that the positive part estimator, $\tilde{f}^{+}(u)=\max[\tilde{f}(u),0]$ has MSE at most equal to $\MSE[\tilde{f}(u)]$. Further modifications that ensure integration to unity can be applied as described in \citet{glad2003}.
\end{remark}

The GEL-based kernel density estimator incorporates the information embedded in the moment restriction \eqref{Eq:GEL.constraints} replacing the sample EDF weights $n^{-1}$ in the construction of $\tilde{f}(u)$ \eqref{Eq:gelkde.known.beta.unweighted} by the
implied probabilities $\tilde{\pi}_{i}$, $i=1,\ldots,n$; \textit{viz}. 
\begin{equation} \label{Eq:gelkde.known.beta.gel.weighted}
\tilde{f}_{\rho}(u)  = \tsum_{i=1}^{n}\tilde{\pi}_{i}k_{b}(u-u_{i}) 
\end{equation}

\begin{remark}
The GEL-based kernel density estimator $\tilde{f}_{\rho}(u)$ \eqref{Eq:gelkde.known.beta.gel.weighted} is the estimator of $f(u)$ obtained from the revised GEL criterion $\sum_{i=1}^{n}[\rho(\eta (f(u)-k_{b}(u-u_{i}))+\lambda^{\transp}g_{i}(\beta))-\rho(0)]/n$ with the implicit moment condition $\E[k_{b}(u-u_{i})]=f(u)$ and associated auxiliary parameter $\eta$; see \citet[Section 3]{smith2011}.
\end{remark}

\begin{remark}
If the validity of the moment restriction \eqref{Eq:GEL.constraints} is in doubt, a pre-test can be conducted using the GEL-based criterion \eqref{Eq:GEL.saddle.point.problem} paralleling the classical likelihood ratio test; see, e.g., 
\citet{kitamura1997}, \citet{imbens1998} and \citet{smith1997,smith2011}.  For example, under the null hypothesis that \eqref{Eq:GEL.constraints} holds for some unique $\beta_{0}\in\mathcal{B}$, the normalised GEL criterion \eqref{Eq:GEL.saddle.point.problem} evaluated at the estimated parameters, $2nP_{n}(\hat{\beta},\hat{\lambda})$, is asymptotically chi-square distributed with $d_{g}-d_{\beta}$ degrees of freedom. The parametric null hypothesis of known $\beta_{0}=\beta^{0}$ can be tested at the $\alpha$ level using the critical region $\{2nP_{n}(\beta^{0},\tilde{\lambda})\geq \chi_{d_{\beta}}^{2}(\alpha)\}$.
\end{remark}

To describe the properties of GEL-based kernel density estimator $\tilde{f}_{\rho }(u)$ \eqref{Eq:gelkde.known.beta.gel.weighted}, the shorthand notation, e.g., $\E[g_{i}|u]=\E[g(z,\beta_{0})|\{z: u(z,\beta_{0})=u\}]$, for conditional expectations given $u$ is adopted. 

\begin{theorem}\label{Thm:GEL.KDE.MSE.known.beta}
If Supplement \ref{Supp:Proofs}: Assumptions \ref{Assumption.GEL.1}--\ref{Assumption.GEL.3} and \ref{Assumption.KDE.default}(a)(i) and (c) are satisfied, then $\tilde{f}_{\rho}(u)=\tilde{f}(u)+\littleO[p](1)$ for all $u$ such that $f(u)<\infty$. 
If, in addition, Assumption \ref{Assumption.reformulation.bijection} is satisfied, then 
\begin{align}\label{Thm:GEL.KDE.MSE.known.beta.Bias}
\mkern-10mu\E[\tilde{f}_{\rho}(u)] & = \E[\tilde{f}(u)]
+  n^{-1}c_{\rho}\left(-\E[g_{i}^{\transp}\Omega^{-1}g_{i}|u]
+ \E[g_{i}^{\transp}\Omega^{-1}g_{i}g_{i}^{\transp}]\Omega^{-1}\E[g_{i}|u]
+d_{g}\right)f(u) + \littleO(n^{-1}),  \\
\label{Thm:GEL.KDE.MSE.known.beta.Variance}
\mkern-10mu\Var[\tilde{f}_{\rho}(u)]  &= \Var[\tilde{f}(u)]  - n^{-1}\E[g_{i}|u]^{\transp}\Omega^{-1}\E[g_{i}|u]f(u)^{2}
+ \littleO(n^{-1}). 
\end{align}
\end{theorem}
Thus, the estimators $\tilde{f}$ and $\tilde{f}_{\rho}$ are asymptotically first-order equivalent, and the asymptotically optimal bandwidth for $\tilde{f}_{\rho}$ is identical to that of $\tilde{f}$, i.e., $b^{\ast}$.

Whenever $c_{\rho}=0$, as is the case for (G)EL with $\rho_{3}=-2$ , e.g., EL, the $n^{-1}$ bias term in \eqref{Thm:GEL.KDE.MSE.known.beta.Bias} vanishes. In general, provided the bandwidth does not go to zero faster than $n^{-1/(2r)}$, and certainly when $b=b^{\ast}\sim n^{-1/(4r+1)}$, this bias term is at most third order. Its contribution to MISE is via the integrated squared bias (ISB) 
\begin{align*}
\ISB[\tilde{f}_{\rho}]  & = \ISB[\tilde{f}] 
+ n^{-1}b^{2r}c_{\rho}2(2r)!^{-1}\mu_{2r}(k)\int\left(-\E[g_{i}^{\transp}\Omega^{-1}g_{i}|u]\right. \\
&\quad\left.+ \E[g_{i}^{\transp}\Omega^{-1}g_{i}g_{i}^{\transp}]\Omega^{-1}\E[g_{i}|u] +d_{g}\right)f^{(2r)}(u)f(u)\mathrm{d}u
+\littleO(n^{-1}b^{2r}\vee n^{-2}),
\end{align*}
with the $\bigO(n^{-1}b^{2r})$ term generally non-zero and either positive or negative. With the asymptotically optimal bandwidth, $n^{-1}(b^{\ast})^{2r}\sim n^{-3/2+1/(8r+2)}$, which approaches $n^{-3/2}$ arbitrarily closely as $r$ increases, whereas the leading terms in $\MISE[\tilde{f}(\cdot;b^{\ast})]$ becomes arbitrarily close to $n^{-1}$.

As long as $\E[g_{i}|u]\neq 0$, the GEL-based estimator $\tilde{f}_{\rho}$ enjoys a second-order reduction in variance due to the $n^{-1}$ term in \eqref{Thm:GEL.KDE.MSE.known.beta.Variance}, which does not depend on the choice of GEL carrier function $\rho(\cdot)$. Hence 
\begin{equation*}
\MISE[\tilde{f}_{\rho}]=\MISE[\tilde{f}] - n^{-1}\tint \E[g_{i}|u]^{\transp}\Omega^{-1}\E[g_{i}|u]f(u)^{2}\mathrm{d}u + \littleO(n^{-1}).
\end{equation*}
While this reduction is negligible asymptotically, the leading term in $\MISE[\tilde{f}]$ approaches zero only a little more slowly than $n^{-1}$. Hence the effect could be substantial in small samples.

\subsection{Unknown \texorpdfstring{$\beta_{0}$}{beta 0}}

Suppose now that $\beta_{0}$ is unknown. Then, after substitution of the estimators $\hat{u}_{i}=u(z_{i},\hat{\beta})$ for $u_{i}$, $i=1,\ldots,n$, in $\tilde{f}$ and $\tilde{f}_{\rho}$ in \eqref{Eq:gelkde.known.beta.unweighted} and \eqref{Eq:gelkde.known.beta.gel.weighted}, the analogous estimators of $f(u)$ are 
\begin{align}
\label{Eq:gelkde.estimated.beta.unweighted}
\hat{f}(u) & = n^{-1}\tsum_{i=1}^{n}k_{b}(u-\hat{u}_{i}), \\
\label{Eq:gelkde.estimated.beta.gel.weighted}
\hat{f}_{\rho}(u) & = \tsum_{i=1}^{n}\hat{\pi}_{i}k_{b}(u-\hat{u}_{i}),
\end{align}
respectively. Because $u_{i}$, $i=1,\ldots,n$, are not directly observable, the behaviour of the estimation error $\hat{u}_{i}-u_{i}$, $i=1,\ldots,n$, needs to be constrained with additional restrictions imposed on $k$ and $b$. Assumption \ref{Assumption.U.and.K.diff.int.0} gives a set of mild sufficient conditions, see, e.g., \citet{vanryzin1969} and \citet{ahmad1992}; similar conditions have also been considered
in, e.g., \citet{cheng2005} and \citet{kiwitt2008}. 

\begin{assumption}\label{Assumption.U.and.K.diff.int.0}
(a) $k$ is H\"{o}lder continuous with exponent $0<\tau\leq1$; 
(b) there exists $d(z)\geq0$ with $\E[d(z)^{\tau}]<\infty$ such that, for some $0<\alpha\leq1$, $\abs{u(z,\beta)-u(z,\beta_{0})}\leq d(z)\norm{\beta-\beta_{0}}^{\alpha}$ for all $z$ and for all $\beta\in\mathcal{N}$; 
(c) $b\to0$ and $n^{\alpha\tau/2}b^{1+\tau}\to\infty$ as $n\to\infty$.
\end{assumption}

The uniform $\alpha$-H\"{o}lder condition Assumption \ref{Assumption.U.and.K.diff.int.0}(b) on $u(z,\beta)$, also known as a Lipschitz condition of order $\alpha$, is an appropriate way to quantify the `degree of continuity' of $u(z,\beta)$; see \citet[pp.42--45]{zygmund2003}. Many kernels used in practice are Lipschitz continuous, and hence satisfy Assumption \ref{Assumption.U.and.K.diff.int.0}(a) with $\tau=1$. For example, a kernel that satisfies Assumption \ref{Assumption.U.and.K.diff.int.0}(a) for any $0<\tau\leq\gamma$  but not for $\gamma<\tau\leq1$ is $k(x) = (1+\gamma)(1-\abs{x})^{\gamma}/2$ if $\abs{x}\leq 1$ and $0$ otherwise, yielding the Bartlett (triangular) kernel if $\gamma=1$. Assumption \ref{Assumption.U.and.K.diff.int.0}(c) is important as it prevents the bandwidth from being too small. Intuitively, if $b$ is very  small, the kernel $k_{b}(u-\hat{u}_{i})$ is very narrowly centered around the incorrect value $\hat{u}_{i}$ potentially  excluding the true value $u_{i}$; see, e.g., \citet[Figure 2.5]{silverman1986} for a generic illustration. Assumption \ref{Assumption.U.and.K.diff.int.0}(c) requires $nb^{4}\to\infty$ regardless of the values of $\tau$ and $\alpha$ and $b=n^{-1/4}$ is the fastest rate achievable when $\alpha=\tau=1$. Note that the optimal bandwidth $b^{\ast}$ is excluded if $[\alpha(4r+1)-2]\tau <2$.

Under these conditions, Theorem \ref{Thm:GEL.KDE.MSE.estimated.beta.Consistency} establishes that the differences between the kernel density estimators $\hat{f}$ \eqref{Eq:gelkde.estimated.beta.unweighted} and $\hat{f}_{\rho }$ \eqref{Eq:gelkde.estimated.beta.gel.weighted} and their counterparts $\tilde{f}$ \eqref{Eq:gelkde.known.beta.unweighted} and $\tilde{f}_{\rho}$ \eqref{Eq:gelkde.known.beta.gel.weighted} based on observable $u_{i}$, $i=1,\ldots,n$, are negligible asymptotically.

\begin{theorem}\label{Thm:GEL.KDE.MSE.estimated.beta.Consistency}
If Supplement \ref{Supp:Proofs}: Assumptions \ref{Assumption.GEL.1}--\ref{Assumption.GEL.3} and \ref{Assumption.U.and.K.diff.int.0} are satisfied, then 
$\hat{f}(u)=\tilde{f}(u)+\littleO[p](1)$ and 
$\hat{f}_{\rho}(u) = \sum_{i=1}^{n}\hat{\pi}_{i}k_{b}(u-u_{i})+\littleO[p](1)$
for all $u$. 
If, in addition, Assumption \ref{Assumption.KDE.default}(a)(i) holds, $\hat{f}_{\rho}(u) = \tilde{f}(u)+\littleO[p](1)$ a.e.
\end{theorem}

To obtain higher order expansions for the mean and variance of $\hat{f}(u)$ \eqref{Eq:gelkde.estimated.beta.unweighted} and $\hat{f}_{\rho}(u)$ \eqref{Eq:gelkde.estimated.beta.gel.weighted} requires a further strengthening of the assumptions. Let $\nabla u(z,\beta)$ and $\nabla^{2}u(z,\beta)$ denote respectively the $d_{\beta}$-vector and $d_{\beta}\times d_{\beta}$ matrix of the first and second derivatives of $u(z,\beta)$ with respect to $\beta$. Also let $\nabla u_{i}=\nabla u(z_{i},\beta _{0})$ and $\nabla^{2}u_{i}=\nabla^{2}u(z_{i},\beta _{0})$.

\begin{assumption}\label{Assumption.U.and.K.diff.int} 
(a) $k$ is twice differentiable and $k^{(2)}$ is H\"{o}lder continuous with exponent $0<\tau\leq1$, $k$, $k^{(1)}$, and $k^{(2)}$ are absolutely integrable;  $\lim_{\abs{x}\to\infty}\abs{x^{s}k^{(s-1)}(x)}=0$, $s=1,2,3$, and
$\int k(x)\mathrm{d}x=1$; 
(b) $u(z,\beta)$ is twice differentiable for all $\beta\in\mathcal{N}$, 
$\E[\norm{\nabla u_{i}}^{4}]<\infty$, $\E[\norm{\nabla^{2}u_{i}}^{4}]<\infty$, and there exists  $d(z)\geq0$ with $\E[d(z)^{4}]<\infty$ such that, for some $0<\alpha\leq1$, $\norm{\nabla^{2}u(z,\beta)-\nabla^{2}u(z,\beta_{0})}\leq d(z)\norm{\beta-\beta_{0}}^{\alpha}$ for all $z$ and for all $\beta\in\mathcal{N}$;
(c) $b\to0$ as $n\to\infty$, $n^{\tau/2}b^{3+\tau}\to\infty$, and $n^{\alpha/2}b^{5/4}\to\infty$; 
(d)(i) $f$ is twice differentiable;
(ii) $\E[\nabla u_{i}|u]$, $\E[\nabla^{\transp}u_{i}Hg_{i}|u]$,	and $\E[\nabla^{2}u_{i}|u]$ are differentiable in $u$ and 
$\E[\nabla u_{i}\nabla^{\transp}u_{i}|u]$ is twice differentiable in $u$; 
(iii) $\mathrm{d}\{\E[\nabla u_{i}|u] f(u)\}/\mathrm{d}u$, $\mathrm{d}\{\E[\nabla^{\transp}u_{i}Hg_{i}|u]f(u)\}/\mathrm{d}u$, $\mathrm{d}\{\E[\nabla^{2}u_{i}|u]f(u)\}/\mathrm{d}u$, and $\mathrm{d}^{2}\{\E[\nabla u_{i}\nabla^{\transp}u_{i}|u]f(u)\}/\mathrm{d}u^{2}$ are absolutely integrable functions of $u$. 
\end{assumption}

Assumption \ref{Assumption.U.and.K.diff.int}(a)(b) implies Assumption \ref{Assumption.U.and.K.diff.int.0}(a)(b) holds with $\alpha=\tau=1$ with the requirement in Assumption \ref{Assumption.U.and.K.diff.int.0}(c) rendered as $n^{1/2}b^{2}\to\infty$.  Note that Assumption \ref{Assumption.U.and.K.diff.int}(a) also implies Assumption  \ref{Assumption.KDE.default}(a)(i). Assumption \ref{Assumption.U.and.K.diff.int}(d) imposes additional smoothness and integrability conditions on $f$ and $u(z,\beta)$. Assumption \ref{Assumption.U.and.K.diff.int}(c) is much stronger than Assumption \ref{Assumption.U.and.K.diff.int.0}(c) requiring $nb^{8}\to\infty$ regardless of the values of $\tau$ and $\alpha$ thereby prohibiting the asymptotically optimal bandwidth $b^{\ast}$ when $k$ is a second order kernel. For $r\geq2$, $b^{\ast}$ is permissible as long as $\tau>6/(4r-1)$ and $\alpha>5/(8r+2)$. Note that, if $\alpha>5/16$, $n^{\tau/2}b^{3+\tau}\to\infty$ implies $n^{\alpha/2}b^{5/4}\to\infty$.

\begin{theorem}\label{Thm:GEL.KDE.MSE.estimated.beta}
If Supplement \ref{Supp:Proofs}: Assumptions \ref{Assumption.GEL.1}--\ref{Assumption.GEL.3}, \ref{Assumption.reformulation.bijection}, and \ref{Assumption.U.and.K.diff.int} are satisfied, then 
$\E[\hat{f}(u)]=\E[\tilde{f}(u)]+n^{-1}\delta(u)+\littleO(n^{-1})$ 
and $\E[\hat{f}_{\rho}(u)]= \E[\tilde{f}(u)]+n^{-1}\delta(u)+n^{-1}\delta_{\rho}(u)+\littleO(n^{-1})$, 
where
\begin{align}
\notag
\delta(u) & = \mathrm{d}\{\E[\nabla^{\transp}u_{i}Hg_{i}|u]f(u)\}/\mathrm{d}u - \zeta_{\lambda}^{\transp}H^{\transp}[\mathrm{d}\{\E[\nabla u_{i}|u] f(u)\}/\mathrm{d}u]\\
\label{Thm:GEL.KDE.MSE.estimated.beta:delta}
&\quad +\tfrac{1}{2}\trace(\Sigma [\mathrm{d}^{2}\{\E[\nabla u_{i}\nabla^{\transp}u_{i}|u]f(u)\}/\mathrm{d}u^{2}
 -\mathrm{d}\{\E[\nabla^{2}u_{i}|u]f(u)\}/\mathrm{d}u])\\
\intertext{and}
\label{Thm:GEL.KDE.MSE.estimated.beta:delta.rho}
\delta_{\rho}(u)   & = (-c_{\rho}\E[g_{i}^{\transp}Pg_{i}|u] +	c_{\rho}(d_{g}-d_{\beta}) +\zeta_{\lambda}^{\transp}P\E[g_{i}|u] )f(u).
\end{align}
Also
\begin{align}
\notag
\Var[\hat{f}(u)]  &= \Var[\tilde{f}(u)] 
+ n^{-1}[\mathrm{d}\{\E[\nabla u_{i}|u] f(u)\}/\mathrm{d}u]^{\transp}\Sigma[\mathrm{d}\{\E[\nabla u_{i}|u] f(u)\}/\mathrm{d}u]\\
\label{Thm:GEL.KDE.MSE.estimated.beta:VAR.f.hat}
&\quad + n^{-1}2[\mathrm{d}\{\E[\nabla u_{i}|u] f(u)\}/\mathrm{d}u]^{\transp}H\E[g_{i}|u]f(u)  + \littleO(n^{-1}), \\
\label{Thm:GEL.KDE.MSE.estimated.beta:VAR.f.hat.rho}
\Var[\hat{f}_{\rho}(u)] &= \Var[\hat{f}(u)] -n^{-1}\E[g_{i}|u]^{\transp}P\E[g_{i}|u]f(u)^{2} + \littleO(n^{-1}).
\end{align}
\end{theorem}

\begin{remark}
The general conclusion of Theorem  \ref{Thm:GEL.KDE.MSE.estimated.beta} for both bias and variance is identical to that of  Theorem \ref{Thm:GEL.KDE.MSE.known.beta}, i.e., the estimation effects of substituting $\hat{u}_{i}$ for $u_{i}$, $i=1,\ldots,n$, and the GEL implied probabilities $\hat{\pi}_{i}$ for $\tilde{\pi}_{i}$, $i=1,\ldots,n$, are both of order $n^{-1}$. The bias term in $\hat{f}$ induced by estimation is similar to that for $\tilde{f}$ in Theorem \ref{Thm:GEL.KDE.MSE.known.beta} except that $P$ in \eqref{Thm:GEL.KDE.MSE.estimated.beta:delta.rho} 
 replaces $\Omega^{-1}$ in \eqref{Thm:GEL.KDE.MSE.known.beta.Bias} 
 and two extra terms enter via $\zeta_{\lambda}$, \textit{viz}. $-a$ and $\E[G_{i}Hg_{i}]$ in \eqref{Eq:GEL.stochastic.expansion.2nd.order.bias.term}. These latter terms appear in the higher order asymptotic bias $n^{-1}H(-a+\E[G_{i}Hg_{i}])$ for the infeasible GEL estimator  based on the optimal moment indicator vector $G^{\transp}\Omega^{-1}g(z,\beta)$, see \citet[Theorem 4.2]{newey2004}, and are inherited by all GEL estimators. Unlike Theorem \ref{Thm:GEL.KDE.MSE.known.beta} for the known $\beta_{0}$ case, this term no longer vanishes for a particular choice of a carrier function $\rho$. The replacement of $\Omega^{-1}$ by $P$ represents the loss of information occasioned by the estimation of $\beta_{0}$. In a number of cases, the term  $\E[g_{i}|u]^{\transp}P\E[g_{i}|u]$ may vanish, see, e.g., Supplement \ref{Supp:Examples}: Example \ref{Example:GEL.with.constant}. This of course always occurs for an exactly identified model $d_{g}=d_{\beta}$ since $\hat{\pi}_{i}=n^{-1}$ and $\hat{f}_{\rho}$ \eqref{Eq:gelkde.estimated.beta.gel.weighted} and $\hat{f}$ \eqref{Eq:gelkde.estimated.beta.unweighted} are identical. However, see Supplement \ref{Supp:Examples}: Example \ref{Sec.Example:normal.over.GG}, in general $\hat{f}_{\rho}$ may still enjoy a second-order reduction in variance due to the systematic use of overidentifying moment information \eqref{Eq:GEL.constraints}.
\end{remark}

The extra bias term $\delta (u)$ \eqref{Thm:GEL.KDE.MSE.estimated.beta:delta} for $\hat{f}_{\rho}$ and those terms appearing in $\Var[\hat{f}(u)]$ \eqref{Thm:GEL.KDE.MSE.estimated.beta:VAR.f.hat} primarily arise due to the substitution of $\hat{u}_{i}$ for $u_{i}$, $i=1,\ldots,n$. Supplement \ref{Supp:Examples}: Examples \ref{Example:reg.on.a.const} and \ref{Example:GEL.with.constant} examine these terms in more detail for regression on a constant and (G)EL with a constant and zero mean condition respectively.
Here, although $\int [\mathrm{d}\{\E[\nabla u_{i}|u] f(u)\}/\mathrm{d}u]^{\transp}\Sigma[\mathrm{d}\{\E[\nabla u_{i}|u] f(u)\}/\mathrm{d}u]\mathrm{d}u$ is non-negative, the term $\int [\mathrm{d}\{\E[\nabla u_{i}|u] f(u)\}/\mathrm{d}u]^{\transp}H\E[g_{i}|u]f(u) \mathrm{d}u$  can be negative, as can be the ISB term due to the additional $\delta(u)$ \eqref{Thm:GEL.KDE.MSE.estimated.beta:delta}.

\subsection{Bias Correction}
While the contribution from the $n^{-1}$ bias terms to MISE is of a lower order than the contribution from the variance terms, the effect of bias can be substantial in small and moderate samples, potentially offsetting any reduction in variance. The direction of the bias cannot of course be known  \textit{a priori}. Hence it may be advisable to bias-correct the density
estimates by estimating and subtracting the $n^{-1}$ bias term.

To be more specific, the bias-corrected estimates are defined as 
\begin{align*}
\hat{f}^{bc}(u) & = \hat{f}(u) - n^{-1}\hat{\delta}(u) \\
\intertext{and}
\hat{f}_{\rho}^{bc}(u) & = \hat{f}_{\rho}(u) - n^{-1}\hat{\delta}(u) - n^{-1}\hat{\delta}_{\rho}(u),
\end{align*}
where $\hat{\delta}(u)$ and $\hat{\delta}_{\rho}(u)$ are suitable (asymptotically) unbiased estimators of $\delta (u)$ \eqref{Thm:GEL.KDE.MSE.estimated.beta:delta} and $\delta_{\rho }(u)$ \eqref{Thm:GEL.KDE.MSE.estimated.beta:delta.rho}. The implied probabilities $\hat{\pi}_{i}$, $i=1,\ldots,n$, can be used to obtain efficient estimators of the component quantities entering $\delta(u)$ and $\delta_{\rho }(u)$ with the modifications described in \citet{glad2003} applied to ensure that the bias-corrected estimate is a density.

\begin{remark}
When $\beta_{0}$ is known, bias-correction requires the estimation of the $n^{-1}$ term in \eqref{Thm:GEL.KDE.MSE.known.beta.Bias} unless $c_{\rho }=0$, i.e., $\rho_{3}=-2$.
\end{remark}

\section{GEL-Based Distribution Function Estimation}\label{Sec:GEL.based.CDF.KE}
The results for distribution function estimation parallel those given in Section \ref{Sec:GEL.based.KDE} for density estimation but can be shown to hold under much weaker conditions, and so are given here separately.

\subsection{Known \texorpdfstring{$\beta_{0}$}{beta}}
When $u_{i}$, $i=1,\ldots,n$, are observed, the c.d.f. $F$ of $u(z,\beta_{0})$ can be estimated by 
\begin{equation}\label{Eq:gelkde.known.beta.unweighted.cdf}
\widetilde{F}(u)=n^{-1}\tsum_{i=1}^{n}K((u-u_{i})/b), 
\end{equation}
with $K(u)=\int_{-\infty}^{u}k(x)\mathrm{d}x$; see \citet{nadaraya1964b} and \citet{watson1964}. The kernel distribution function estimator \eqref{Eq:gelkde.known.beta.unweighted.cdf}  can be obtained by integrating \eqref{Eq:gelkde.known.beta.unweighted} or motivated as a smoothed version of the EDF.

Assumption \ref{Assumption.KDE.default}(a)(i) is sufficient for $\widetilde{F}$ to be an asymptotically unbiased and consistent estimator of $F$ at all continuity points of $F$ if $b\to0$ as $n\to\infty$. In addition, if $F$ is continuous then $\widetilde{F}$ converges to $F$ uniformly with probability $1$ (w.p.$1$.); see \citet{yamato1973}. If $k$ satisfies Assumption \ref{Assumption.KDE.default}(a)(ii) with $\mu_{2r+2}(k)<\infty$ for some $r\geq1$, $f$ satisfies Assumption \ref{Assumption.KDE.default}(b) with $s=2r+1$, and $b\to0$
as $n\to\infty$ (Assumption \ref{Assumption.KDE.default}(c) is not required here), then 
\begin{align*}
\E[\widetilde{F}(u)] & = F(u) + (2r)!^{-1}\mu_{2r}(k)f^{(2r-1)}(u)b^{2r}+\bigO(b^{2r+2}), \\
\Var[\widetilde{F}(u)] & = n^{-1}F(u)(1- F(u)) - n^{-1}bf(u)\psi(k) +\bigO(n^{-1}b^{2+\I{r>1}}), 
\end{align*}
where $\psi(k) = 2\int xK(x)k(x)\mathrm{d}x$. Hence 
\begin{equation}\label{KDFE.observed.MISE.asy.approximation}
\MISE[\widetilde{F}(\cdot;b)] = n^{-1}V_{F} - n^{-1}b\psi(k) 
+(2r)!^{-2}\mu_{2r}(k)^{2}R(f^{(2r-1)})b^{4r} +\bigO(n^{-1}b^{2+\I{r>1}}\vee b^{4r+2}),
\end{equation}
where $V_{F}=\tint F(u)(1- F(u))\mathrm{d}u$. 

Provided $\psi(k)>0$, the asymptotically optimal bandwidth minimising the leading terms in \eqref{KDFE.observed.MISE.asy.approximation} is $b^{\ast}=\varsigma n^{-1/(4r-1)}$, where 
 $\varsigma=[(2r)!^{2}\psi(k)/(4r\mu_{2r}(k)^{2}R(f^{(2r-1)}))]^{1/(4r-1)}$, and the asymptotically optimal MISE is 
\begin{equation*}
\MISE[\widetilde{F}(\cdot;b^{\ast})] = n^{-1}V_{F}
 - \varsigma\psi(k)[1 -(4r)^{-1}]n^{-4r/(4r-1)} +\bigO(n^{-(4r+1+\I{r>1})/(4r-1)}).
\end{equation*}

\begin{remark}
The leading term $n^{-1}V_{F}$ in \eqref{KDFE.observed.MISE.asy.approximation} is the integrated variance and, hence, the MISE of EDF. Thus, whenever $\psi(k)>0$ and $b$ approaches zero at least as fast as $n^{-1/(4r-1)}$, kernel smoothing provides a second order asymptotic improvement in MISE relative to the EDF. Smoothness of the kernel estimates and the reduction in MISE are
the two main reasons to prefer the kernel distribution function estimator \eqref{Eq:gelkde.known.beta.unweighted.cdf} over the EDF. The condition $\psi(k)>0$ is satisfied if $k$ is a symmetric second order kernel, since in this case $\psi(k) = \int K(x)(1-K(x))\mathrm{d}x>0$. Although $\psi(k)$ need not be positive in general, this property holds for certain classes of kernels, including Gaussian kernels of arbitrary order; see \citet{oryshchenko2017}. 
\end{remark}

\begin{remark}
If $k$ is of order greater than two, $K$ is not monotone, and the resultant estimates may not themselves be distribution functions. However, if necessary, the estimates can be corrected by rearrangement; see \citet{chernozhukov2009}. The MISE of the rearranged estimator can be at most equal to, and is often strictly smaller, than the MISE of the original estimator.
\end{remark}

The modified GEL kernel distribution function estimator corresponding to $\tilde{f}_{\rho}$ \eqref{Eq:gelkde.known.beta.gel.weighted} which incorporates the information embedded in the moment restrictions \eqref{Eq:GEL.constraints} is 
\begin{equation} \label{Eq:gelkde.known.beta.gel.weighted.cdf}
\widetilde{F}_{\rho}(u)  = \tsum_{i=1}^{n}\tilde{\pi}_{i}K((u-u_{i})/b) 
\end{equation}

\begin{theorem}\label{Thm:GEL.KDFE.MSE.known.beta}
If Supplement \ref{Supp:Proofs}: Assumptions \ref{Assumption.GEL.1}--\ref{Assumption.GEL.3}  and \ref{Assumption.KDE.default}(a)(i) are satisfied and $b\to0$ as $n\to\infty$, then $\widetilde{F}_{\rho}(u)=\widetilde{F}(u)+\littleO[p](1)$ at all points of continuity of $F$. If, in addition, Assumption \ref{Assumption.reformulation.bijection} is satisfied, then
\begin{equation}\label{Thm:GEL.KDFE.MSE.known.beta.Bias}
\E[\widetilde{F}_{\rho}(u)]  = \E[\widetilde{F}(u)]
+\tfrac{c_{\rho}}{n}\tint_{-\infty}^{u}\left( -\E[g_{i}^{\transp}\Omega^{-1}g_{i}|t] +\E[g_{i}^{\transp}\Omega^{-1}g_{i}g_{i}^{\transp}]\Omega^{-1}\E[g_{i}|t]+d_{g}\right)\mathrm{d}F(t)
+ \littleO(n^{-1}).
\end{equation}
If also $\lim_{\abs{x}\to\infty}\abs{x^{2}k(x)}=0$, then  
\begin{equation}
\label{Thm:GEL.KDFE.MSE.known.beta.Variance}
\Var[\widetilde{F}_{\rho}(u)]  = \Var[\widetilde{F}(u)]  
- n^{-1}[\tint_{-\infty}^{u}\E[g_{i}|t]\mathrm{d}F(t)]^{\transp}\Omega^{-1}[\tint_{-\infty}^{u}\E[g_{i}|t]\mathrm{d}F(t)]
+ \littleO(n^{-1}b).
\end{equation}
\end{theorem}

These results are qualitatively similar to Theorem \ref{Thm:GEL.KDE.MSE.known.beta}, the important difference being that the reduction in variance is now first-order asymptotically, whereas the contribution from the $n^{-1}$ bias term in \eqref{Thm:GEL.KDFE.MSE.known.beta.Bias} to MISE is of order $n^{-1}b^{2r}$. \textit{Ceteris paribus}, the asymptotically optimal c.d.f. bandwidth converges to zero at a faster rate than that for density estimation. Hence the additional bias effect can be
expected to be of less importance.

\subsection{Unknown \texorpdfstring{$\beta_{0}$}{beta}}
When $\beta_{0}$ is unknown, the analogues of $\widetilde{F}$ and $\widetilde{F}_{\rho}$ are 
\begin{align}
\label{Eq:gelkde.estimated.beta.unweighted.cdf}
\widehat{F}(u) & = n^{-1}\tsum_{i=1}^{n}K((u-\hat{u}_{i})/b), \\
\label{Eq:gelkde.estimated.beta.gel.weighted.cdf} 
\widehat{F}_{\rho}(u) & = \tsum_{i=1}^{n}\hat{\pi}_{i}K((u-\hat{u}_{i})/b),
\end{align}
respectively.

\begin{theorem}\label{Thm:GEL.KDFE.MSE.estimated.beta.Consistency}
If Supplement \ref{Supp:Proofs}: Assumptions \ref{Assumption.GEL.1}--\ref{Assumption.GEL.3} and \ref{Assumption.KDE.default}(a)(i) are satisfied, Assumption \ref{Assumption.U.and.K.diff.int.0}(b) holds with $\tau=1$ for some $0<\alpha\leq1$, and $b\to0$ and  $n^{\alpha/2}b\to\infty$ as $n\to\infty$, then 
$\widehat{F}(u)=\widetilde{F}(u)+\littleO[p](1)$, 
$\widehat{F}_{\rho}(u)=\widetilde{F}(u)+\littleO[p](1)$, and 
$\widehat{F}_{\rho}(u) = \sum_{i=1}^{n}\hat{\pi}_{i}K((u-u_{i})/b)+\littleO[p](1)$
for all $u$. 
\end{theorem}

Similar to Theorem \ref{Thm:GEL.KDE.MSE.estimated.beta.Consistency}, Theorem \ref{Thm:GEL.KDFE.MSE.estimated.beta.Consistency} establishes that the differences between $\widehat{F}$ \eqref{Eq:gelkde.estimated.beta.unweighted.cdf} and $\widehat{F}_{\rho}$ \eqref{Eq:gelkde.estimated.beta.gel.weighted.cdf} and their counterparts based on observable $u_{i}$, $i=1,\ldots,n$, are negligible asymptotically. No additional requirements are placed on $k$ beyond the standard conditions in \ref{Assumption.KDE.default}(a)(i) and the restriction on the bandwidth is thus weaker than Assumption \ref{Assumption.U.and.K.diff.int.0}(c).

Higher order expansions similar to those in Theorem \ref{Thm:GEL.KDE.MSE.estimated.beta} may be obtained under the following conditions.

\begin{assumption}\label{Assumption.U.and.K.diff.int.CDF} Suppose Assumption \ref{Assumption.U.and.K.diff.int}(b) holds. 
(a) $k$ is differentiable and $k^{(1)}$ is H\"{o}lder continuous with exponent $0<\tau\leq1$, $k$ and $k^{(1)}$ are absolutely integrable, $\lim_{\abs{x}\to\infty}\abs{x^{2}k(x)}=0$,  $\lim_{\abs{x}\to\infty}\abs{x^{2}k^{(1)}(x)}=0$, 
and  $\int k(x)\mathrm{d}x=1$;
(b) $b\to0$ as $n\to\infty$, $n^{\tau/2}b^{2+\tau}\to\infty$, and $n^{\alpha/2}b^{1/4}\to\infty$;
(c) (i) $f(u)$ and $\E[\nabla u_{i}\nabla^{\transp}u_{i}|u]$ are differentiable in $u$; 
(ii) $\mathrm{d}\{\E[\nabla u_{i}\nabla^{\transp}u_{i}|u]f(u)\}/\mathrm{d}u$ is an absolutely integrable function of $u$. 
\end{assumption}

\begin{theorem}\label{Thm:GEL.KDFE.MSE.estimated.beta}
If Supplement \ref{Supp:Proofs}: Assumptions \ref{Assumption.GEL.1}--\ref{Assumption.GEL.3}, \ref{Assumption.reformulation.bijection}, and \ref{Assumption.U.and.K.diff.int.CDF} are satisfied, then as $n\to\infty$,
$\E[\widehat{F}(u)] = \E[\widetilde{F}(u)]+n^{-1}\Delta(u)+\littleO(n^{-1})$ and 
$\E[\widehat{F}_{\rho}(u)] = \E[\widetilde{F}(u)]+n^{-1}\Delta(u)+n^{-1}\Delta_{\rho}(u)+\littleO(n^{-1})$, where
\begin{align}
\notag
\Delta(u) &=  \E[\nabla^{\transp}u_{i}Hg_{i}|u]f(u)
-\zeta_{\lambda}^{\transp}H^{\transp}\E[\nabla u_{i}|u]^{\transp}f(u)   \\
\label{Thm:GEL.KDFE.MSE.estimated.beta:Delta}
&\quad +\tfrac{1}{2}\trace\left(\Sigma[\mathrm{d}\{\E[\nabla u_{i}\nabla^{\transp}u_{i}|u]f(u)\}/\mathrm{d}u
-\E[\nabla^{2}u_{i}|u]f(u)]\right)\\
\label{Thm:GEL.KDFE.MSE.estimated.beta:Delta.Rho}
\intertext{and}
\Delta_{\rho}(u) & = \tint_{-\infty}^{u}(-c_{\rho}\E[g_{i}^{\transp}Pg_{i}|t]+ c_{\rho}(d_{g}-d_{\beta}) +\zeta_{\lambda}^{\transp}P\E[g_{i}|t] )\mathrm{d}F(t) = \tint_{-\infty}^{u}\delta_{\rho}(t)\mathrm{d}t.
\end{align}
Also 
\begin{align}
\notag
\Var[\widehat{F}(u)] & = \Var[\widetilde{F}(u)] + n^{-1}\E[\nabla u_{i}|u]^{\transp}\Sigma \E[\nabla u_{i}|u]f(u)^{2} \\
\label{Thm:GEL.KDFE.MSE.estimated.beta:Var.F}
&\quad + 2n^{-1}\E[\nabla u_{i}|u]^{\transp}H[\tint_{-\infty}^{u}\E[g_{i}|t]\mathrm{d}F(t)]f(u) +\littleO(n^{-1}),\\
\label{Thm:GEL.KDFE.MSE.estimated.beta:Var.F.rho}
\Var[\widehat{F}_{\rho}(u)] & =  \Var[\widehat{F}(u)]  -n^{-1}[\tint_{-\infty}^{u}\E[g_{i}|t]\mathrm{d}F(t)]^{\transp}P[\tint_{-\infty}^{u}\E[g_{i}|t]\mathrm{d}F(t)]  +\littleO(n^{-1}b).
\end{align}
If, in addition, $\mathrm{d}\{ \E[\nabla u_{i}|u]f(u)\}/\mathrm{d}u$ is absolutely integrable, the remainder term of $\Var[\widehat{F}(u)]$ is  $\littleO(n^{-1}b)$.
\end{theorem}

\begin{remark}
If $\delta (u)$ in Theorem \ref{Thm:GEL.KDE.MSE.estimated.beta} is defined, then $\Delta(u)=\int_{-\infty}^{u}\delta(t)\mathrm{d}t$, but there is no requirement that $\Delta(u)$ is absolutely continuous in Theorem \ref{Thm:GEL.KDFE.MSE.estimated.beta}. Otherwise, the interpretation is exactly the same as in Theorem \ref{Thm:GEL.KDE.MSE.estimated.beta}. In particular, the main qualitative conclusions in Supplement \ref{Supp:Examples}: Examples \ref{Example:GEL.with.constant} and \ref{Sec.Example:normal.over.GG} still hold.
\end{remark}

\section{Simulation Evidence}\label{Sec:Monte.Carlo} 
\subsection{Preliminaries}
Consider the inverse hyperbolic sine (IHS) transformation model 
\begin{equation}\label{Eq:IHS.transformation.linear.in.X.model}
\arsinh(\theta_{0} y)/\theta_{0} = \delta_{0} + \gamma_{0}x + u, \qquad  \E[u|x]=0;
\end{equation}
here $\beta = (\delta,\gamma,\theta)^{\transp}$ and $z=(y,x)^{\transp}$.  The IHS transformation has been proposed in \citet[p.158]{johnson1949} as an alternative to the Box-Cox power transform, $(y^{\lambda}-1)/\lambda$, $y\geq 0$, and developed in \citet{burbidge1988} and \citet{mackinnon1990}; see also, e.g., \citet{ramirez1994}, \citet{brown2015} and the references therein for recent applications in statistics and econometrics, and \citet{tsai2017} for comparisons with other transformations. When $\theta =0$, the IHS transform is defined as the limiting value, $\lim_{\theta\to0}\arsinh(\theta y)/\theta=y$, which corresponds to the Box-Cox transform with $\lambda=1$; when $\theta\neq0$, the shapes of the IHS transforms are similar to those of the Box-Cox with $\lambda<1$. The advantage of the IHS transform is that it is a smooth function of $y\in\R$ and $\theta\in\R$ with values at $\theta=0$ defined as the corresponding limits.

The infeasible optimal instruments in the IHS transformation model \eqref{Eq:IHS.transformation.linear.in.X.model} are 
\begin{equation*}\label{Eq:IHS.transformation.linear.in.X.model.optimal.instruments}
S(x,\beta_{0}) = (-1, \:\: -x, \:\: \E[\tanh(\theta_{0}(u +\delta_{0}+\gamma_{0}x))|x]/\theta_{0}^{2}-(\delta_{0}+\gamma_{0}x)/\theta_{0})^{\transp};
\end{equation*}
see \citet{robinson1991}. The last element of $S(x;\beta_{0})$, $s_{3}(x;\beta_{0})$, depends on the conditional distribution of $u$ given $x$, and, in general, there is little reason to argue for a particular scalar function of  $x$ as a good approximant.  For example, if $u|x\sim N(0,\sigma ^{2})$, based on $\tanh (x)\simeq 2\Phi ((\pi /2)^{1/2}x)-1$ twice, $s_{3}(x;\beta_{0})$
is approximately $\tanh\left(\theta_{0}(\delta_{0}+\gamma_{0}x)/(\pi\theta_{0}^{2}\sigma^{2}/2+1)^{1/2}\right)/\theta_{0}^{2} -(\delta_{0}+\gamma_{0}x)/\theta_{0}$ which suggests the use of odd degree polynomials in $x$ as instruments; other and better approximations are of course available.

In all cases the true parameters are $\delta_{0}=1$, $\gamma_{0}=2$ and $\theta_{0}=0.08$ which yield a signal-to-noise ratio of $\gamma_{0}^{2}/(1+\gamma_{0}^{2})=4/5=0.8$ somewhat more stringent than that of $16/17=0.941$ in \citet[Section 7]{robinson1991}.

\subsection{Design}
Given the uncertainty concerning the conditional distribution $u|x$ the approach adopted here is to simply compare estimators based on moment conditions $\E[g(z,\beta_{0})]=0$ \eqref{Eq:GEL.constraints} where 
\begin{equation*}
g(z,\beta) = u(z,\beta)(1,\: x,\: \ldots, x^{d_{g}-1})^{\transp}, 
\end{equation*}
for $d_{g}=3$ (exactly identified), $4$ and $5$ (over-identified).

Three data generating processes for $(x,u)$ are considered.

\vspace*{\topsep}
\noindent\textsc{Scenario 1.} $x$ and $u$ are distributed as independent standard normal $N(0,1)$; cf. \citet[Section 7, case (ii)]{robinson1991}.

\begin{remark}
Scenario 1 satisfies the conditions of Supplement \ref{Supp:Examples}: Example \ref{Example:GEL.with.constant}. Hence 
$\IVar[\hat{f}_{\rho}]=\IVar[\hat{f}]+\littleO(n^{-1})$ and the relative integrated variance (IVar)
\begin{equation}\label{Eq:MC.Case1.Rel.IVar.hat.f.to.tilde.f}
\IVar[\hat{f}]/\IVar[\tilde{f}] = 1
-\tfrac{b}{4\pi^{1/2}R(k)}
+ \tfrac{b}{\tau^{\transp}D\tau R(k)}\tint\left(\mathrm{d}\{(\tau_{0|u}(u)-\tau_{0})f(u)\}/\mathrm{d}u\right)^{2}\mathrm{d}u
+\littleO(b),
\end{equation}
where $\tau_{0|u}(u) = \E[\tanh(\theta_{0}(u +\delta_{0}+\gamma_{0} x))|u]/\theta_{0}^{2}-(\delta_{0}+u)/\theta_{0}$, 
$\tau_{j}=\E[x^{j}s_{3}(x,\beta_{0})]$, $j=0,1,2,\ldots$, $\tau=(\tau_{0},\tau_{1},\ldots,\tau_{d_{g}-1})^{\transp}$, and 
$D = M^{-1}-\mathop{\mathrm{diag}}(I_{2},0)$ 
 with $M=\{M_{ij}\}_{i,j=1}^{d_{g}}$, $M_{ij}=\E[x^{i+j-2}]$, $i,j=1,\ldots,d_{g}$.
The term $-b/(4\pi^{1/2}R(k))$ does not depend on the number of moment conditions $d_{g}$ and is the asymptotic reduction in integrated variance due to the constraint that the mean of $u$ is zero; see also Supplement \ref{Supp:Examples}: Example \ref{Example:reg.on.a.const}. The second term in $b$ is non-negative and represents the increase in integrated variance due to  estimation of $\gamma _{0}$ and $\theta _{0}$; it decreases as the number of moment condition increases; e.g. for $d_{g}=4$, $5$, $10$, $20$, $\tau^{\transp}D\tau =9.8092$, $9.8514$, $9.9857$ and $9.9859$, respectively.
\end{remark}

\begin{figure}[htbp]\centering\small
\begin{tabular}{@{}cc@{}}
\multicolumn{2}{c}{}\\
(\#2) Skewed unimodal, $\nu=8$ & (\#3) Skewed bimodal, $\nu=8$  \\
\includegraphics[width=0.48\linewidth]{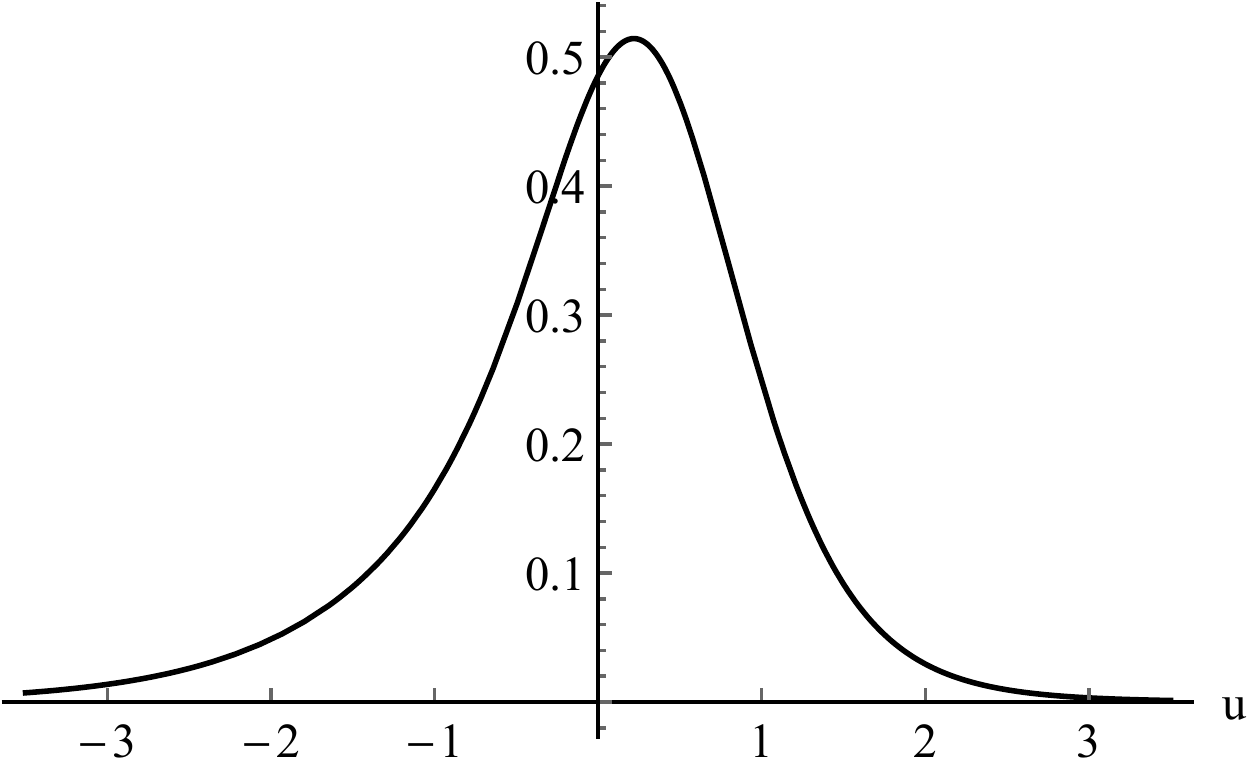}  & 
\includegraphics[width=0.48\linewidth]{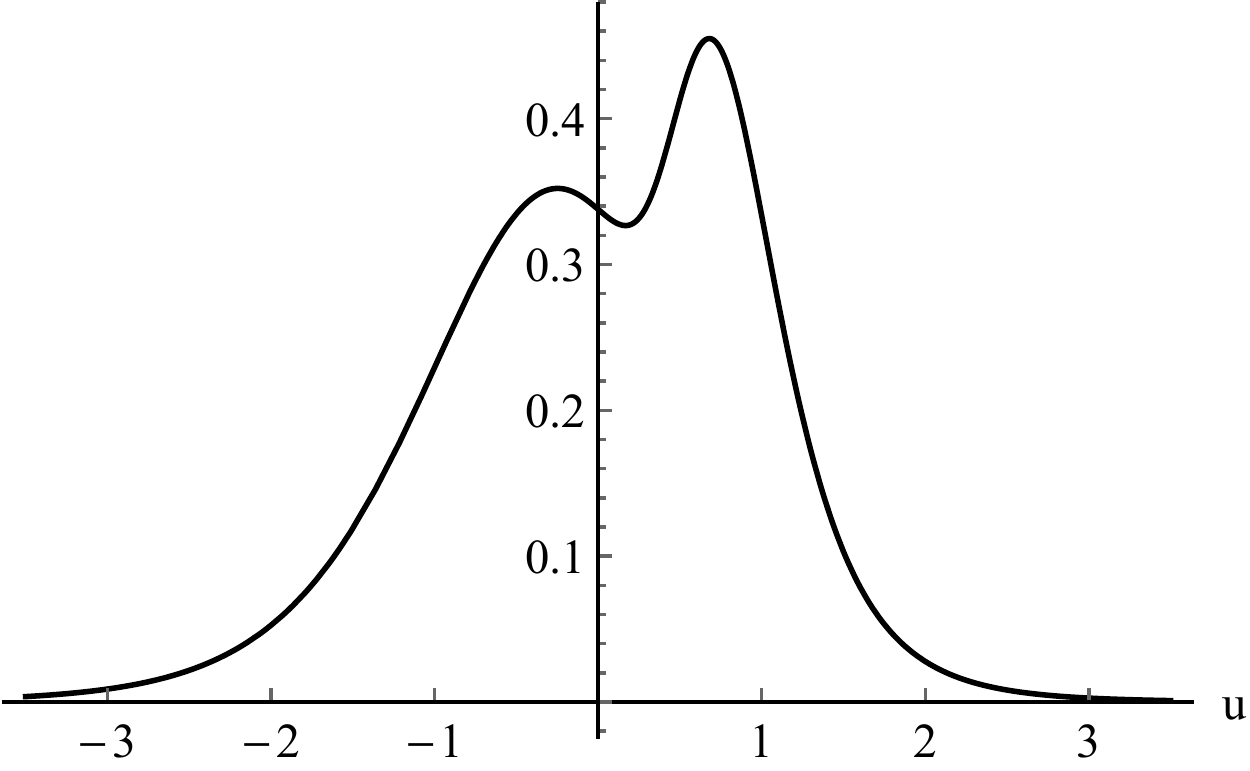} \\[2pt]
$f_{NM}$: $\frac{1}{5}N(-\frac{3}{4},1)+\frac{1}{5}N(-\frac{1}{4},(\frac{2}{3})^{2})+\frac{3}{5}N(\frac{1}{3},(\frac{5}{9})^{2})$ & 
$f_{NM}$: $\frac{3}{4}N(-\frac{1}{3},1)+\frac{1}{4}N(1,(\frac{1}{3})^{2})$ \\[2pt] 
$R(f^{(4)})^{-1/9}=0.595872$,  $R(f^{(3)})^{-1/7}=0.738786$.  &  
$R(f^{(4)})^{-1/9}=0.372818$,  $R(f^{(3)})^{-1/7}=0.499502$.  \\  
\end{tabular}
\caption{Selected mixture densities (scaled)}
\label{Fig:MCnoncTmixDens}
\end{figure}

\vspace*{\topsep}
\noindent\textsc{Scenarios 2 and 3.}  $x$ and $u$ have joint density $f_{ux}(u,x)=xf_{NM}(ux)f_{x}(x)$ where $x$ is a generalised gamma random variable, \citet{stacy1962}, with parameters $p=2$, $d=\nu$ and $a=(2/\nu)^{1/2}$ for some $\nu>4$ and $f_{NM}$ is the normal mixture density with $m$ components, \textit{viz}. $f_{NM}(w) = \sum_{j=1}^{m}\omega_{j}\phi_{\sigma_{j}}(w-\mu_{j})$, $-\infty<\mu _{j}<\infty$, $\sigma_{j}>0$, $j=1,\ldots,m$,  $\sum_{j=1}^{m}\omega_{j}=1$, and $\sum_{j=1}^{m}\omega_{j}\mu_{j}=0$, i.e., $\E[w]=0$. 
Here $\phi(x)$ denotes the standard normal p.d.f. and $\phi_{\sigma}(x)=\phi(x/\sigma)/\sigma$. 
The joint density $f_{ux}$ is the density of $u=w/x$ and $x$ where $w$ and $x$ are independent. The conditional density of $u$ given $x$ is 
$f_{u|x}(u|x) = xf_{NM}(ux) = \sum_{j=1}^{m}\omega_{j}\phi_{\sigma_{j}/x}(u-\mu_{j}/x)$. 
Hence, $\E[u|x]=0$ and 
$\E[u_{i}^{2}|x] = \sum_{j=1}^{m}\omega_{j}(\sigma_{j}^{2}+\mu_{j}^{2})/x^{2}$.
The marginal density of $u$ is a mixture of noncentral $t$ densities 
$f_{u}(u) = \sum_{j=1}^{m}\omega_{j}t_{\nu}(u/\sigma_{j};\: \mu_{i}/\sigma_{i})/\sigma_{j}$
where $t_{\nu}(\cdot;\eta)$ is the density of a noncentral $t$-distributed random variable with $\nu$ degrees of freedom and noncentrality parameter $\eta$ allowing a wide variety of shapes for $f_{u}$ by varying the mixture $f_{NM}$. The skewed unimodal and bimodal densities shown in Figure \ref{Fig:MCnoncTmixDens} describe the NM densities for Scenarios 2 and 3 respectively, i.e., the mixture densities \citet[\#2 and \#8]{marron1992} centered to have zero mean.

\subsection{Kernel Functions and Bandwidths}
Fourth order Gaussian-based kernels, $k(x)=(3-x^{2})\phi(x)/2$ and $K(x)=\Phi(x)+x\phi (x)/2$, $\Phi(x)=\int_{-\infty}^{x}\phi(u)\mathrm{d}u$, are employed; see \citet[Section 2]{wand1990} and \citet{oryshchenko2017} respectively. Thus the choices of the asymptotically optimal bandwidths $(27/4\sqrt{\pi})^{1/9}R(f^{(4)})^{-1/9}n^{-1/9}$ and $(7/2\sqrt{\pi })^{1/7}R(f^{(3)})^{-1/7}n^{-1/7}$ for p.d.f. and c.d.f. estimation respectively are permitted, thereby satisfying Assumptions \ref{Assumption.U.and.K.diff.int}(c) and \ref{Assumption.U.and.K.diff.int.CDF}(b). The practical issue of estimating the derivatives of $f$ required for the computation of $R(f^{(j)})$, $j=3,4$, is ignored and the respective true values used. For the standard normal distribution these are $R(\phi^{(3)})=15/(16\sqrt{\pi})$ and $R(\phi^{(4)})=105/(32\sqrt{\pi})$; for the mixture distributions, approximate values are shown in Figure \ref{Fig:MCnoncTmixDens}.

\subsection{Results}
The study compares the performance of GEL-based kernel density p.d.f. and c.d.f. estimators. The GEL parameter estimators are CUE, EL and ET, the most notable special cases of the GEL family. For each estimator the mean and variance were computed on a grid $1000$ of points between $-5$ and $5$ and are reported as the integrated squared bias and integrated variance relative 
to those of the corresponding infeasible estimator based on the true $u$, i.e., $\tilde{f}$ and $\widetilde{F}$.

Tables \ref{Tab:Supp:mcIHSmainD1PDFTabB1}, \ref{Tab:Supp:mcIHSmainD2PDFTabB1} and \ref{Tab:Supp:mcIHSmainD3PDFTabB1} report results for Scenarios 1, 2 and 3 respectively. The ISB, IVar and MISE (all $\times10^5$) for the infeasible $\tilde{f}$ and $\widetilde{F}$ are presented. Rows ISB, IVar, and MISE are the ISB, IVar, and MISE of $\hat{f}$, $\hat{f}_{\rho}$ ($\widehat{F}$, $\widehat{F}_{\rho}$) relative to the infeasible $\tilde{f}$ ($\widetilde{F}$), respectively; row `vs $d_{g}=3$' is the MISE of $\hat{f}$, $\hat{f}_{\rho}$ ($\hat{F}$, $\widehat{F}_{\rho}$) relative to the corresponding value for $d_{g}=3$; row `w. vs unw.' is the MISE of $\hat{f}_{\rho}$ ($\widehat{F}_{\rho}$) relative to $\hat{f}$ ($\widehat{F}$). Rows MISE, `vs $d_{g}=3$',  and `w. vs unw.' examine the significance of the paired $t$-statistics in a two-sided test for equality of the respective ISE means, e.g., $\int (\hat{f}(u)-f(u))^{2}\mathrm{d}u$; the symbol \dag\ indicates that the $p$-value is between $0.01$ and $0.05$ whereas \ddag\ that it is less than $0.01$ and in all other cases the $p$-value is greater than $0.05$. Values of  relative MISE less than $1$ are emphasised in bold.

Sample sizes $n=100$, $500$, $1000$, and $2000$ are examined.

All computations were carried out in MATLAB; the relevant code and additional results, including the properties of GEL estimators, are available from the first named author upon request. All results are based on  $10,000$ random draws.

\renewcommand{\arraystretch}{0.9}
\setlength{\tabcolsep}{5pt}
\begin{table}[p]\centering\footnotesize
\caption{Performance of GEL-based residual density and distribution function estimators in the IHS transformation model, $\arsinh(0.08y)/0.08 = 1+2x+u$, in Scenario 1}
\label{Tab:Supp:mcIHSmainD1PDFTabB1}
\begin{tabular}{ll|rr|rr|rrrr|rrrr}\hline\hline
 & &\multirow{3}{*}{$\tilde{f}$} & \multirow{3}{*}{$\widetilde{F}$}&\multicolumn{2}{c}{$d_{g}=3$}\vline&\multicolumn{4}{c}{$d_{g}=4$}\vline&\multicolumn{4}{c}{$d_{g}=5$}\\ \cline{5-14} 
 & & & & & & & & & & & & & \\[-0.9em] &  & & & $\hat{f}$ & $\widehat{F}$  & $\hat{f}$ & $\hat{f}_{\rho}$ & $\widehat{F}$ & $\widehat{F}_{\rho}$ & $\hat{f}$ & $\hat{f}_{\rho}$ & $\widehat{F}$ & $\widehat{F}_{\rho}$ \\ \hline 
\multicolumn{13}{c}{$n=100$}\\ \hline
CUE&$\ISB$&31.3&8.9&0.70&3.41&0.13&0.20&0.47&1.12&0.12&0.37&0.56&1.85\\ 
 &$\IVar$&449.0&402.3&1.29&0.71&0.88&0.88&0.46&0.43&0.87&0.86&0.47&0.42\\ 
 &$\MISE$&480.2&411.2&1.26\,\textbf{\ddag}&\textbf{0.78}\,\textbf{\ddag}&\textbf{0.83}\,\textbf{\ddag}&\textbf{0.83}\,\textbf{\ddag}&\textbf{0.46}\,\textbf{\ddag}&\textbf{0.45}\,\textbf{\ddag}&\textbf{0.82}\,\textbf{\ddag}&\textbf{0.83}\,\textbf{\ddag}&\textbf{0.47}\,\textbf{\ddag}&\textbf{0.45}\,\textbf{\ddag}\\ 
\multicolumn{2}{r}{vs $d_{g}=3$}\vline & & & &&\textbf{0.66}\,\textbf{\ddag}&\textbf{0.66}\,\textbf{\ddag}&\textbf{0.60}\,\textbf{\ddag}&\textbf{0.58}\,\textbf{\ddag}&\textbf{0.65}\,\textbf{\ddag}&\textbf{0.66}\,\textbf{\ddag}&\textbf{0.61}\,\textbf{\ddag}&\textbf{0.59}\,\textbf{\ddag}\\ 
\multicolumn{2}{r}{w. vs unw.}\vline & & & &&\multicolumn{2}{r}{\textbf{0.999}}& \multicolumn{2}{r}{\textbf{0.965}\,\textbf{\ddag}}\vline&\multicolumn{2}{r}{1.010\,\textbf{\ddag}}& \multicolumn{2}{r}{\textbf{0.964}\,\textbf{\ddag}}\\ 
\hdashline[2pt/2pt]EL &$\ISB$& & & & &0.16&0.17&0.56&0.64&0.17&0.20&0.71&0.87\\ 
&$\IVar$& & & & &0.84&0.84&0.45&0.42&0.87&0.89&0.51&0.45\\ 
&$\MISE$& & & & &\textbf{0.80}\,\textbf{\ddag}&\textbf{0.80}\,\textbf{\ddag}&\textbf{0.45}\,\textbf{\ddag}&\textbf{0.42}\,\textbf{\ddag}&\textbf{0.83}\,\textbf{\ddag}&\textbf{0.85}\,\textbf{\ddag}&\textbf{0.51}\,\textbf{\ddag}&\textbf{0.46}\,\textbf{\ddag}\\ 
\multicolumn{2}{r}{vs $d_{g}=3$}\vline & & & &&\textbf{0.64}\,\textbf{\ddag}&\textbf{0.64}\,\textbf{\ddag}&\textbf{0.58}\,\textbf{\ddag}&\textbf{0.55}\,\textbf{\ddag}&\textbf{0.66}\,\textbf{\ddag}&\textbf{0.68}\,\textbf{\ddag}&\textbf{0.66}\,\textbf{\ddag}&\textbf{0.60}\,\textbf{\ddag}\\ 
\multicolumn{2}{r}{w. vs unw.}\vline & & & &&\multicolumn{2}{r}{1.001}& \multicolumn{2}{r}{\textbf{0.930}\,\textbf{\ddag}}\vline&\multicolumn{2}{r}{1.024}& \multicolumn{2}{r}{\textbf{0.906}\,\textbf{\ddag}}\\ 
\hdashline[2pt/2pt]ET &$\ISB$& & & & &0.15&0.20&0.55&1.01&0.14&0.34&0.66&1.70\\ 
&$\IVar$& & & & &0.83&0.89&0.43&0.47&0.85&0.86&0.48&0.53\\ 
&$\MISE$& & & & &\textbf{0.79}\,\textbf{\ddag}&\textbf{0.85}\,\textbf{\ddag}&\textbf{0.44}\,\textbf{\ddag}&\textbf{0.48}\,\textbf{\ddag}&\textbf{0.81}\,\textbf{\ddag}&\textbf{0.84}\,\textbf{\ddag}&\textbf{0.49}\,\textbf{\ddag}&\textbf{0.88}\\ 
\multicolumn{2}{r}{vs $d_{g}=3$}\vline & & & &&\textbf{0.63}\,\textbf{\ddag}&\textbf{0.68}\,\textbf{\ddag}&\textbf{0.56}\,\textbf{\ddag}&\textbf{0.62}\,\textbf{\ddag}&\textbf{0.64}\,\textbf{\ddag}&\textbf{0.67}\,\textbf{\ddag}&\textbf{0.64}\,\textbf{\ddag}&1.16\\ 
\multicolumn{2}{r}{w. vs unw.}\vline & & & &&\multicolumn{2}{r}{1.071\,\textbf{\dag}}& \multicolumn{2}{r}{1.092}\vline&\multicolumn{2}{r}{1.037}& \multicolumn{2}{r}{1.789}\\ 
\hline
\multicolumn{13}{c}{$n=500$}\\ \hline
CUE&$\ISB$&10.0&1.8&0.37&1.53&0.46&0.36&0.29&0.28&0.40&0.27&0.27&0.33\\ 
 &$\IVar$&119.8&88.3&0.99&0.61&0.87&0.87&0.46&0.45&0.88&0.88&0.47&0.46\\ 
 &$\MISE$&129.8&90.2&\textbf{0.94}\,\textbf{\ddag}&\textbf{0.63}\,\textbf{\ddag}&\textbf{0.84}\,\textbf{\ddag}&\textbf{0.83}\,\textbf{\ddag}&\textbf{0.45}\,\textbf{\ddag}&\textbf{0.45}\,\textbf{\ddag}&\textbf{0.84}\,\textbf{\ddag}&\textbf{0.83}\,\textbf{\ddag}&\textbf{0.47}\,\textbf{\ddag}&\textbf{0.46}\,\textbf{\ddag}\\ 
\multicolumn{2}{r}{vs $d_{g}=3$}\vline & & & &&\textbf{0.89}\,\textbf{\ddag}&\textbf{0.88}\,\textbf{\ddag}&\textbf{0.72}\,\textbf{\ddag}&\textbf{0.71}\,\textbf{\ddag}&\textbf{0.90}\,\textbf{\ddag}&\textbf{0.89}\,\textbf{\ddag}&\textbf{0.74}\,\textbf{\ddag}&\textbf{0.73}\,\textbf{\ddag}\\ 
\multicolumn{2}{r}{w. vs unw.}\vline & & & &&\multicolumn{2}{r}{\textbf{0.991}\,\textbf{\ddag}}& \multicolumn{2}{r}{\textbf{0.988}\,\textbf{\ddag}}\vline&\multicolumn{2}{r}{\textbf{0.988}\,\textbf{\ddag}}& \multicolumn{2}{r}{\textbf{0.986}\,\textbf{\ddag}}\\ 
\hdashline[2pt/2pt]EL &$\ISB$& & & & &0.45&0.45&0.29&0.30&0.41&0.40&0.28&0.29\\ 
&$\IVar$& & & & &0.87&0.87&0.46&0.45&0.88&0.88&0.47&0.46\\ 
&$\MISE$& & & & &\textbf{0.83}\,\textbf{\ddag}&\textbf{0.84}\,\textbf{\ddag}&\textbf{0.45}\,\textbf{\ddag}&\textbf{0.45}\,\textbf{\ddag}&\textbf{0.84}\,\textbf{\ddag}&\textbf{0.84}\,\textbf{\ddag}&\textbf{0.47}\,\textbf{\ddag}&\textbf{0.46}\,\textbf{\ddag}\\ 
\multicolumn{2}{r}{vs $d_{g}=3$}\vline & & & &&\textbf{0.89}\,\textbf{\ddag}&\textbf{0.89}\,\textbf{\ddag}&\textbf{0.72}\,\textbf{\ddag}&\textbf{0.72}\,\textbf{\ddag}&\textbf{0.89}\,\textbf{\ddag}&\textbf{0.90}\,\textbf{\ddag}&\textbf{0.74}\,\textbf{\ddag}&\textbf{0.73}\,\textbf{\ddag}\\ 
\multicolumn{2}{r}{w. vs unw.}\vline & & & &&\multicolumn{2}{r}{1.002\,\textbf{\ddag}}& \multicolumn{2}{r}{\textbf{0.993}\,\textbf{\ddag}}\vline&\multicolumn{2}{r}{1.003\,\textbf{\dag}}& \multicolumn{2}{r}{\textbf{0.979}\,\textbf{\ddag}}\\ 
\hdashline[2pt/2pt]ET &$\ISB$& & & & &0.45&0.39&0.29&0.28&0.40&0.31&0.27&0.30\\ 
&$\IVar$& & & & &0.87&0.87&0.46&0.45&0.88&0.88&0.47&0.46\\ 
&$\MISE$& & & & &\textbf{0.83}\,\textbf{\ddag}&\textbf{0.83}\,\textbf{\ddag}&\textbf{0.45}\,\textbf{\ddag}&\textbf{0.45}\,\textbf{\ddag}&\textbf{0.84}\,\textbf{\ddag}&\textbf{0.83}\,\textbf{\ddag}&\textbf{0.46}\,\textbf{\ddag}&\textbf{0.46}\,\textbf{\ddag}\\ 
\multicolumn{2}{r}{vs $d_{g}=3$}\vline & & & &&\textbf{0.89}\,\textbf{\ddag}&\textbf{0.88}\,\textbf{\ddag}&\textbf{0.72}\,\textbf{\ddag}&\textbf{0.71}\,\textbf{\ddag}&\textbf{0.89}\,\textbf{\ddag}&\textbf{0.89}\,\textbf{\ddag}&\textbf{0.74}\,\textbf{\ddag}&\textbf{0.73}\,\textbf{\ddag}\\ 
\multicolumn{2}{r}{w. vs unw.}\vline & & & &&\multicolumn{2}{r}{\textbf{0.996}\,\textbf{\ddag}}& \multicolumn{2}{r}{\textbf{0.991}\,\textbf{\ddag}}\vline&\multicolumn{2}{r}{\textbf{0.994}\,\textbf{\ddag}}& \multicolumn{2}{r}{\textbf{0.986}\,\textbf{\ddag}}\\ 
\hline
\multicolumn{13}{c}{$n=1000$}\\ \hline
CUE&$\ISB$&6.1&0.9&0.48&1.03&0.62&0.55&0.41&0.33&0.58&0.46&0.36&0.28\\ 
 &$\IVar$&66.1&45.6&0.99&0.62&0.89&0.89&0.48&0.48&0.90&0.90&0.49&0.49\\ 
 &$\MISE$&72.2&46.5&\textbf{0.95}\,\textbf{\ddag}&\textbf{0.63}\,\textbf{\ddag}&\textbf{0.87}\,\textbf{\ddag}&\textbf{0.86}\,\textbf{\ddag}&\textbf{0.48}\,\textbf{\ddag}&\textbf{0.47}\,\textbf{\ddag}&\textbf{0.87}\,\textbf{\ddag}&\textbf{0.86}\,\textbf{\ddag}&\textbf{0.49}\,\textbf{\ddag}&\textbf{0.48}\,\textbf{\ddag}\\ 
\multicolumn{2}{r}{vs $d_{g}=3$}\vline & & & &&\textbf{0.91}\,\textbf{\ddag}&\textbf{0.91}\,\textbf{\ddag}&\textbf{0.76}\,\textbf{\ddag}&\textbf{0.75}\,\textbf{\ddag}&\textbf{0.92}\,\textbf{\ddag}&\textbf{0.91}\,\textbf{\ddag}&\textbf{0.78}\,\textbf{\ddag}&\textbf{0.77}\,\textbf{\ddag}\\ 
\multicolumn{2}{r}{w. vs unw.}\vline & & & &&\multicolumn{2}{r}{\textbf{0.992}\,\textbf{\ddag}}& \multicolumn{2}{r}{\textbf{0.990}\,\textbf{\ddag}}\vline&\multicolumn{2}{r}{\textbf{0.988}\,\textbf{\ddag}}& \multicolumn{2}{r}{\textbf{0.988}\,\textbf{\ddag}}\\ 
\hdashline[2pt/2pt]EL &$\ISB$& & & & &0.62&0.62&0.40&0.40&0.59&0.58&0.37&0.36\\ 
&$\IVar$& & & & &0.89&0.89&0.48&0.48&0.89&0.89&0.49&0.48\\ 
&$\MISE$& & & & &\textbf{0.86}\,\textbf{\ddag}&\textbf{0.86}\,\textbf{\ddag}&\textbf{0.48}\,\textbf{\ddag}&\textbf{0.48}\,\textbf{\ddag}&\textbf{0.87}\,\textbf{\ddag}&\textbf{0.87}\,\textbf{\ddag}&\textbf{0.49}\,\textbf{\ddag}&\textbf{0.48}\,\textbf{\ddag}\\ 
\multicolumn{2}{r}{vs $d_{g}=3$}\vline & & & &&\textbf{0.91}\,\textbf{\ddag}&\textbf{0.91}\,\textbf{\ddag}&\textbf{0.76}\,\textbf{\ddag}&\textbf{0.76}\,\textbf{\ddag}&\textbf{0.91}\,\textbf{\ddag}&\textbf{0.91}\,\textbf{\ddag}&\textbf{0.77}\,\textbf{\ddag}&\textbf{0.77}\,\textbf{\ddag}\\ 
\multicolumn{2}{r}{w. vs unw.}\vline & & & &&\multicolumn{2}{r}{1.001\,\textbf{\dag}}& \multicolumn{2}{r}{\textbf{0.996}\,\textbf{\ddag}}\vline&\multicolumn{2}{r}{1.001}& \multicolumn{2}{r}{\textbf{0.989}\,\textbf{\ddag}}\\ 
\hdashline[2pt/2pt]ET &$\ISB$& & & & &0.62&0.58&0.40&0.36&0.58&0.50&0.36&0.30\\ 
&$\IVar$& & & & &0.89&0.89&0.48&0.48&0.89&0.89&0.49&0.48\\ 
&$\MISE$& & & & &\textbf{0.86}\,\textbf{\ddag}&\textbf{0.86}\,\textbf{\ddag}&\textbf{0.48}\,\textbf{\ddag}&\textbf{0.47}\,\textbf{\ddag}&\textbf{0.87}\,\textbf{\ddag}&\textbf{0.86}\,\textbf{\ddag}&\textbf{0.48}\,\textbf{\ddag}&\textbf{0.48}\,\textbf{\ddag}\\ 
\multicolumn{2}{r}{vs $d_{g}=3$}\vline & & & &&\textbf{0.91}\,\textbf{\ddag}&\textbf{0.91}\,\textbf{\ddag}&\textbf{0.76}\,\textbf{\ddag}&\textbf{0.76}\,\textbf{\ddag}&\textbf{0.91}\,\textbf{\ddag}&\textbf{0.91}\,\textbf{\ddag}&\textbf{0.77}\,\textbf{\ddag}&\textbf{0.76}\,\textbf{\ddag}\\ 
\multicolumn{2}{r}{w. vs unw.}\vline & & & &&\multicolumn{2}{r}{\textbf{0.996}\,\textbf{\ddag}}& \multicolumn{2}{r}{\textbf{0.993}\,\textbf{\ddag}}\vline&\multicolumn{2}{r}{\textbf{0.993}\,\textbf{\ddag}}& \multicolumn{2}{r}{\textbf{0.990}\,\textbf{\ddag}}\\ 
\hline
\multicolumn{13}{c}{$n=2000$}\\ \hline
CUE&$\ISB$&3.5&0.4&0.55&0.62&0.74&0.69&0.53&0.45&0.71&0.62&0.49&0.37\\ 
 &$\IVar$&36.6&23.0&1.02&0.65&0.92&0.92&0.52&0.52&0.93&0.93&0.53&0.52\\ 
 &$\MISE$&40.1&23.5&\textbf{0.98}\,\textbf{\ddag}&\textbf{0.65}\,\textbf{\ddag}&\textbf{0.90}\,\textbf{\ddag}&\textbf{0.90}\,\textbf{\ddag}&\textbf{0.52}\,\textbf{\ddag}&\textbf{0.51}\,\textbf{\ddag}&\textbf{0.91}\,\textbf{\ddag}&\textbf{0.90}\,\textbf{\ddag}&\textbf{0.53}\,\textbf{\ddag}&\textbf{0.52}\,\textbf{\ddag}\\ 
\multicolumn{2}{r}{vs $d_{g}=3$}\vline & & & &&\textbf{0.92}\,\textbf{\ddag}&\textbf{0.92}\,\textbf{\ddag}&\textbf{0.80}\,\textbf{\ddag}&\textbf{0.79}\,\textbf{\ddag}&\textbf{0.93}\,\textbf{\ddag}&\textbf{0.92}\,\textbf{\ddag}&\textbf{0.81}\,\textbf{\ddag}&\textbf{0.80}\,\textbf{\ddag}\\ 
\multicolumn{2}{r}{w. vs unw.}\vline & & & &&\multicolumn{2}{r}{\textbf{0.994}\,\textbf{\ddag}}& \multicolumn{2}{r}{\textbf{0.994}\,\textbf{\ddag}}\vline&\multicolumn{2}{r}{\textbf{0.990}\,\textbf{\ddag}}& \multicolumn{2}{r}{\textbf{0.992}\,\textbf{\ddag}}\\ 
\hdashline[2pt/2pt]EL &$\ISB$& & & & &0.74&0.74&0.52&0.52&0.71&0.71&0.48&0.48\\ 
&$\IVar$& & & & &0.92&0.92&0.52&0.52&0.92&0.92&0.52&0.52\\ 
&$\MISE$& & & & &\textbf{0.90}\,\textbf{\ddag}&\textbf{0.90}\,\textbf{\ddag}&\textbf{0.52}\,\textbf{\ddag}&\textbf{0.52}\,\textbf{\ddag}&\textbf{0.90}\,\textbf{\ddag}&\textbf{0.90}\,\textbf{\ddag}&\textbf{0.52}\,\textbf{\ddag}&\textbf{0.52}\,\textbf{\ddag}\\ 
\multicolumn{2}{r}{vs $d_{g}=3$}\vline & & & &&\textbf{0.92}\,\textbf{\ddag}&\textbf{0.92}\,\textbf{\ddag}&\textbf{0.80}\,\textbf{\ddag}&\textbf{0.80}\,\textbf{\ddag}&\textbf{0.92}\,\textbf{\ddag}&\textbf{0.93}\,\textbf{\ddag}&\textbf{0.81}\,\textbf{\ddag}&\textbf{0.80}\,\textbf{\ddag}\\ 
\multicolumn{2}{r}{w. vs unw.}\vline & & & &&\multicolumn{2}{r}{1.000}& \multicolumn{2}{r}{\textbf{0.999}}\vline&\multicolumn{2}{r}{1.001}& \multicolumn{2}{r}{\textbf{0.996}\,\textbf{\ddag}}\\ 
\hdashline[2pt/2pt]ET &$\ISB$& & & & &0.74&0.71&0.53&0.48&0.71&0.65&0.49&0.41\\ 
&$\IVar$& & & & &0.92&0.92&0.52&0.52&0.92&0.92&0.52&0.52\\ 
&$\MISE$& & & & &\textbf{0.90}\,\textbf{\ddag}&\textbf{0.90}\,\textbf{\ddag}&\textbf{0.52}\,\textbf{\ddag}&\textbf{0.52}\,\textbf{\ddag}&\textbf{0.90}\,\textbf{\ddag}&\textbf{0.90}\,\textbf{\ddag}&\textbf{0.52}\,\textbf{\ddag}&\textbf{0.52}\,\textbf{\ddag}\\ 
\multicolumn{2}{r}{vs $d_{g}=3$}\vline & & & &&\textbf{0.92}\,\textbf{\ddag}&\textbf{0.92}\,\textbf{\ddag}&\textbf{0.80}\,\textbf{\ddag}&\textbf{0.79}\,\textbf{\ddag}&\textbf{0.93}\,\textbf{\ddag}&\textbf{0.92}\,\textbf{\ddag}&\textbf{0.80}\,\textbf{\ddag}&\textbf{0.80}\,\textbf{\ddag}\\ 
\multicolumn{2}{r}{w. vs unw.}\vline & & & &&\multicolumn{2}{r}{\textbf{0.997}\,\textbf{\ddag}}& \multicolumn{2}{r}{\textbf{0.996}\,\textbf{\ddag}}\vline&\multicolumn{2}{r}{\textbf{0.994}\,\textbf{\ddag}}& \multicolumn{2}{r}{\textbf{0.994}\,\textbf{\ddag}}\\ 
\hline
\hline 
\end{tabular}

\vspace*{-0.8\baselineskip}
\begin{flushleft} 
\textbf{Notes:} see text.
\end{flushleft}
\end{table}

\begin{table}[p]\centering\footnotesize
\caption{Performance of GEL-based residual density and distribution function estimators in the IHS transformation model, $\arsinh(0.08y)/0.08 = 1+2x+u$, in Scenario 2}
\label{Tab:Supp:mcIHSmainD2PDFTabB1}
\begin{tabular}{ll|rr|rr|rrrr|rrrr}\hline\hline
 & &\multirow{3}{*}{$\tilde{f}$} & \multirow{3}{*}{$\widetilde{F}$}&\multicolumn{2}{c}{$d_{g}=3$}\vline&\multicolumn{4}{c}{$d_{g}=4$}\vline&\multicolumn{4}{c}{$d_{g}=5$}\\ \cline{5-14} 
 & & & & & & & & & & & & & \\[-0.9em] &  & & & $\hat{f}$ & $\widehat{F}$  & $\hat{f}$ & $\hat{f}_{\rho}$ & $\widehat{F}$ & $\widehat{F}_{\rho}$ & $\hat{f}$ & $\hat{f}_{\rho}$ & $\widehat{F}$ & $\widehat{F}_{\rho}$ \\ \hline 
\multicolumn{13}{c}{$n=100$}\\ \hline
CUE&$\ISB$&29.3&3.8&0.73&5.71&0.85&1.13&2.65&5.14&0.99&1.52&3.65&7.65\\ 
 &$\IVar$&822.0&427.7&1.12&0.75&0.93&0.94&0.54&0.50&0.93&0.94&0.55&0.49\\ 
 &$\MISE$&852.5&432.5&1.11\,\textbf{\ddag}&\textbf{0.79}\,\textbf{\ddag}&\textbf{0.93}\,\textbf{\ddag}&\textbf{0.94}\,\textbf{\ddag}&\textbf{0.56}\,\textbf{\ddag}&\textbf{0.54}\,\textbf{\ddag}&\textbf{0.93}\,\textbf{\ddag}&\textbf{0.95}\,\textbf{\ddag}&\textbf{0.57}\,\textbf{\ddag}&\textbf{0.55}\,\textbf{\ddag}\\ 
\multicolumn{2}{r}{vs $d_{g}=3$}\vline & & & &&\textbf{0.84}\,\textbf{\ddag}&\textbf{0.85}\,\textbf{\ddag}&\textbf{0.70}\,\textbf{\ddag}&\textbf{0.68}\,\textbf{\ddag}&\textbf{0.84}\,\textbf{\ddag}&\textbf{0.86}\,\textbf{\ddag}&\textbf{0.72}\,\textbf{\ddag}&\textbf{0.70}\,\textbf{\ddag}\\ 
\multicolumn{2}{r}{w. vs unw.}\vline & & & &&\multicolumn{2}{r}{1.013\,\textbf{\ddag}}& \multicolumn{2}{r}{\textbf{0.963}\,\textbf{\ddag}}\vline&\multicolumn{2}{r}{1.026\,\textbf{\ddag}}& \multicolumn{2}{r}{\textbf{0.966}\,\textbf{\ddag}}\\ 
\hdashline[2pt/2pt]EL &$\ISB$& & & & &0.71&0.78&1.95&2.59&0.70&0.83&1.99&3.03\\ 
&$\IVar$& & & & &0.92&0.93&0.53&0.50&0.93&0.96&0.56&0.51\\ 
&$\MISE$& & & & &\textbf{0.91}\,\textbf{\ddag}&\textbf{0.93}\,\textbf{\ddag}&\textbf{0.54}\,\textbf{\ddag}&\textbf{0.52}\,\textbf{\ddag}&\textbf{0.92}\,\textbf{\ddag}&\textbf{0.95}\,\textbf{\ddag}&\textbf{0.57}\,\textbf{\ddag}&\textbf{0.54}\,\textbf{\ddag}\\ 
\multicolumn{2}{r}{vs $d_{g}=3$}\vline & & & &&\textbf{0.82}\,\textbf{\ddag}&\textbf{0.84}\,\textbf{\ddag}&\textbf{0.68}\,\textbf{\ddag}&\textbf{0.66}\,\textbf{\ddag}&\textbf{0.83}\,\textbf{\ddag}&\textbf{0.86}\,\textbf{\ddag}&\textbf{0.72}\,\textbf{\ddag}&\textbf{0.68}\,\textbf{\ddag}\\ 
\multicolumn{2}{r}{w. vs unw.}\vline & & & &&\multicolumn{2}{r}{1.015\,\textbf{\ddag}}& \multicolumn{2}{r}{\textbf{0.956}\,\textbf{\ddag}}\vline&\multicolumn{2}{r}{1.029\,\textbf{\ddag}}& \multicolumn{2}{r}{\textbf{0.934}\,\textbf{\ddag}}\\ 
\hdashline[2pt/2pt]ET &$\ISB$& & & & &0.77&0.97&2.37&4.19&0.86&1.28&2.99&6.22\\ 
&$\IVar$& & & & &0.91&0.95&0.51&0.59&0.92&0.94&0.55&0.63\\ 
&$\MISE$& & & & &\textbf{0.90}\,\textbf{\ddag}&\textbf{0.95}&\textbf{0.53}\,\textbf{\ddag}&\textbf{0.62}\,\textbf{\ddag}&\textbf{0.92}\,\textbf{\ddag}&\textbf{0.96}&\textbf{0.58}\,\textbf{\ddag}&\textbf{0.80}\\ 
\multicolumn{2}{r}{vs $d_{g}=3$}\vline & & & &&\textbf{0.82}\,\textbf{\ddag}&\textbf{0.86}\,\textbf{\ddag}&\textbf{0.67}\,\textbf{\ddag}&\textbf{0.80}&\textbf{0.83}\,\textbf{\ddag}&\textbf{0.87}\,\textbf{\ddag}&\textbf{0.73}\,\textbf{\ddag}&1.03\\ 
\multicolumn{2}{r}{w. vs unw.}\vline & & & &&\multicolumn{2}{r}{1.054\,\textbf{\dag}}& \multicolumn{2}{r}{1.174}\vline&\multicolumn{2}{r}{1.053\,\textbf{\ddag}}& \multicolumn{2}{r}{1.390}\\ 
\hline
\multicolumn{13}{c}{$n=500$}\\ \hline
CUE&$\ISB$&10.0&0.8&0.80&4.25&0.88&0.86&0.96&1.36&0.87&0.84&1.15&1.92\\ 
 &$\IVar$&210.2&89.9&1.08&0.82&0.93&0.93&0.53&0.52&0.93&0.93&0.53&0.51\\ 
 &$\MISE$&220.4&90.9&1.07\,\textbf{\ddag}&\textbf{0.85}\,\textbf{\ddag}&\textbf{0.93}\,\textbf{\ddag}&\textbf{0.93}\,\textbf{\ddag}&\textbf{0.54}\,\textbf{\ddag}&\textbf{0.52}\,\textbf{\ddag}&\textbf{0.93}\,\textbf{\ddag}&\textbf{0.93}\,\textbf{\ddag}&\textbf{0.54}\,\textbf{\ddag}&\textbf{0.53}\,\textbf{\ddag}\\ 
\multicolumn{2}{r}{vs $d_{g}=3$}\vline & & & &&\textbf{0.87}\,\textbf{\ddag}&\textbf{0.87}\,\textbf{\ddag}&\textbf{0.63}\,\textbf{\ddag}&\textbf{0.62}\,\textbf{\ddag}&\textbf{0.87}\,\textbf{\ddag}&\textbf{0.87}\,\textbf{\ddag}&\textbf{0.63}\,\textbf{\ddag}&\textbf{0.62}\,\textbf{\ddag}\\ 
\multicolumn{2}{r}{w. vs unw.}\vline & & & &&\multicolumn{2}{r}{1.001}& \multicolumn{2}{r}{\textbf{0.980}\,\textbf{\ddag}}\vline&\multicolumn{2}{r}{1.001}& \multicolumn{2}{r}{\textbf{0.980}\,\textbf{\ddag}}\\ 
\hdashline[2pt/2pt]EL &$\ISB$& & & & &0.87&0.89&0.87&0.94&0.84&0.85&0.86&0.98\\ 
&$\IVar$& & & & &0.93&0.98&0.53&0.72&0.93&0.93&0.53&0.52\\ 
&$\MISE$& & & & &\textbf{0.93}\,\textbf{\ddag}&\textbf{0.97}&\textbf{0.53}\,\textbf{\ddag}&\textbf{0.72}&\textbf{0.92}\,\textbf{\ddag}&\textbf{0.93}\,\textbf{\ddag}&\textbf{0.53}\,\textbf{\ddag}&\textbf{0.52}\,\textbf{\ddag}\\ 
\multicolumn{2}{r}{vs $d_{g}=3$}\vline & & & &&\textbf{0.87}\,\textbf{\ddag}&\textbf{0.91}\,\textbf{\dag}&\textbf{0.63}\,\textbf{\ddag}&\textbf{0.85}&\textbf{0.87}\,\textbf{\ddag}&\textbf{0.87}\,\textbf{\ddag}&\textbf{0.63}\,\textbf{\ddag}&\textbf{0.62}\,\textbf{\ddag}\\ 
\multicolumn{2}{r}{w. vs unw.}\vline & & & &&\multicolumn{2}{r}{1.049}& \multicolumn{2}{r}{1.343}\vline&\multicolumn{2}{r}{1.005\,\textbf{\ddag}}& \multicolumn{2}{r}{\textbf{0.978}\,\textbf{\ddag}}\\ 
\hdashline[2pt/2pt]ET &$\ISB$& & & & &0.88&0.87&0.93&1.16&0.85&0.82&1.03&1.50\\ 
&$\IVar$& & & & &0.93&0.93&0.53&0.52&0.92&0.93&0.52&0.51\\ 
&$\MISE$& & & & &\textbf{0.93}\,\textbf{\ddag}&\textbf{0.93}\,\textbf{\ddag}&\textbf{0.53}\,\textbf{\ddag}&\textbf{0.52}\,\textbf{\ddag}&\textbf{0.92}\,\textbf{\ddag}&\textbf{0.92}\,\textbf{\ddag}&\textbf{0.53}\,\textbf{\ddag}&\textbf{0.52}\,\textbf{\ddag}\\ 
\multicolumn{2}{r}{vs $d_{g}=3$}\vline & & & &&\textbf{0.87}\,\textbf{\ddag}&\textbf{0.87}\,\textbf{\ddag}&\textbf{0.63}\,\textbf{\ddag}&\textbf{0.62}\,\textbf{\ddag}&\textbf{0.86}\,\textbf{\ddag}&\textbf{0.87}\,\textbf{\ddag}&\textbf{0.62}\,\textbf{\ddag}&\textbf{0.61}\,\textbf{\ddag}\\ 
\multicolumn{2}{r}{w. vs unw.}\vline & & & &&\multicolumn{2}{r}{1.001\,\textbf{\ddag}}& \multicolumn{2}{r}{\textbf{0.982}\,\textbf{\ddag}}\vline&\multicolumn{2}{r}{1.002\,\textbf{\ddag}}& \multicolumn{2}{r}{\textbf{0.984}\,\textbf{\ddag}}\\ 
\hline
\multicolumn{13}{c}{$n=1000$}\\ \hline
CUE&$\ISB$&6.5&0.5&0.84&2.78&0.94&0.92&0.89&0.99&0.91&0.87&0.94&1.19\\ 
 &$\IVar$&115.0&45.7&1.09&0.81&0.94&0.94&0.54&0.53&0.94&0.94&0.54&0.52\\ 
 &$\MISE$&121.7&46.3&1.07\,\textbf{\ddag}&\textbf{0.83}\,\textbf{\ddag}&\textbf{0.94}\,\textbf{\ddag}&\textbf{0.94}\,\textbf{\ddag}&\textbf{0.54}\,\textbf{\ddag}&\textbf{0.53}\,\textbf{\ddag}&\textbf{0.94}\,\textbf{\ddag}&\textbf{0.94}\,\textbf{\ddag}&\textbf{0.54}\,\textbf{\ddag}&\textbf{0.53}\,\textbf{\ddag}\\ 
\multicolumn{2}{r}{vs $d_{g}=3$}\vline & & & &&\textbf{0.88}\,\textbf{\ddag}&\textbf{0.88}\,\textbf{\ddag}&\textbf{0.65}\,\textbf{\ddag}&\textbf{0.64}\,\textbf{\ddag}&\textbf{0.88}\,\textbf{\ddag}&\textbf{0.87}\,\textbf{\ddag}&\textbf{0.65}\,\textbf{\ddag}&\textbf{0.64}\,\textbf{\ddag}\\ 
\multicolumn{2}{r}{w. vs unw.}\vline & & & &&\multicolumn{2}{r}{\textbf{0.999}\,\textbf{\ddag}}& \multicolumn{2}{r}{\textbf{0.982}\,\textbf{\ddag}}\vline&\multicolumn{2}{r}{\textbf{0.999}\,\textbf{\ddag}}& \multicolumn{2}{r}{\textbf{0.980}\,\textbf{\ddag}}\\ 
\hdashline[2pt/2pt]EL &$\ISB$& & & & &0.94&0.95&0.87&0.89&0.92&0.92&0.84&0.88\\ 
&$\IVar$& & & & &0.94&0.94&0.54&0.53&0.94&0.94&0.54&0.53\\ 
&$\MISE$& & & & &\textbf{0.94}\,\textbf{\ddag}&\textbf{0.94}\,\textbf{\ddag}&\textbf{0.54}\,\textbf{\ddag}&\textbf{0.54}\,\textbf{\ddag}&\textbf{0.94}\,\textbf{\ddag}&\textbf{0.94}\,\textbf{\ddag}&\textbf{0.54}\,\textbf{\ddag}&\textbf{0.53}\,\textbf{\ddag}\\ 
\multicolumn{2}{r}{vs $d_{g}=3$}\vline & & & &&\textbf{0.88}\,\textbf{\ddag}&\textbf{0.88}\,\textbf{\ddag}&\textbf{0.65}\,\textbf{\ddag}&\textbf{0.64}\,\textbf{\ddag}&\textbf{0.87}\,\textbf{\ddag}&\textbf{0.88}\,\textbf{\ddag}&\textbf{0.65}\,\textbf{\ddag}&\textbf{0.64}\,\textbf{\ddag}\\ 
\multicolumn{2}{r}{w. vs unw.}\vline & & & &&\multicolumn{2}{r}{1.001\,\textbf{\ddag}}& \multicolumn{2}{r}{\textbf{0.982}\,\textbf{\ddag}}\vline&\multicolumn{2}{r}{1.003\,\textbf{\ddag}}& \multicolumn{2}{r}{\textbf{0.984}\,\textbf{\ddag}}\\ 
\hdashline[2pt/2pt]ET &$\ISB$& & & & &0.94&0.93&0.88&0.93&0.91&0.88&0.89&1.01\\ 
&$\IVar$& & & & &0.94&0.94&0.54&0.53&0.94&0.94&0.53&0.52\\ 
&$\MISE$& & & & &\textbf{0.94}\,\textbf{\ddag}&\textbf{0.94}\,\textbf{\ddag}&\textbf{0.54}\,\textbf{\ddag}&\textbf{0.53}\,\textbf{\ddag}&\textbf{0.94}\,\textbf{\ddag}&\textbf{0.94}\,\textbf{\ddag}&\textbf{0.54}\,\textbf{\ddag}&\textbf{0.53}\,\textbf{\ddag}\\ 
\multicolumn{2}{r}{vs $d_{g}=3$}\vline & & & &&\textbf{0.88}\,\textbf{\ddag}&\textbf{0.88}\,\textbf{\ddag}&\textbf{0.65}\,\textbf{\ddag}&\textbf{0.64}\,\textbf{\ddag}&\textbf{0.87}\,\textbf{\ddag}&\textbf{0.87}\,\textbf{\ddag}&\textbf{0.65}\,\textbf{\ddag}&\textbf{0.63}\,\textbf{\ddag}\\ 
\multicolumn{2}{r}{w. vs unw.}\vline & & & &&\multicolumn{2}{r}{\textbf{1.000}}& \multicolumn{2}{r}{\textbf{0.982}\,\textbf{\ddag}}\vline&\multicolumn{2}{r}{1.000}& \multicolumn{2}{r}{\textbf{0.983}\,\textbf{\ddag}}\\ 
\hline
\multicolumn{13}{c}{$n=2000$}\\ \hline
CUE&$\ISB$&4.2&0.3&0.93&1.88&0.99&0.97&0.96&0.97&0.96&0.92&0.94&1.01\\ 
 &$\IVar$&64.9&23.7&1.09&0.81&0.95&0.95&0.55&0.54&0.95&0.95&0.55&0.54\\ 
 &$\MISE$&69.1&24.0&1.08\,\textbf{\ddag}&\textbf{0.82}\,\textbf{\ddag}&\textbf{0.95}\,\textbf{\ddag}&\textbf{0.95}\,\textbf{\ddag}&\textbf{0.56}\,\textbf{\ddag}&\textbf{0.55}\,\textbf{\ddag}&\textbf{0.95}\,\textbf{\ddag}&\textbf{0.95}\,\textbf{\ddag}&\textbf{0.55}\,\textbf{\ddag}&\textbf{0.54}\,\textbf{\ddag}\\ 
\multicolumn{2}{r}{vs $d_{g}=3$}\vline & & & &&\textbf{0.88}\,\textbf{\ddag}&\textbf{0.88}\,\textbf{\ddag}&\textbf{0.68}\,\textbf{\ddag}&\textbf{0.67}\,\textbf{\ddag}&\textbf{0.88}\,\textbf{\ddag}&\textbf{0.87}\,\textbf{\ddag}&\textbf{0.68}\,\textbf{\ddag}&\textbf{0.66}\,\textbf{\ddag}\\ 
\multicolumn{2}{r}{w. vs unw.}\vline & & & &&\multicolumn{2}{r}{\textbf{0.999}\,\textbf{\ddag}}& \multicolumn{2}{r}{\textbf{0.983}\,\textbf{\ddag}}\vline&\multicolumn{2}{r}{\textbf{0.998}\,\textbf{\ddag}}& \multicolumn{2}{r}{\textbf{0.979}\,\textbf{\ddag}}\\ 
\hdashline[2pt/2pt]EL &$\ISB$& & & & &0.99&0.99&0.95&0.95&0.97&0.97&0.91&0.93\\ 
&$\IVar$& & & & &0.95&0.95&0.56&0.55&0.95&0.95&0.55&0.54\\ 
&$\MISE$& & & & &\textbf{0.95}\,\textbf{\ddag}&\textbf{0.95}\,\textbf{\ddag}&\textbf{0.56}\,\textbf{\ddag}&\textbf{0.55}\,\textbf{\ddag}&\textbf{0.95}\,\textbf{\ddag}&\textbf{0.95}\,\textbf{\ddag}&\textbf{0.56}\,\textbf{\ddag}&\textbf{0.54}\,\textbf{\ddag}\\ 
\multicolumn{2}{r}{vs $d_{g}=3$}\vline & & & &&\textbf{0.88}\,\textbf{\ddag}&\textbf{0.88}\,\textbf{\ddag}&\textbf{0.68}\,\textbf{\ddag}&\textbf{0.67}\,\textbf{\ddag}&\textbf{0.88}\,\textbf{\ddag}&\textbf{0.88}\,\textbf{\ddag}&\textbf{0.68}\,\textbf{\ddag}&\textbf{0.66}\,\textbf{\ddag}\\ 
\multicolumn{2}{r}{w. vs unw.}\vline & & & &&\multicolumn{2}{r}{\textbf{1.000}}& \multicolumn{2}{r}{\textbf{0.983}\,\textbf{\ddag}}\vline&\multicolumn{2}{r}{\textbf{1.000}}& \multicolumn{2}{r}{\textbf{0.978}\,\textbf{\ddag}}\\ 
\hdashline[2pt/2pt]ET &$\ISB$& & & & &0.99&0.98&0.96&0.95&0.95&0.93&0.92&0.94\\ 
&$\IVar$& & & & &0.95&0.95&0.55&0.54&0.95&0.95&0.55&0.54\\ 
&$\MISE$& & & & &\textbf{0.95}\,\textbf{\ddag}&\textbf{0.95}\,\textbf{\ddag}&\textbf{0.56}\,\textbf{\ddag}&\textbf{0.55}\,\textbf{\ddag}&\textbf{0.95}\,\textbf{\ddag}&\textbf{0.94}\,\textbf{\ddag}&\textbf{0.55}\,\textbf{\ddag}&\textbf{0.54}\,\textbf{\ddag}\\ 
\multicolumn{2}{r}{vs $d_{g}=3$}\vline & & & &&\textbf{0.88}\,\textbf{\ddag}&\textbf{0.88}\,\textbf{\ddag}&\textbf{0.68}\,\textbf{\ddag}&\textbf{0.67}\,\textbf{\ddag}&\textbf{0.87}\,\textbf{\ddag}&\textbf{0.87}\,\textbf{\ddag}&\textbf{0.67}\,\textbf{\ddag}&\textbf{0.66}\,\textbf{\ddag}\\ 
\multicolumn{2}{r}{w. vs unw.}\vline & & & &&\multicolumn{2}{r}{\textbf{0.999}\,\textbf{\ddag}}& \multicolumn{2}{r}{\textbf{0.983}\,\textbf{\ddag}}\vline&\multicolumn{2}{r}{\textbf{0.999}\,\textbf{\ddag}}& \multicolumn{2}{r}{\textbf{0.980}\,\textbf{\ddag}}\\ 
\hline
\hline 
\end{tabular}

\vspace*{-0.8\baselineskip}
\begin{flushleft}
\textbf{Notes:} see text. 
\end{flushleft}
\end{table}

\begin{table}[p]\centering\footnotesize
\caption{Performance of GEL-based residual density and distribution function estimators in the IHS transformation model, $\arsinh(0.08y)/0.08 = 1+2x+u$, in Scenario 3}
\label{Tab:Supp:mcIHSmainD3PDFTabB1}
\begin{tabular}{ll|rr|rr|rrrr|rrrr}\hline\hline
 & &\multirow{3}{*}{$\tilde{f}$} & \multirow{3}{*}{$\widetilde{F}$}&\multicolumn{2}{c}{$d_{g}=3$}\vline&\multicolumn{4}{c}{$d_{g}=4$}\vline&\multicolumn{4}{c}{$d_{g}=5$}\\ \cline{5-14} 
 & & & & & & & & & & & & & \\[-0.9em] &  & & & $\hat{f}$ & $\widehat{F}$  & $\hat{f}$ & $\hat{f}_{\rho}$ & $\widehat{F}$ & $\widehat{F}_{\rho}$ & $\hat{f}$ & $\hat{f}_{\rho}$ & $\widehat{F}$ & $\widehat{F}_{\rho}$ \\ \hline 
\multicolumn{13}{c}{$n=100$}\\ \hline
CUE&$\ISB$&23.9&1.3&7.34&41.60&4.25&5.01&14.61&24.59&4.68&6.19&17.12&33.44\\ 
 &$\IVar$&1546.0&485.4&1.19&0.76&1.01&1.01&0.53&0.50&1.01&1.02&0.55&0.50\\ 
 &$\MISE$&1570.5&487.0&1.28\,\textbf{\ddag}&\textbf{0.88}\,\textbf{\ddag}&1.06\,\textbf{\ddag}&1.08\,\textbf{\ddag}&\textbf{0.57}\,\textbf{\ddag}&\textbf{0.56}\,\textbf{\ddag}&1.07\,\textbf{\ddag}&1.10\,\textbf{\ddag}&\textbf{0.59}\,\textbf{\ddag}&\textbf{0.59}\,\textbf{\ddag}\\ 
\multicolumn{2}{r}{vs $d_{g}=3$}\vline & & & &&\textbf{0.82}\,\textbf{\ddag}&\textbf{0.84}\,\textbf{\ddag}&\textbf{0.65}\,\textbf{\ddag}&\textbf{0.65}\,\textbf{\ddag}&\textbf{0.83}\,\textbf{\ddag}&\textbf{0.86}\,\textbf{\ddag}&\textbf{0.67}\,\textbf{\ddag}&\textbf{0.68}\,\textbf{\ddag}\\ 
\multicolumn{2}{r}{w. vs unw.}\vline & & & &&\multicolumn{2}{r}{1.018\,\textbf{\ddag}}& \multicolumn{2}{r}{\textbf{0.986}\,\textbf{\ddag}}\vline&\multicolumn{2}{r}{1.033\,\textbf{\ddag}}& \multicolumn{2}{r}{\textbf{0.999}}\\ 
\hdashline[2pt/2pt]EL &$\ISB$& & & & &4.08&4.08&13.99&15.96&4.28&4.38&13.51&17.07\\ 
&$\IVar$& & & & &0.99&1.01&0.51&0.49&1.00&1.04&0.54&0.51\\ 
&$\MISE$& & & & &1.04\,\textbf{\ddag}&1.05\,\textbf{\ddag}&\textbf{0.55}\,\textbf{\ddag}&\textbf{0.53}\,\textbf{\ddag}&1.05\,\textbf{\ddag}&1.09\,\textbf{\ddag}&\textbf{0.57}\,\textbf{\ddag}&\textbf{0.55}\,\textbf{\ddag}\\ 
\multicolumn{2}{r}{vs $d_{g}=3$}\vline & & & &&\textbf{0.81}\,\textbf{\ddag}&\textbf{0.82}\,\textbf{\ddag}&\textbf{0.62}\,\textbf{\ddag}&\textbf{0.61}\,\textbf{\ddag}&\textbf{0.82}\,\textbf{\ddag}&\textbf{0.85}\,\textbf{\ddag}&\textbf{0.65}\,\textbf{\ddag}&\textbf{0.64}\,\textbf{\ddag}\\ 
\multicolumn{2}{r}{w. vs unw.}\vline & & & &&\multicolumn{2}{r}{1.014\,\textbf{\ddag}}& \multicolumn{2}{r}{\textbf{0.972}\,\textbf{\ddag}}\vline&\multicolumn{2}{r}{1.034\,\textbf{\ddag}}& \multicolumn{2}{r}{\textbf{0.972}}\\ 
\hdashline[2pt/2pt]ET &$\ISB$& & & & &4.14&4.71&14.22&22.01&4.53&5.69&14.89&27.85\\ 
&$\IVar$& & & & &0.99&1.01&0.50&0.49&1.00&1.02&0.66&0.84\\ 
&$\MISE$& & & & &1.03\,\textbf{\ddag}&1.06\,\textbf{\ddag}&\textbf{0.54}\,\textbf{\ddag}&\textbf{0.54}\,\textbf{\ddag}&1.05\,\textbf{\ddag}&1.09\,\textbf{\ddag}&\textbf{0.78}&1.91\\ 
\multicolumn{2}{r}{vs $d_{g}=3$}\vline & & & &&\textbf{0.81}\,\textbf{\ddag}&\textbf{0.83}\,\textbf{\ddag}&\textbf{0.62}\,\textbf{\ddag}&\textbf{0.62}\,\textbf{\ddag}&\textbf{0.82}\,\textbf{\ddag}&\textbf{0.85}\,\textbf{\ddag}&\textbf{0.90}&2.25\\ 
\multicolumn{2}{r}{w. vs unw.}\vline & & & &&\multicolumn{2}{r}{1.027\,\textbf{\ddag}}& \multicolumn{2}{r}{1.006}\vline&\multicolumn{2}{r}{1.040\,\textbf{\ddag}}& \multicolumn{2}{r}{2.459}\\ 
\hline
\multicolumn{13}{c}{$n=500$}\\ \hline
CUE&$\ISB$&9.6&0.4&2.39&13.99&1.54&1.58&2.61&3.94&1.60&1.69&2.84&5.19\\ 
 &$\IVar$&379.2&100.3&1.10&0.76&1.00&1.00&0.51&0.50&1.00&1.00&0.51&0.50\\ 
 &$\MISE$&388.9&100.8&1.13\,\textbf{\ddag}&\textbf{0.81}\,\textbf{\ddag}&1.01\,\textbf{\ddag}&1.01\,\textbf{\ddag}&\textbf{0.52}\,\textbf{\ddag}&\textbf{0.51}\,\textbf{\ddag}&1.01\,\textbf{\ddag}&1.02\,\textbf{\ddag}&\textbf{0.52}\,\textbf{\ddag}&\textbf{0.51}\,\textbf{\ddag}\\ 
\multicolumn{2}{r}{vs $d_{g}=3$}\vline & & & &&\textbf{0.89}\,\textbf{\ddag}&\textbf{0.90}\,\textbf{\ddag}&\textbf{0.64}\,\textbf{\ddag}&\textbf{0.63}\,\textbf{\ddag}&\textbf{0.90}\,\textbf{\ddag}&\textbf{0.90}\,\textbf{\ddag}&\textbf{0.65}\,\textbf{\ddag}&\textbf{0.64}\,\textbf{\ddag}\\ 
\multicolumn{2}{r}{w. vs unw.}\vline & & & &&\multicolumn{2}{r}{1.002\,\textbf{\ddag}}& \multicolumn{2}{r}{\textbf{0.981}\,\textbf{\ddag}}\vline&\multicolumn{2}{r}{1.004\,\textbf{\ddag}}& \multicolumn{2}{r}{\textbf{0.982}\,\textbf{\ddag}}\\ 
\hdashline[2pt/2pt]EL &$\ISB$& & & & &1.54&1.55&2.55&2.65&1.57&1.59&2.46&2.59\\ 
&$\IVar$& & & & &1.00&1.00&0.51&0.50&1.00&1.07&0.51&0.80\\ 
&$\MISE$& & & & &1.01\,\textbf{\ddag}&1.01\,\textbf{\ddag}&\textbf{0.51}\,\textbf{\ddag}&\textbf{0.50}\,\textbf{\ddag}&1.01\,\textbf{\ddag}&1.09&\textbf{0.52}\,\textbf{\ddag}&\textbf{0.81}\\ 
\multicolumn{2}{r}{vs $d_{g}=3$}\vline & & & &&\textbf{0.89}\,\textbf{\ddag}&\textbf{0.89}\,\textbf{\ddag}&\textbf{0.64}\,\textbf{\ddag}&\textbf{0.62}\,\textbf{\ddag}&\textbf{0.89}\,\textbf{\ddag}&\textbf{0.96}&\textbf{0.64}\,\textbf{\ddag}&\textbf{1.00}\\ 
\multicolumn{2}{r}{w. vs unw.}\vline & & & &&\multicolumn{2}{r}{1.002\,\textbf{\ddag}}& \multicolumn{2}{r}{\textbf{0.979}\,\textbf{\ddag}}\vline&\multicolumn{2}{r}{1.074}& \multicolumn{2}{r}{1.563}\\ 
\hdashline[2pt/2pt]ET &$\ISB$& & & & &1.53&1.56&2.59&3.40&1.58&1.64&2.65&4.11\\ 
&$\IVar$& & & & &1.00&1.00&0.51&0.49&1.00&1.00&0.50&0.49\\ 
&$\MISE$& & & & &1.01\,\textbf{\ddag}&1.01\,\textbf{\ddag}&\textbf{0.51}\,\textbf{\ddag}&\textbf{0.50}\,\textbf{\ddag}&1.01\,\textbf{\ddag}&1.01\,\textbf{\ddag}&\textbf{0.51}\,\textbf{\ddag}&\textbf{0.51}\,\textbf{\ddag}\\ 
\multicolumn{2}{r}{vs $d_{g}=3$}\vline & & & &&\textbf{0.89}\,\textbf{\ddag}&\textbf{0.89}\,\textbf{\ddag}&\textbf{0.64}\,\textbf{\ddag}&\textbf{0.62}\,\textbf{\ddag}&\textbf{0.89}\,\textbf{\ddag}&\textbf{0.90}\,\textbf{\ddag}&\textbf{0.64}\,\textbf{\ddag}&\textbf{0.63}\,\textbf{\ddag}\\ 
\multicolumn{2}{r}{w. vs unw.}\vline & & & &&\multicolumn{2}{r}{1.002\,\textbf{\ddag}}& \multicolumn{2}{r}{\textbf{0.982}\,\textbf{\ddag}}\vline&\multicolumn{2}{r}{1.004\,\textbf{\ddag}}& \multicolumn{2}{r}{\textbf{0.985}\,\textbf{\ddag}}\\ 
\hline
\multicolumn{13}{c}{$n=1000$}\\ \hline
CUE&$\ISB$&6.6&0.2&1.86&8.15&1.33&1.34&1.66&2.16&1.37&1.39&1.80&2.79\\ 
 &$\IVar$&206.9&50.4&1.09&0.74&1.00&1.00&0.51&0.50&1.00&1.00&0.51&0.50\\ 
 &$\MISE$&213.5&50.6&1.12\,\textbf{\ddag}&\textbf{0.77}\,\textbf{\ddag}&1.01\,\textbf{\ddag}&1.01\,\textbf{\ddag}&\textbf{0.52}\,\textbf{\ddag}&\textbf{0.51}\,\textbf{\ddag}&1.01\,\textbf{\ddag}&1.01\,\textbf{\ddag}&\textbf{0.52}\,\textbf{\ddag}&\textbf{0.51}\,\textbf{\ddag}\\ 
\multicolumn{2}{r}{vs $d_{g}=3$}\vline & & & &&\textbf{0.90}\,\textbf{\ddag}&\textbf{0.90}\,\textbf{\ddag}&\textbf{0.67}\,\textbf{\ddag}&\textbf{0.66}\,\textbf{\ddag}&\textbf{0.91}\,\textbf{\ddag}&\textbf{0.91}\,\textbf{\ddag}&\textbf{0.67}\,\textbf{\ddag}&\textbf{0.66}\,\textbf{\ddag}\\ 
\multicolumn{2}{r}{w. vs unw.}\vline & & & &&\multicolumn{2}{r}{1.001\,\textbf{\ddag}}& \multicolumn{2}{r}{\textbf{0.980}\,\textbf{\ddag}}\vline&\multicolumn{2}{r}{1.001\,\textbf{\ddag}}& \multicolumn{2}{r}{\textbf{0.978}\,\textbf{\ddag}}\\ 
\hdashline[2pt/2pt]EL &$\ISB$& & & & &1.34&1.35&1.65&1.64&1.36&1.37&1.68&1.74\\ 
&$\IVar$& & & & &1.00&1.13&0.51&1.09&1.00&1.00&0.51&0.50\\ 
&$\MISE$& & & & &1.01\,\textbf{\ddag}&1.14&\textbf{0.52}\,\textbf{\ddag}&1.09&1.01\,\textbf{\ddag}&1.01\,\textbf{\ddag}&\textbf{0.52}\,\textbf{\ddag}&\textbf{0.50}\,\textbf{\ddag}\\ 
\multicolumn{2}{r}{vs $d_{g}=3$}\vline & & & &&\textbf{0.90}\,\textbf{\ddag}&1.02&\textbf{0.67}\,\textbf{\ddag}&1.41&\textbf{0.90}\,\textbf{\ddag}&\textbf{0.90}\,\textbf{\ddag}&\textbf{0.67}\,\textbf{\ddag}&\textbf{0.65}\,\textbf{\ddag}\\ 
\multicolumn{2}{r}{w. vs unw.}\vline & & & &&\multicolumn{2}{r}{1.124}& \multicolumn{2}{r}{2.108}\vline&\multicolumn{2}{r}{1.001\,\textbf{\dag}}& \multicolumn{2}{r}{\textbf{0.973}\,\textbf{\ddag}}\\ 
\hdashline[2pt/2pt]ET &$\ISB$& & & & &1.34&1.34&1.66&1.91&1.36&1.37&1.76&2.32\\ 
&$\IVar$& & & & &1.00&1.00&0.51&0.50&1.00&1.00&0.51&0.49\\ 
&$\MISE$& & & & &1.01\,\textbf{\ddag}&1.01\,\textbf{\ddag}&\textbf{0.51}\,\textbf{\ddag}&\textbf{0.50}\,\textbf{\ddag}&1.01\,\textbf{\ddag}&1.01\,\textbf{\ddag}&\textbf{0.51}\,\textbf{\ddag}&\textbf{0.50}\,\textbf{\ddag}\\ 
\multicolumn{2}{r}{vs $d_{g}=3$}\vline & & & &&\textbf{0.90}\,\textbf{\ddag}&\textbf{0.90}\,\textbf{\ddag}&\textbf{0.67}\,\textbf{\ddag}&\textbf{0.65}\,\textbf{\ddag}&\textbf{0.90}\,\textbf{\ddag}&\textbf{0.90}\,\textbf{\ddag}&\textbf{0.66}\,\textbf{\ddag}&\textbf{0.65}\,\textbf{\ddag}\\ 
\multicolumn{2}{r}{w. vs unw.}\vline & & & &&\multicolumn{2}{r}{1.001\,\textbf{\ddag}}& \multicolumn{2}{r}{\textbf{0.979}\,\textbf{\ddag}}\vline&\multicolumn{2}{r}{1.001\,\textbf{\ddag}}& \multicolumn{2}{r}{\textbf{0.979}\,\textbf{\ddag}}\\ 
\hline
\multicolumn{13}{c}{$n=2000$}\\ \hline
CUE&$\ISB$&4.0&0.1&1.60&4.53&1.24&1.23&1.34&1.54&1.25&1.25&1.41&1.86\\ 
 &$\IVar$&113.2&25.6&1.10&0.74&1.00&1.00&0.52&0.51&1.00&1.00&0.52&0.51\\ 
 &$\MISE$&117.2&25.8&1.11\,\textbf{\ddag}&\textbf{0.76}\,\textbf{\ddag}&1.01\,\textbf{\ddag}&1.01\,\textbf{\ddag}&\textbf{0.53}\,\textbf{\ddag}&\textbf{0.52}\,\textbf{\ddag}&1.01\,\textbf{\ddag}&1.01\,\textbf{\ddag}&\textbf{0.52}\,\textbf{\ddag}&\textbf{0.51}\,\textbf{\ddag}\\ 
\multicolumn{2}{r}{vs $d_{g}=3$}\vline & & & &&\textbf{0.91}\,\textbf{\ddag}&\textbf{0.91}\,\textbf{\ddag}&\textbf{0.69}\,\textbf{\ddag}&\textbf{0.68}\,\textbf{\ddag}&\textbf{0.91}\,\textbf{\ddag}&\textbf{0.91}\,\textbf{\ddag}&\textbf{0.69}\,\textbf{\ddag}&\textbf{0.67}\,\textbf{\ddag}\\ 
\multicolumn{2}{r}{w. vs unw.}\vline & & & &&\multicolumn{2}{r}{\textbf{0.999}\,\textbf{\ddag}}& \multicolumn{2}{r}{\textbf{0.981}\,\textbf{\ddag}}\vline&\multicolumn{2}{r}{\textbf{0.999}\,\textbf{\ddag}}& \multicolumn{2}{r}{\textbf{0.977}\,\textbf{\ddag}}\\ 
\hdashline[2pt/2pt]EL &$\ISB$& & & & &1.25&1.25&1.34&1.35&1.25&1.26&1.36&1.38\\ 
&$\IVar$& & & & &1.00&1.00&0.52&0.51&1.00&1.11&0.52&0.91\\ 
&$\MISE$& & & & &1.01\,\textbf{\ddag}&1.01\,\textbf{\ddag}&\textbf{0.53}\,\textbf{\ddag}&\textbf{0.52}\,\textbf{\ddag}&1.01\,\textbf{\ddag}&1.12&\textbf{0.53}\,\textbf{\ddag}&\textbf{0.92}\\ 
\multicolumn{2}{r}{vs $d_{g}=3$}\vline & & & &&\textbf{0.91}\,\textbf{\ddag}&\textbf{0.91}\,\textbf{\ddag}&\textbf{0.69}\,\textbf{\ddag}&\textbf{0.68}\,\textbf{\ddag}&\textbf{0.91}\,\textbf{\ddag}&1.00&\textbf{0.69}\,\textbf{\ddag}&1.20\\ 
\multicolumn{2}{r}{w. vs unw.}\vline & & & &&\multicolumn{2}{r}{\textbf{0.999}\,\textbf{\ddag}}& \multicolumn{2}{r}{\textbf{0.980}\,\textbf{\ddag}}\vline&\multicolumn{2}{r}{1.105}& \multicolumn{2}{r}{1.743}\\ 
\hdashline[2pt/2pt]ET &$\ISB$& & & & &1.24&1.24&1.34&1.43&1.24&1.25&1.39&1.62\\ 
&$\IVar$& & & & &1.00&1.00&0.52&0.51&1.00&1.00&0.52&0.51\\ 
&$\MISE$& & & & &1.01\,\textbf{\ddag}&1.01\,\textbf{\ddag}&\textbf{0.53}\,\textbf{\ddag}&\textbf{0.52}\,\textbf{\ddag}&1.01\,\textbf{\ddag}&1.01\,\textbf{\ddag}&\textbf{0.52}\,\textbf{\ddag}&\textbf{0.51}\,\textbf{\ddag}\\ 
\multicolumn{2}{r}{vs $d_{g}=3$}\vline & & & &&\textbf{0.91}\,\textbf{\ddag}&\textbf{0.91}\,\textbf{\ddag}&\textbf{0.69}\,\textbf{\ddag}&\textbf{0.68}\,\textbf{\ddag}&\textbf{0.91}\,\textbf{\ddag}&\textbf{0.91}\,\textbf{\ddag}&\textbf{0.69}\,\textbf{\ddag}&\textbf{0.67}\,\textbf{\ddag}\\ 
\multicolumn{2}{r}{w. vs unw.}\vline & & & &&\multicolumn{2}{r}{\textbf{0.999}\,\textbf{\ddag}}& \multicolumn{2}{r}{\textbf{0.980}\,\textbf{\ddag}}\vline&\multicolumn{2}{r}{\textbf{0.999}\,\textbf{\ddag}}& \multicolumn{2}{r}{\textbf{0.977}\,\textbf{\ddag}}\\ 
\hline
\hline 
\end{tabular}

\vspace*{-0.8\baselineskip}
\begin{flushleft}
\textbf{Notes:} see text. 
\end{flushleft}
\end{table}
\setlength{\tabcolsep}{6pt} 
\renewcommand{\arraystretch}{1}

\subsubsection{Scenario 1}
The first $\sim b$ term in eq. \eqref{Eq:MC.Case1.Rel.IVar.hat.f.to.tilde.f} is approximately $-0.321n^{-1/9}$, which for $n=100$, $500$, $1000$, and $2000$ is approximately $-0.192$, $-0.161$, $-0.149$, and $-0.138$ respectively. The second $\sim b$ term is
approximately $0.04728n^{-1/9}$ for $d_{g}=4$ and $0.04708n^{-1/9}$ for $d_{g}=5$, which offsets the reduction in variance slightly. The predicted relative IVar of $\hat{f}$ and $\hat{f}_{\rho}$ up to order $\littleO(b)$ is thus $0.836$, $0.863$, $0.873$ and $0.882$ for $n=100$, $500$, $1000$, and $2000$ respectively and is identical within three digit precision for $d_{g}=4$ and $5$.

The results reported in Table \ref{Tab:Supp:mcIHSmainD1PDFTabB1} confirm these predictions. In fact, the reduction in variance is even larger than expected in small and medium samples due to the $\littleO(b)$ effects. Furthermore, estimators $\hat{f}$ and $\hat{f}_{\rho}$ have smaller ISB relative to $\tilde{f}$. A comparison of $\hat{f}$ and $\hat{f}_{\rho}$ between $d_{g}=3$ (just-identified) and $d_{g}=4,5$ (over-identified) for moderate and larger sample sizes emphasises further the contribution of additional moment information. Hence $\hat{f}$ and $\hat{f}_{\rho}$ enjoy a reduction in MISE of as much as $21\%$ for $n=100$ and $10\%$ for $n=2000$ relative to $\tilde{f}$. The benefits are even more pronounced for c.d.f. estimation, where the reduction in MISE can be as much as $56\%$ for $n=100$ and around $53\%$ in moderate samples. There are also small but statistically significant benefits to re-weighting which are mostly due to the smaller biases of $\hat{f}_{\rho}$ and $\widehat{F}
_{\rho}$ relative to $\hat{f}$ and $\widehat{F}$ at moderate and larger sample sizes. There is some deterioration in ISB, IVar and, thus, MISE with increases in $d_{g}$ which can be contributed to the increased importance of outliers.

Finally, while in moderate and large samples the performances of CUE, EL, and ET are virtually identical, in small samples ET can be unstable with larger $d_{g}$.

\subsubsection{Scenarios 2 and 3}
Scenarios 2 and 3 with densities of $(x,u)$ which are heavy-tailed and also, e.g., skewed and bimodal, illustrate the many difficulties for both GEL estimation and kernel p.d.f. and c.d.f. estimation which are absent in the relatively benign Scenario 1.

The performance of CUE in small samples is generally worse than that of EL and ET. It ranks last by MSE in both scenarios with $n=100$ and $500$, except Scenario 3 with $n=100$ where ET underperforms. In a number of cases increasing with $d_{g}$ the optimisation routine for ET failed. Somewhat surprisingly, although it is known to be sensitive to outliers, EL appears to deliver good results in the simulation experiments. It ranks first by MSE in Scenario 3 with $d_{g}=5$ and alternates with ET otherwise. These differences become very small with $n=1,000$ and greater.


The conclusion about the inferior performance of CUE in small samples holds true for CUE-based kernel density p.d.f. and c.d.f. estimators as well; see Tables \ref{Tab:Supp:mcIHSmainD2PDFTabB1} and \ref{Tab:Supp:mcIHSmainD3PDFTabB1}, in particular, the ISBs of $\hat{f}$ and $\hat{f}_{\rho}$ with $d_{g}=4,5$ in Table \ref{Tab:Supp:mcIHSmainD2PDFTabB1}.
However, the ranking of EL and ET-based kernel density p.d.f. and c.d.f. estimators by MISE does not always correspond to the ranking of the underlying EL and ET estimators of $\beta_{0}$ by MSE. 
In particular, the sensitivity of EL to outliers adversely affects the estimators $\hat{f}_{\rho}$ and $\widehat{F}_{\rho}$ via the implied probabilities in Scenario 3 with $n=500$ and greater; see Table \ref{Tab:Supp:mcIHSmainD3PDFTabB1}. ET and CUE perform better in those cases. 

Unlike Scenario 1, in Scenario 3 none of the feasible kernel density estimators have smaller MISE than their infeasible counterparts for the sample sizes considered. In Scenario 2, with less complicated distributional features, these estimators do achieve a reduction in MISE with $d_{g}=4,5$. The same is true for the feasible kernel c.d.f. estimators in Scenario 2 with $d_{g}=3,4,5$, and more often than not in Scenario 3 as well, with the few exceptions mentioned above. 
Importantly, it is generally beneficial to increase the number of moment conditions beyond those necessary to identify the parameters except when stability of GEL estimators of $\beta_{0}$ is likely to deteriorate. 


Finally, the benefits of re-weighting are present, but not universal, and as expected, are quite small; cf. Supplement \ref{Supp:Examples}: Example \ref{Sec.Example:normal.over.GG}.

\section{Summary and Conclusions}\label{sec:conclusions}
Large sample results and simulation evidence reported in this paper suggest that it is generally sensible to apply either the standard or re-weighted kernel estimators to estimate the p.d.f. or c.d.f. of a scalar residual $u(z,\beta_{0})$ in a variety of situations, provided error associated with the estimation of $\beta_{0}$ satisfies some mild regularity conditions and care is taken to ensure the bandwidth is not too small. If the assumptions on $u(z,\beta)$ prove difficult to verify in practice, using fourth or higher order kernels and the corresponding asymptotically optimal bandwidths will generally assist with ensuring the appropriate regularity conditions hold.

Incorporating information from overidentifying moment conditions by re-weighting the estimators using GEL implied probabilities offers efficiency gains which are realised in regular situations. However, if the model is highly nonlinear and the distribution of the data is heavy-tailed or contaminated with outliers, the methods proposed in this paper, including GEL, should be applied with some caution in very small samples. Robustified hybrid estimators such as the exponentially tilted empirical likelihood,
see, e.g., \citet{schennach2007}, may prove useful in these circumstances.

While the results in this paper were presented only for the scalar-valued $u(z,\beta)$, generalisations to the vector case are relatively straightforward provided an analogue of the bijection Assumption \ref{Assumption.reformulation.bijection} holds.

An issue for future research to usefully address is the construction of tests for overidentifying moment conditions or parametric restrictions based on the differences between the kernel p.d.f. estimators $\hat{f}_{\rho}$ and $\hat{f}$ or $\tilde{f}_{\rho}$ and $\tilde{f}$ for known $\beta_{0}$. Test statistics of the Bickel-Rosenblatt type based on the integrated squared difference $\int(\hat{f}_{\rho}(u)-\hat{f}(u))^{2}\mathrm{d}u$, \citet{bickel1973}, \citet{fan1994,fan1998}, or the integrated absolute difference, \citet{cao2005}, would be of interest. Alternatively, Kolmogorov-Smirnov or Cram\'{e}r-von Mises-type tests could be constructed based on the differences between kernel c.d.f. estimators.


\phantomsection
\addcontentsline{toc}{section}{References}
\small
\setlength{\bibsep}{0pt plus 0.3ex}

\normalsize
\clearpage
\setcounter{page}{1}
\setcounter{section}{0}

\title{\scshape Supplement A to \textquotedblleft Improved Density and Distribution Function Estimation\textquotedblright: Proofs} 
\author{\protect\begin{tabular}{cp{10mm}c}
Vitaliy Oryshchenko\textsuperscript{*}
&  & Richard J. Smith \\ 
Department of Economics  & & c{\it e}mmap, U.C.L and I.F.S.\\
University of Manchester & & Faculty of Economics, University of Cambridge \\
 & & Department of Economics, University of Melbourne \\
 & & ONS Economic Statistics Centre of Excellence
 \protect\end{tabular}}
\date{\today}
\maketitle

\phantomsection
\addcontentsline{toc}{section}{Supplement A: Proofs}
\textcolor{white}{ 
\customlabel{Supp:Proofs}{A}
}
\vspace*{-\baselineskip} 
\renewcommand{\thepage}{[A.\arabic{page}]}
\renewcommand{\thesection}{A.\arabic{section}}
\renewcommand{\theequation}{A.\arabic{equation}}
\renewcommand{\theassumption}{A.\arabic{assumption}}
\renewcommand{\theremark}{A.\arabic{remark}}
\renewcommand{\thelemma}{A.\arabic{lemma}}
\renewcommand{\thecorollary}{A.\arabic{corollary}}
Throughout the Appendix, $0<C<\infty$ and $0\leq \omega\leq1$ will denote generic constants that may be different in different uses. CS, T, and H refer to the Cauchy-Schwarz, triangle, and H\"{o}lder inequalities, respectively with LIE and WLLN the law of iterated expectations and Khintchine's i.i.d. weak law of large numbers. MVT is the mean value theorem.

In addition, $\interior(\cdot)$ denotes the interior of $\cdot$, w.p.(a.)1 with probability (approaching) 1, and $\mathcal{N}$ is an open neighbourhood of $\beta_{0}$.

\section{GEL Stochastic Expansions}\label{App:GEL.stochastic.expansions}
The following identification and regularity conditions are imposed.
\begin{assumption}\label{Assumption.GEL.1}
(a) $\beta_{0}\in\mathcal{B}$ is the unique solution to $\E[g(z,\beta)]=0$; 
(b) $\mathcal{B}$ is compact; 
(c) $g(z,\beta)$ is continuous at each $\beta\in\mathcal{B}$ w.p.1;
(d) $\E[\sup_{\beta\in\mathcal{B}}\norm{g(z,\beta)}^{2}]<\infty$;
(e) $\Omega$ is nonsingular; 
(f) $\rho(v)$ is twice continuously differentiable in a neighbourhood of zero.
\end{assumption}
Assumption \ref{Assumption.GEL.1} is \citet[Assumption 1]{newey2004} and is sufficient for the consistency of $\hat{\beta}$. Moreover, $\hat{\lambda}=\argmax_{\lambda\in\Lambda_{n}(\hat{\beta})}P_{n}(\hat{\beta},\lambda)$ exists w.p.a.1 and $\hat{\lambda}=\bigO[p](n^{-1/2})$; see \citet[Theorem 3.1]{newey2004}.

\begin{assumption}\label{Assumption.GEL.2}
(a) $\beta_{0}\in\interior(\mathcal{B})$;
(b) $g(z,\beta)$ is continuously differentiable for $\beta\in\mathcal{N}$ and \\
$\E[\sup_{\beta\in\mathcal{N}}\norm{\nabla g(z,\beta)}]<\infty$; 
(c) $\mathop{\mathrm{rank}}(G)=d_{\beta}$.
\end{assumption}
Assumption \ref{Assumption.GEL.2} is \citet[Assumption 2]{newey2004}. If Assumptions \ref{Assumption.GEL.1} and \ref{Assumption.GEL.2} hold then $n^{1/2}((\hat{\beta}-\beta_{0})^{\transp},\:\hat{\lambda}^{\transp})^{\transp}\xrightarrow{d}N\left(0,\mathop{\mathrm{diag}}(\Sigma,P)\right)$; see \citet[Theorem 3.2]{newey2004}.

Let $\nabla^{2}g(z,\beta)$ denote a vector of all distinct second order partial derivatives with respect to $\beta$.
\begin{assumption}\label{Assumption.GEL.3} \hspace*{-0.7em}
(a) $\E[\norm{g(z,\beta_{0})}^{6}]<\infty$;
(b) $g(z,\beta)$ is twice differentiable for $\beta\in\mathcal{N}$, 
$\E[\norm{\nabla g(z,\beta_{0})}^{4}]$ $<\infty$, 
$\E[\norm{\nabla^{2}g(z,\beta_{0})}^{2}]<\infty$;
(c) there exists $d(z)\geq0$ with $\E[d(z)^{2}]<\infty$ such that $\norm{\nabla^{2}g(z,\beta)-\nabla^{2}g(z,\beta_{0})}\leq d(z)\norm{\beta-\beta_{0}}$ for all $z$ and $\beta\in\mathcal{N}$;
(d) $\rho(v)$ is four times differentiable with Lipschitz fourth derivative in a neighbourhood of zero.
\end{assumption}
Cf. \citet[Assumption 3]{newey2004}.

Write $\tilde{g}=n^{-1}\sum_{i=1}^{n}g_{i}$, $\widetilde{G}=n^{-1}\sum_{i=1}^{n}G_{i}-G$, and $\widetilde{\Omega}=n^{-1}\sum_{i=1}^{n}g_{i}g_{i}^{\transp}-\Omega$. Also let $g_{i}^{j}=\partial g(z_{i},\beta_{0})/\partial\beta_{j}$ and $G_{i}^{j}=\partial^{2}g(z_{i},\beta_{0})/\partial\beta_{j}\partial\beta^{\transp}$, $j=1,\ldots,d_{\beta}$. 
From the proof of Theorem 3.4 in \citet{newey2004}, GEL estimators satisfy the following stochastic expansion
\begin{equation}\label{Eq:GEL.stochastic.expansion}
\begin{bmatrix}\hat{\beta}-\beta_{0}\\\hat{\lambda}\end{bmatrix}
= -\begin{bmatrix}H\\P\end{bmatrix}\tilde{g} + \begin{bmatrix}-\Sigma & H\\ H^{\transp} & P\end{bmatrix}\tilde{\zeta}+\bigO[p](n^{-3/2}),
\end{equation}
where 
\begin{align*}
\tilde{\zeta} & = \left\{\begin{bmatrix}0& \widetilde{G}^{\transp} \\\widetilde{G} &\widetilde{\Omega}\end{bmatrix}
-\frac{1}{2}\sum_{j=1}^{d_{\beta}}[H\tilde{g}]_{j}\begin{bmatrix} 
0 & E[G_{i}^{j}]^{\transp} 
\\ E[G_{i}^{j}] & E[g_{i}^{j}g_{i}^{\transp}+g_{i}g_{i}^{j\transp}] \end{bmatrix} 
\right. \\ & \qquad \left.
-\frac{1}{2}\sum_{j=1}^{d_{g}}[P\tilde{g}]_{j}\begin{bmatrix}
E[\partial^{2}g_{ij}/\partial\beta\partial\beta^{\transp}] & 
E[G_{i}^{\transp}\bv_{j}g_{i}^{\transp}+g_{ij}G_{i}^{\transp}] \\ 
E[g_{i}\bv_{j}^{\transp}G_{i}+g_{ij}G_{i}] & -\rho_{3}E[g_{ij}g_{i}g_{i}^{\transp}]
\end{bmatrix}
\right\}\begin{bmatrix}H\\P\end{bmatrix}\tilde{g}.
\end{align*}

\begin{remark} Write $\tilde{\zeta}=(\tilde{\zeta}_{\beta}^{\transp},\tilde{\zeta}_{\lambda}^{\transp})^{\transp}$ partitioned conformably with $\beta$ and $\lambda$. Then $\E[\tilde{\zeta}_{\beta}]=0$ and $\E[\tilde{\zeta}_{\lambda}]=\zeta_{\lambda}$ given in eq. \eqref{Eq:GEL.stochastic.expansion.2nd.order.bias.term}. 
If $\beta_{0}$ is known, the stochastic expansion for $\tilde{\lambda}$ is identical to that in eq. \eqref{Eq:GEL.stochastic.expansion} except $H$ is set to zero and $\Omega^{-1}$ replaces $P$, i.e.,
$\tilde{\lambda}=-\Omega^{-1}\tilde{g} + \Omega^{-1}\tilde{\zeta}_{\lambda}+\bigO[p](n^{-3/2})$, where 
$\tilde{\zeta}_{\lambda} = \widetilde{\Omega}\Omega^{-1}\tilde{g}
+\rho_{3}\sum_{j=1}^{d_{\beta}}[\Omega^{-1}\tilde{g}]_{j}\E[g_{ij}g_{i}g_{i}^{\transp}]\Omega^{-1}\tilde{g}/2$. Thus, in expectation, the first two terms in eq. \eqref{Eq:GEL.stochastic.expansion.2nd.order.bias.term} are eliminated and 
$\E[\tilde{\zeta}_{\lambda}] = n^{-1}c_{\rho}\E[g_{i}g_{i}^{\transp}\Omega^{-1}g_{i}]$.
\end{remark}

\begin{remark}
When $\beta_{0}$ is known, Assumptions \ref{Assumption.GEL.3}(b,c) can be relaxed to $g(z,\beta)$ is continuously differentiable for $\beta\in\mathcal{N}$, $\E[\sup_{\beta\in\mathcal{N}}\norm{\nabla g(z,\beta_{0})}]<\infty$, and there exists $d(z)\geq0$ with $\E[d(z)]<\infty$ such that $\norm{\nabla g(z,\beta)-\nabla g(z,\beta_{0})}\leq d(z)\norm{\beta-\beta_{0}}$ for all $z$ and  $\beta\in\mathcal{N}$. The Lipschitz condition in Assumptions \ref{Assumption.GEL.3}(b,c,d) can also be relaxed to $\alpha$-H\"{o}lder for some $0<\alpha\leq1$ and changing the remainder terms from $\bigO(n^{-3/2})$ to $\bigO(n^{-1-\alpha/2})$.
\end{remark}

\begin{remark}
The two-step GMM estimator is defined as $\hat{\beta}_{GMM}=\argmin_{\beta\in\mathcal{B}}\hat{g}(\beta)^{\transp}\widehat{\Omega}(\tilde{\beta})^{-1}\hat{g}(\beta)$ where $\tilde{\beta}$ is a $\sqrt{n}$-consistent preliminary estimator of $\beta_{0}$. If the preliminary estimator $\tilde{\beta}$ is first order efficient, i.e., $\tilde{\beta}-\beta_{0}=-H\tilde{g}+\bigO[p](n^{-1})$, then, if Assumptions \ref{Assumption.GEL.1}--\ref{Assumption.GEL.3} hold, all GMM estimators $\hat{\beta}_{GMM}$ admit the same expansion to order $\bigO[p](n^{-3/2})$; see \citet[Section 3]{newey2004}. Moreover, defining $\hat{\lambda}_{GMM}=-\widehat{\Omega}(\tilde{\beta})^{-1}\hat{g}(\hat{\beta}_{GMM})$, the expansion is 
\begin{equation*}
\begin{bmatrix}\hat{\beta}_{GMM}-\beta_{0}\\\hat{\lambda}_{GMM}\end{bmatrix}
= -\begin{bmatrix}H\\P\end{bmatrix}\tilde{g} + \begin{bmatrix}-\Sigma & H\\ H^{\transp} & P\end{bmatrix}\tilde{\zeta}^{GMM}+\bigO[p](n^{-3/2}),
\end{equation*}
where
\begin{align*}
\tilde{\zeta}^{GMM} & = \left\{\begin{bmatrix}0& \widetilde{G}^{\transp} \\\widetilde{G} &\widetilde{\Omega}-\sum_{j=1}^{d_{\beta}}\E[g_{i}^{j}g_{i}^{\transp}+g_{i}g_{i}^{j\transp}]\bv_{j}^{\transp}H\tilde{g}\end{bmatrix}
-\frac{1}{2}\sum_{j=1}^{d_{\beta}}[H\tilde{g}]_{j}\begin{bmatrix} 
0 & E[G_{i}^{j}]^{\transp} 
\\ E[G_{i}^{j}] & 0 \end{bmatrix} 
\right. \\ & \qquad \left.
-\frac{1}{2}\sum_{j=1}^{d_{g}}[P\tilde{g}]_{j}\begin{bmatrix}
E[\partial^{2}g_{ij}/\partial\beta\partial\beta^{\transp}] & 
0 \\  0 & 0
\end{bmatrix}
\right\}\begin{bmatrix}H\\P\end{bmatrix}\tilde{g}.
\end{align*}
Writing $\tilde{\zeta}^{GMM}=(\tilde{\zeta}_{\beta}^{GMM\transp},\tilde{\zeta}_{\lambda}^{GMM\transp})^{\transp}$  partitioned conformably with $\beta$ and $\lambda$, $\zeta_{\beta}^{GMM}=\E[\tilde{\zeta}_{\beta}^{GMM}]=\E[G_{i}^{\transp}Pg_{i}]$ and $\zeta_{\lambda}^{GMM}=\E[\tilde{\zeta}_{\lambda}]=-a+\E[G_{i}Hg_{i}]+\E[g_{i}g_{i}^{\transp}Pg_{i}]$. Hence, the second order bias of $\hat{\beta}_{GMM}$, \citet[Theorem 4.2]{newey2004}, is given by
\begin{equation*}
\E[\hat{\beta}_{GMM}]-\beta _{0}=-n^{-1}\Sigma\zeta_{\beta}^{GMM}+n^{-1}H\zeta_{\lambda}^{GMM}+\bigO(n^{-3/2}),
\end{equation*}
the notable difference with GEL being the additional term $-n^{-1}\Sigma\zeta _{\beta}^{GMM}$ with the term $n^{-1}H\zeta_{\lambda}^{GMM}$ identical to CUE.
\end{remark}

\section{Preliminary Lemmas}

\begin{lemma}\label{Lemma.Implied.probabilities} If Assumptions \ref{Assumption.GEL.1}--\ref{Assumption.GEL.3} are satisfied, then 
\begin{equation}\label{Eq:GEL.implied.prob.expansion}
n\hat{\pi}_{i}=1 -g_{i}^{\transp}P\tilde{g}   -\tfrac{\rho_{3}}{2}(g_{i}^{\transp}P\tilde{g})^{2}
+ g_{i}^{\transp}\begin{bmatrix}H^{\transp}&P\end{bmatrix}\tilde{\zeta} 
 +\tilde{g}^{\transp}PG_{i}H\tilde{g}
+c_{\rho}\tilde{g}^{\transp}P\tilde{g}
+ \littleO[p](n^{-1})
\end{equation}
uniformly $i=1,\ldots,n$.
\end{lemma}

\noindent\textsc{Proof.} Let $\hat{v}_{i}=\hat{\lambda}^{\transp}g(z_{i},\hat{\beta})$. 
A third order Taylor expansion of $\rho^{(1)}(\hat{v}_{i})$ around $0$ yields 
\begin{equation*}
\rho^{(1)}(\hat{v}_{i}) = -1-\hat{v}_{i}+\tfrac{\rho_{3}}{2}\hat{v}_{i}^{2}+\tfrac{\rho_{4}}{6}\hat{v}_{i}^{3}(1+\littleO[p](1))
\end{equation*}
noting $\abs{\hat{v}_{i}}\xrightarrow{p}0$ uniformly $i=1,\ldots,n$ by \citet[Lemma A1]{newey2004}. 
A Taylor expansion from eq. \eqref{Eq:GEL.stochastic.expansion} of $g(z_{i},\hat{\beta})$ about $\beta_{0}$ yields 
$g(z_{i},\hat{\beta})=g_{i}-G_{i}H\tilde{g}+\littleO[p](n^{-1/2})$ uniformly $i=1,\ldots,n$ by \citet[Lemma 3]{owen1990}. Hence, substituting, using eq. \eqref{Eq:GEL.stochastic.expansion},
\begin{equation*}
\rho^{(1)}(\hat{v}_{i}) = -1
+g_{i}^{\transp}P\tilde{g} - g_{i}^{\transp}\begin{bmatrix}H^{\transp} & P\end{bmatrix}\tilde{\zeta}
-\tilde{g}^{\transp}H^{\transp}G_{i}^{\transp}P\tilde{g} 
+\tfrac{\rho_{3}}{2}(g_{i}^{\transp}P\tilde{g})^{2}
+\littleO[p](n^{-1}).
\end{equation*}
From a similar expansion, using $n^{-1}\sum_{i=1}^{n}g(z_{i},\hat{\beta})=\Omega P\tilde{g}+\bigO[p](n^{-1})$,  
eq. \eqref{Eq:GEL.stochastic.expansion}, and $P\Omega P=P$, 
\begin{equation*}
n^{-1}\tsum_{i=1}^{n}\rho^{(1)}(\hat{v}_{i}) 
= -1-\hat{\lambda}^{\transp}\Omega P\tilde{g}+\tfrac{\rho_{3}}{2}\hat{\lambda}^{\transp}\Omega\hat{\lambda} 
+\bigO[p](n^{-3/2})
= -1+c_{\rho}\tilde{g}^{\transp}P\tilde{g}+\bigO[p](n^{-3/2}).
\end{equation*}
Hence, 
$[\tsum_{i=1}^{n}\rho^{(1)}(\hat{v}_{i})]^{-1}
= -n^{-1}[1+c_{\rho}\tilde{g}^{\transp}P\tilde{g}+\bigO[p](n^{-3/2})]$ and 
\begin{equation*}
n\hat{\pi}_{i} = 
1
-g_{i}^{\transp}P\tilde{g} 
+ g_{i}^{\transp}\begin{bmatrix}H^{\transp} & P\end{bmatrix}\tilde{\zeta}
+\tilde{g}^{\transp}H^{\transp}G_{i}^{\transp}P\tilde{g} 
-\tfrac{\rho_{3}}{2}(g_{i}^{\transp}P\tilde{g})^{2}
+c_{\rho}\tilde{g}^{\transp}P\tilde{g}
+\littleO[p](n^{-1})
\end{equation*}
uniformly $i=1,\ldots,n$. \hfill $\blacksquare$

\begin{corollary}[Known $\beta_{0}$]\label{Corollary.Implied.probabilities}  If Assumptions \ref{Assumption.GEL.1}--\ref{Assumption.GEL.3} are satisfied, then 
\begin{equation}\label{Eq:GEL.implied.prob.expansion.known.beta}
n\tilde{\pi}_{i}=1 -g_{i}^{\transp}\Omega^{-1}\tilde{g}  -\tfrac{\rho_{3}}{2}(g_{i}^{\transp}\Omega^{-1}\tilde{g})^{2}
+ g_{i}^{\transp}\Omega^{-1}\tilde{\zeta}_{\lambda}
+c_{\rho}\tilde{g}^{\transp}\Omega^{-1}\tilde{g}
+ \littleO[p](n^{-1})
\end{equation}
uniformly $i=1,\ldots,n$.
\end{corollary}

Let $a(z)$ denote a real scalar function of $z$ such that $\E[a(z)^{2}]<\infty$. Write $a_{i}=a(z_{i})$, $i=1,\ldots,n$.

\begin{lemma}\label{Lemma.Implied.probabilities.moments} If Assumptions \ref{Assumption.GEL.1}--\ref{Assumption.GEL.3} are satisfied, then 
\begin{equation}\label{Lemma.implied.prob.expectations.prod}
\E[(n\hat{\pi}_{i}-1)a_{i}] = n^{-1}\left(
-c_{\rho}\E[a_{i}g_{i}^{\transp}Pg_{i}] 
+ \E[a_{i}g_{i}^{\transp}]P\zeta_{\lambda} 
+c_{\rho}(d_{g}-d_{\beta})\E[a_{i}]\right)
+ \littleO(n^{-1})
\end{equation}
uniformly $i=1,\ldots,n$. For $i\neq j$, 
\begin{equation}\label{Lemma.implied.prob.expectations.x.prod.0}
\E[(n\hat{\pi}_{i}-1)a_{i}a_{j}] = \E[(n\hat{\pi}_{i}-1)a_{i}]\E[a_{j}]-n^{-1}\E[a_{i}g_{i}^{\transp}]P\E[g_{j}a_{j}] +\bigO(n^{-2}),
\end{equation}
\begin{equation}\label{Lemma.implied.prob.expectations.x.prod}
\E[(n\hat{\pi}_{i}-1)(n\hat{\pi}_{j}-1)a_{i}a_{j}] = n^{-1}\E[a_{i}g_{i}^{\transp}]P\E[g_{j}a_{j}] +\bigO(n^{-2}).
\end{equation}
Let $\bar{a}=n^{-1}\sum_{i=1}^{n}a_{i}$ and $\hat{a}=\sum_{i=1}^{n}\hat{\pi}_{i}a_{i}$. Then, 
\begin{equation}\label{Lemma.implied.prob.variance}
\Var[\hat{a}] = \Var[\bar{a}] - n^{-1}\E[a_{i}g_{i}^{\transp}]P\E[g_{j}a_{j}] + \bigO(n^{-2}).
\end{equation}
\end{lemma}

\noindent\textsc{Proof}. The first result follows from the expansion for $\hat{\pi}_{i}$ in Lemma \ref{Lemma.Implied.probabilities}. In particular, noting $\E[g_{i}]=0$ and $\E[a_{i}\littleO[p](n^{-1})]=\littleO(n^{-1})$ by uniformity of $\littleO[p](n^{-1})$, then, by independence, 
\begin{align*}
\E[(n\hat{\pi}_{i}-1)a_{i}] & =
 -n^{-1}\E[a_{i}g_{i}^{\transp}Pg_{i}] 
  -\tfrac{\rho_{3}}{2}n^{-1}\E[a_{i}g_{i}^{\transp}P\E[g_{j}g_{j}^{\transp}]Pg_{i}]
+ \E[a_{i}g_{i}^{\transp}]\begin{bmatrix}H^{\transp}&P\end{bmatrix}\E[\tilde{\zeta}] \\
& \quad +n^{-1}\trace(\E[a_{i}G_{i}]H\E[g_{j}g_{j}^{\transp}]P)
+c_{\rho}n^{-1}\E[a_{i}]\trace(\E[g_{j}g_{j}^{\transp}]P)
+ \littleO(n^{-1})\\
& = -n^{-1}c_{\rho}\E[a_{i}g_{i}^{\transp}Pg_{i}] 
+ n^{-1}\E[a_{i}g_{i}^{\transp}]P\zeta_{\lambda} 
+n^{-1}c_{\rho}(d_{g}-d_{\beta})\E[a_{i}]
+ \littleO(n^{-1})
\end{align*}
uniformly $i=1,\ldots,n$, using $\E[\tilde{\zeta}]=(0^{\transp},n^{-1}\zeta_{\lambda}^{\transp})^{\transp}$, $P\Omega P=P$, $H\Omega P=0$, and $\trace(\Omega P)=d_{g}-d_{\beta}$. 
Eqs. \eqref{Lemma.implied.prob.expectations.x.prod.0} and \eqref{Lemma.implied.prob.expectations.x.prod} follow by a similar argument. 

Finally note that  $\hat{a}-\bar{a}=n^{-1}\sum_{i=1}^{n}(n\hat{\pi}_{i}-1)a_{i}$. Hence, $\Var[\hat{a}] = \Var[\bar{a}]+\Var[\hat{a}-\bar{a}]+2\Cov[\hat{a}-\bar{a},\bar{a}]$. Now, from above, $\E[\hat{a}-\bar{a}]=\bigO(n^{-1})$. Hence, 
\begin{equation*}
\Var[\hat{a}-\bar{a}] =  \E_{i\neq j}[(n\hat{\pi}_{i}-1)(n\hat{\pi}_{j}-1)a_{i}a_{j}] +\bigO(n^{-2}) = n^{-1}\E[a_{i}g_{i}^{\transp}]P\E[g_{j}a_{j}] +\bigO(n^{-2}).
\end{equation*}
Also, 
\begin{align*}
\Cov[\hat{a}-\bar{a},\xi]	
 &= n^{-1}\E[(n\hat{\pi}_{i}-1)a_{i}^{2}]
+(1-n^{-1})\E_{i\neq j}[(n\hat{\pi}_{i}-1)a_{i}a_{j}]
-\E[(n\hat{\pi}_{i}-1)a_{i})]E[a_{j}] \\\tag*{$\blacksquare$}
 &= -n^{-1}\E[a_{i}g_{i}^{\transp}]P\E[g_{j}a_{j}] + \bigO(n^{-2}).
\end{align*}

\begin{corollary}[Known $\beta_{0}$]\label{Corollary.Implied.probabilities.moments} If Assumptions \ref{Assumption.GEL.1}--\ref{Assumption.GEL.3} are satisfied, then 
\begin{equation}\label{Lemma.implied.prob.expectations.prod.known.beta}
\E[(n\tilde{\pi}_{i}-1)a_{i}] = n^{-1}c_{\rho}\left(
-\E[g_{i}^{\transp}\Omega^{-1}g_{i} a_{i}]
+ \E[g_{i}^{\transp}\Omega^{-1}g_{i}g_{i}^{\transp}]\Omega^{-1}\E[g_{i}a_{i}]
+q\E[a_{i}]\right)
+ \littleO(n^{-1})
\end{equation}
uniformly $i=1,\ldots,n$. Lemma \ref{Lemma.Implied.probabilities.moments} remains valid with $\Omega^{-1}$ replacing $P$.
\end{corollary}

Repeated use is made of the following lemma; see \citet[Theorem 1.1.1]{bochner1955} and \citet[Theorem 1A]{parzen1962}. See also \citet[App.A.2.6]{pagan1999}.
\begin{lemma}\label{Lemma:convergence}
Suppose that $f:\R\mapsto\R$ and $k:\R\mapsto\R$ are Borel functions satisfying 
(a) $\int_{-\infty}^{\infty}\abs{f(x)}\mathrm{d}x < \infty$; 
(b) $\sup_{-\infty<x<\infty}\abs{k(x)}<\infty$, $\int_{-\infty}^{\infty}\abs{k(x)}\mathrm{d}x<\infty$, and $\lim_{\abs{x}\to\infty}\abs{xk(x)}=0$.
Then $\tint_{-\infty}^{\infty} b^{-1}\abs{k((y-x)/b)}\abs{f(x)}\mathrm{d}x<\infty$ a.e. and 
\begin{equation}\label{Eq:Lemma:convergence.1}
\lim_{b\downarrow0}\abs{\tint_{-\infty}^{\infty} b^{-1}k((y-x)/b)f(x)\mathrm{d}x - f(y)\tint_{-\infty}^{\infty} k(t)\mathrm{d}t}=0
\end{equation}
at every continuity point $y$ of $f$; if $f$ is uniformly continuous, then convergence is uniform. 
Under the same conditions 
$\lim_{b\downarrow0}\abs{\int_{-\infty}^{\infty} b^{-1}k((y-x)/b)^{r}f(x)\mathrm{d}x - f(y)\int_{-\infty}^{\infty} k(t)^{r}\mathrm{d}t}=0$ at every continuity point $y$ of $f$ for any $r\geq1$.
If $\sup_{-\infty<x<\infty}\abs{f(x)}<\infty$, $\int_{-\infty}^{\infty}\abs{k(x)}\mathrm{d}x<\infty$ is sufficient for \eqref{Eq:Lemma:convergence.1} to hold. 
\hfill$\square$
\end{lemma}

\begin{remark}
If $k$ is H\"{o}lder continuous with exponent $0<\tau\leq1$ and, thus, uniformly continuous, and absolutely integrable, then it is bounded. 
\end{remark}



\section{Proofs of Theorems}\label{S.A.Proofs} 

\noindent\textsc{Proof of Theorem \ref{Thm:GEL.KDE.MSE.known.beta}.} Write 
$\tilde{f}_{\rho}(u)  = \tilde{f}(u) +  n^{-1}\sum_{i=1}^{n}(n\tilde{\pi}_{i}-1)k_{b}(u-u_{i})$. By Corollary \ref{Corollary.Implied.probabilities} and \citet[Lemma 3]{owen1990}, $\max_{1\leq i\leq n}\abs{n\tilde{\pi}_{i}-1}=\littleO[p](1)$. By Lemma \ref{Lemma:convergence}, $\E[\abs{k_{b}(u-u_{i})}]<\infty$ whenever $\abs{f(u)}<\infty$ which holds a.e. Thus,  $k_{b}(u-u_{i})$, $i=1,\ldots,n$, satisfies the conditions for WLLN. Hence, the first conclusion follows. 

From Assumption \ref{Assumption.KDE.default}(a)(i),  $b\E[k_{b}(u-u_{i})^{2}]<\infty$ a.e. 
 By CS, invoking Assumptions \ref{Assumption.GEL.1}(e) and \ref{Assumption.GEL.3}(a), $\E[\abs{g_{i}b^{1/2}k_{b}(u-u_{i})}]<\infty$ and $\E[\abs{g_{i}^{\transp}\Omega^{-1}g_{i}b^{1/2}k_{b}(u-u_{i})}]<\infty$. Hence, by 
Corollary \ref{Corollary.Implied.probabilities.moments},
 setting $a_{i}=b^{1/2}k_{b}(u-u_{i})$, 
\begin{align*}
\E[(n\tilde{\pi}_{i}-1)k_{b}(u-u_{i})]  = n^{-1}c_{\rho}\{&
-\E[g_{i}^{\transp}\Omega^{-1}g_{i}k_{b}(u-u_{i})]
+ \E[g_{i}^{\transp}\Omega^{-1}g_{i}g_{i}^{\transp}]\Omega^{-1}\E[g_{i}k_{b}(u-u_{i})] 
\\ & 
+d_{g}\E[k_{b}(u-u_{i})]\}
+ \littleO(n^{-1}).
\end{align*}
Under Assumption \ref{Assumption.KDE.default}(a)(i), $\E[k_{b}(u-u_{i})]=f(u)+\littleO(1)$. 
Invoking Assumption \ref{Assumption.reformulation.bijection} and the change of variables $z\mapsto(u,v^{\transp})^{\transp}$, 
then, by LIE and Lemma \ref{Lemma:convergence}, 
$\E[g_{i}k_{b}(u-u_{i})] 
= \int \E[g_{i}|t]f(t)k_{b}(u-t)\mathrm{d}t 
= \E[g_{i}|u]f(u) + \littleO(1)$.
Similarly, 
 $\E[g_{i}^{\transp}\Omega^{-1}g_{i}k_{b}(u-u_{i})] = \E[g_{i}^{\transp}\Omega^{-1}g_{i}|u]f(u)+\littleO(1)$. The final result is a  direct consequence of Lemma \ref{Lemma.Implied.probabilities.moments} and the same argument. \hfill$\blacksquare$
\vspace*{0.5\baselineskip}

Set
\begin{align}
\label{Eq.KDE.Diff.delta.1}  
\hat{\delta}_{1}(u) & = n^{-1}\tsum_{i=1}^{n}\left[k_{b}(u-\hat{u}_{i}) -k_{b}(u-u_{i})\right];\\
\label{Eq.KDE.Diff.delta.2}
\hat{\delta}_{2}(u) & = n^{-1}\tsum_{i=1}^{n}(n\hat{\pi}_{i}-1)\left[k_{b}(u-\hat{u}_{i}) - k_{b}(u-u_{i})\right];\\
\label{Eq.KDE.Diff.delta.3}
\hat{\delta}_{3}(u) & = n^{-1}\tsum_{i=1}^{n}(n\hat{\pi}_{i}-1)k_{b}(u-u_{i}).
\end{align}
Note $\hat{f}(u) = \tilde{f}(u) + \hat{\delta}_{1}(u)$ and 
$\hat{f}_{\rho}(u) = \hat{f}(u) +\hat{\delta}_{2}(u)+\hat{\delta}_{3}(u)$.

\vspace*{0.5\baselineskip}

\noindent\textsc{Proof of Theorem \ref{Thm:GEL.KDE.MSE.estimated.beta.Consistency}}.
Under Assumptions \ref{Assumption.GEL.1} and \ref{Assumption.GEL.2}, $\hat{\beta}\in\mathcal{N}$ w.p.a.1 and $n^{1/2}(\hat{\beta}-\beta_{0})=\bigO[p](1)$. First, by Assumption \ref{Assumption.U.and.K.diff.int.0}(a,b), from eq. \eqref{Eq.KDE.Diff.delta.1},   
\begin{align*}
\abs{\hat{\delta}_{1}(u)} 
&\leq \tfrac{1}{n}\tsum_{i=1}^{n}\abs{k_{b}(u-\hat{u}_{i}) -k_{b}(u-u_{i})}
\leq \tfrac{C}{nb^{1+\tau}}\tsum_{i=1}^{n}\abs{\hat{u}_{i} -u_{i}}^{\tau} \\
& \leq \tfrac{C}{n^{\alpha\tau/2}b^{1+\tau}}\norm{n^{1/2}(\hat{\beta}-\beta_{0})}^{\alpha\tau} \tfrac{1}{n}\tsum_{i=1}^{n}d(z_{i})^{\tau}
=\littleO[p](1)
\end{align*}
since $n^{-1}\sum_{i=1}^{n}d(z_{i})^{\tau}=\bigO[p](1)$ by WLLN and $n^{\alpha\tau/2}b^{1+\tau}\to\infty$ from 
Assumption \ref{Assumption.U.and.K.diff.int.0}(c). 
Next, $\max_{1\leq i\leq n}\abs{n\hat{\pi}_{i}-1}=\littleO[p](1)$ by Lemma \ref{Lemma.Implied.probabilities} and \citet[Lemma 3]{owen1990}, from eq. \eqref{Eq.KDE.Diff.delta.2}, 
\begin{align*}
\abs{\hat{\delta}_{2}(u)} & \leq \tfrac{1}{n}\tsum_{i=1}^{n}\abs{(n\hat{\pi}_{i}-1)\left[k_{b}(u-\hat{u}_{i}) - k_{b}(u-u_{i})\right]}\\
& \leq \tfrac{C}{n^{\alpha\tau/2}b^{1+\tau}}\norm{n^{1/2}(\hat{\beta}-\beta_{0})}^{\alpha\tau}(\max_{1\leq i\leq n}\abs{n\hat{\pi}_{i}-1})\tfrac{1}{n}\tsum_{i=1}^{n}d(z_{i})^{\tau}
=\littleO[p](1).
\end{align*}
Hence, the first conclusion follows. 
The final result follows from eq. \eqref{Eq.KDE.Diff.delta.3}, by noting that also 
\begin{equation*}
\abs{\hat{\delta}_{3}(u)} \leq \tfrac{1}{n}\tsum_{i=1}^{n}\abs{(n\hat{\pi}_{i}-1)k_{b}(u-u_{i})}
\leq (\max_{1\leq i\leq n}\abs{n\hat{\pi}_{i}-1}) \tfrac{1}{n}\tsum_{i=1}^{n}\abs{k_{b}(u-u_{i})}
=\littleO[p](1)
\end{equation*}
by WLLN since $\E[\abs{k_{b}(u-u_{i})}]<\infty$ a.e. by Lemma \ref{Lemma:convergence}.
\hfill$\blacksquare$
\vspace*{0.5\baselineskip}

\noindent\textsc{Proof of Theorem \ref{Thm:GEL.KDE.MSE.estimated.beta}}. \textbf{Preliminaries.}
From a second order Taylor expansion around $\beta_{0}$,
\begin{align*}
k_{b}(u-\hat{u}_{i})  & = k_{b}(u-u_{i})
-k_{b}^{(1)}(u-u_{i})\nabla^{\transp}u_{i}(\hat{\beta}-\beta_{0}) \\
& +\tfrac{1}{2}(\hat{\beta}-\beta_{0})^{\transp}
[ k_{b}^{(2)}(u-\bar{u}_{i})\nabla \bar{u}_{i}\nabla^{\transp} \bar{u}_{i}
-k_{b}^{(1)}(u-\bar{u}_{i})\nabla^{2}\bar{u}_{i} ]	
(\hat{\beta}-\beta_{0}),
\end{align*}
where $k_{b}^{(j)}(x)=k^{(j)}(x/b)/b^{j+1}$, $j=1,2$, and $\bar{u}_{i}=u(z_{i},\bar{\beta})$, $i=1,\ldots,n$, with $\bar{\beta}$ on the line segment joining $\hat{\beta}$ and $\beta_{0}$;  and $\nabla\bar{u}_{i}$ and $\nabla^{2}\bar{u}_{i}$, $i=1,\ldots,n$, are defined analogously. Note that 
$\norm{\bar{\beta}-\beta_{0}}\leq \norm{\hat{\beta}-\beta_{0}}=\bigO[p](n^{-1/2})$. 
Assumption \ref{Assumption.U.and.K.diff.int}(b) and twice differentiability of $u(z,\beta)$ for $\beta\in\mathcal{N}$ implies  there exist $d_{0}(z)\geq0$ with $\E[d_{0}(z)^{4}]<\infty$ and $d_{1}(z)\geq0$ with $\E[d_{1}(z)^{4}]<\infty$
such that $\abs{u(z,\beta)-u(z,\beta_{0})}\leq d_{0}(z)\norm{\beta-\beta_{0}}$ and 
$\norm{\nabla u(z,\beta)-\nabla u(z,\beta_{0})}\leq d_{1}(z)\norm{\beta-\beta_{0}}$ for all $z$ and $\beta\in\mathcal{N}$. 
Thus, by T, 
$\norm{\nabla \bar{u}_{i}\nabla^{\transp} \bar{u}_{i}-\nabla u_{i}\nabla^{\transp}u_{i}} 
\leq 2d_{1}(z_{i})\norm{\nabla u_{i}}\norm{\hat{\beta}-\beta_{0}} +d_{1}(z_{i})^{2}\norm{\hat{\beta}-\beta_{0}}^{2}$. By \citet[Lemma 3]{owen1990}, $\max_{1\leq i\leq n}d_{1}(z_{i})^{2}=\littleO[p](n^{1/2})$ and 
$\max_{1\leq i\leq n}d_{1}(z_{i})\norm{\nabla u_{i}}=\littleO[p](n^{1/2})$. 
Hence, $\norm{\nabla \bar{u}_{i}\nabla^{\transp} \bar{u}_{i}-\nabla u_{i}\nabla^{\transp}u_{i}} 
\leq n^{-1/2}[d_{1}(z_{i})\norm{\nabla u_{i}}+\littleO[p](1)]\bigO[p](1)$.
By CS, from Assumption \ref{Assumption.U.and.K.diff.int}(b), for $0<\tau\leq1$, 
$\E[d_{0}(z_{i})^{\tau}\norm{\nabla u_{i}}^{2}]<\infty$ and $\E[d_{1}(z_{i})^{2}\norm{\nabla u_{i}}^{2}]<\infty$. 
Thus, $\E[\abs{b^{-1/2}k^{(2)}((u-u_{i})/b)}d_{1}(z_{i})\norm{\nabla u_{i}}]<\infty$ since 
$\E[b^{-1}k^{(2)}((u-u_{i})/b)^{2}]<\infty$ also by CS and using Lemma \ref{Lemma:convergence}. Hence, by T, and noting 
 $n^{\tau/2}b^{3+\tau}\to\infty$, $0<\tau\leq1$, from Assumption \ref{Assumption.U.and.K.diff.int}(c),
\begin{align*}
&\norm{\tfrac{1}{n}\tsum_{i=1}^{n}(k_{b}^{(2)}(u-\bar{u}_{i})\nabla \bar{u}_{i}\nabla^{\transp} \bar{u}_{i}
-k_{b}^{(2)}(u-u_{i})\nabla u_{i}\nabla^{\transp}u_{i})} \\
&\qquad \leq \tfrac{C}{n^{\tau/2}b^{3+\tau}}\norm{n^{1/2}(\hat{\beta}-\beta_{0})}^{\tau} \tfrac{1}{n}\tsum_{i=1}^{n}d_{0}(z_{i})^{\tau}\norm{\nabla u_{i}}^{2} \\
&\qquad\quad +\tfrac{1}{n}\tsum_{i=1}^{n}\left[\tfrac{1}{n^{1/2}b^{5/2}}\abs{b^{-1/2}k^{(2)}((u-u_{i})/b)}
+\tfrac{\littleO[p](1)}{n^{\tau/2}b^{3+\tau}}\right]
\left[d_{1}(z_{i})\norm{\nabla u_{i}} +\littleO[p](1)\right]\bigO[p](1) 
 = \littleO[p](1).
\end{align*}

Assumption \ref{Assumption.U.and.K.diff.int}(a) implies $k^{(1)}$ is Lipschitz, and hence, invoking 
Assumption \ref{Assumption.U.and.K.diff.int}(b),  for all mean values $\bar{\beta}$ between $\hat{\beta}$ and $\beta_{0}$, 
$\abs{k_{b}^{(1)}(u-\bar{u}_{i}) -k_{b}^{(1)}(u-u_{i})}\leq b^{-3}Cd_{0}(z_{i})\norm{\hat{\beta}-\beta_{0}}$ w.p.a.1.  By Assumption \ref{Assumption.U.and.K.diff.int}(a) and Lemma \ref{Lemma:convergence}, $\E[b^{-1}\vert k^{(1)}((u-u_{i})/b)\vert^{4/3}]<\infty$ a.e., and as $\E[d(z_{i})^{4}]<\infty$, $\E[\abs{b^{-3/4}k^{(1)}((u-u_{i})/b)}d(z_{i})]<\infty$ using the H\"{o}lder inequality with exponents $4/3$ and $4$.
Therefore, by the same argument as above, 
\begin{align*}
& \norm{\tfrac{1}{n}\tsum_{i=1}^{n}(k_{b}^{(1)}(u-\bar{u}_{i})\nabla^{2}\bar{u}_{i} -k_{b}^{(1)}(u-u_{i})\nabla^{2}u_{i})}
\leq \tfrac{C}{n^{1/2}b^{3}}\norm{n^{1/2}(\hat{\beta}-\beta_{0})} \tfrac{1}{n}\tsum_{i=1}^{n}d_{0}(z_{i})\norm{\nabla^{2}u_{i}}\\
& +\tfrac{1}{n^{\alpha/2}b^{5/4}}\norm{n^{1/2}(\hat{\beta}-\beta_{0})}^{\alpha} \tfrac{1}{n}\tsum_{i=1}^{n}\left[\abs{b^{-3/4}k^{(1)}((u-u_{i})/b)}+\tfrac{C}{n^{1/2}b^{7/4}}d_{0}(z_{i})\norm{n^{1/2}(\hat{\beta}-\beta_{0})}\right]d(z_{i})
= \littleO[p](1).
\end{align*}

Using expansion eq. \eqref{Eq:GEL.stochastic.expansion} and Lemma \ref{Lemma.Implied.probabilities} eq. \eqref{Eq:GEL.implied.prob.expansion}, from eq. \eqref{Eq.KDE.Diff.delta.1},
\begin{align}
\notag
\hat{\delta}_{1}(u) & = \tfrac{1}{n}\tsum_{i=1}^{n}k_{b}^{(1)}(u-u_{i})\nabla^{\transp}u_{i}H\tilde{g} 
- \tfrac{1}{n}\tsum_{i=1}^{n}k_{b}^{(1)}(u-u_{i})\nabla^{\transp}u_{i}\begin{bmatrix}-\Sigma & H\end{bmatrix}\tilde{\zeta}\\
\label{Eq.KDE.Diff.delta.1.expansion}
&\quad +\tfrac{1}{2}\tilde{g}^{\transp}H^{\transp}
\tfrac{1}{n}\tsum_{i=1}^{n}[k_{b}^{(2)}(u-u_{i})\nabla u_{i}\nabla^{\transp} u_{i}
-k_{b}^{(1)}(u-u_{i})\nabla^{2}u_{i}]H\tilde{g}
+\littleO[p](n^{-1}),
\end{align}
from eq. \eqref{Eq.KDE.Diff.delta.2},
\begin{equation}\label{Eq.KDE.Diff.delta.2.expansion}
\hat{\delta}_{2}(u) 
 = -\tfrac{1}{n}\tsum_{i=1}^{n}k_{b}^{(1)}(u-u_{i})\nabla^{\transp}u_{i}H\tilde{g}\tilde{g}^{\transp}Pg_{i}+\littleO[p](n^{-1}),
\end{equation}
and, from eq. \eqref{Eq.KDE.Diff.delta.3},
\begin{align}\notag
\hat{\delta}_{3}(u)& = \tfrac{1}{n}\tsum_{i=1}^{n}[-g_{i}^{\transp}P\tilde{g}   -\tfrac{\rho_{3}}{2}(g_{i}^{\transp}P\tilde{g})^{2} + g_{i}^{\transp}\begin{bmatrix}H^{\transp}&P\end{bmatrix}\tilde{\zeta} 
 +\tilde{g}^{\transp}PG_{i}H\tilde{g}
+c_{\rho}\tilde{g}^{\transp}P\tilde{g}]k_{b}(u-u_{i})  \\
\label{Eq.KDE.Diff.delta.3.expansion}
& \mkern580mu + \littleO[p](n^{-1}).
\end{align}

\vspace*{0.5\baselineskip}
\noindent\textbf{Expectation.} Since $H\Omega H^{\transp}=\Sigma$,  from eq.  \eqref{Eq.KDE.Diff.delta.1.expansion},
\begin{align*}
\E[\hat{\delta}_{1}(u)] & = n^{-1}\E[k_{b}^{(1)}(u-u_{i})\nabla^{\transp}u_{i}Hg_{i}] 
- n^{-1}\E[k_{b}^{(1)}(u-u_{i})\nabla^{\transp}u_{i}]H\zeta_{\lambda} 
\\ & \quad
 +\tfrac{1}{2}n^{-1}\trace\left(\Sigma
\E[k_{b}^{(2)}(u-u_{i})\nabla u_{i}\nabla^{\transp} u_{i}
-k_{b}^{(1)}(u-u_{i})\nabla^{2}u_{i}]\right)
+\littleO(n^{-1}).
\end{align*}
Assumption \ref{Assumption.U.and.K.diff.int}(a) states $\lim_{\abs{x}\to\infty}\abs{x^{2}k^{(1)}(x)}=0$ and implies that $\int k^{(1)}(x)\mathrm{d}x=0$, $\int xk^{(1)}(x)\mathrm{d}x=-1$, and $xk^{(1)}(x)$ satisfies the hypotheses of Lemma \ref{Lemma:convergence}, i.e., it is bounded and absolutely integrable. 
Thus, invoking Assumption \ref{Assumption.U.and.K.diff.int}(d), by MVT and Lemma \ref{Lemma:convergence},
\begin{align}\notag
\E[k_{b}^{(1)}(u-u_{i})\nabla u_{i}] 
& = \tfrac{1}{b}\tint\E[\nabla u_{i}|u-bt] f(u-bt)k^{(1)}(t)\mathrm{d}t \\
\notag
& = \tfrac{1}{b}\E[\nabla u_{i}|u] f(u)\tint k^{(1)}(t)\mathrm{d}t
-\tint (\mathrm{d}\{\E[\nabla u_{i}|u-\omega bt] f(u-\omega bt)\}/\mathrm{d}u)tk^{(1)}(t)\mathrm{d}t\\
\label{Eq:Thm:GEL.KDE.MSE.estimated.beta:E.dk.du}
& =\mathrm{d}\{\E[\nabla u_{i}|u]f(u)\}/\mathrm{d}u + \littleO(1).
\end{align}
Similarly, $\E[k_{b}^{(1)}(u-u_{i})\nabla^{\transp}u_{i}Hg_{i}] = \mathrm{d}\{\E[\nabla^{\transp}u_{i}Hg_{i}|u]f(u)\}/\mathrm{d}u+ \littleO(1)$ and 
$\E[k_{b}^{(1)}(u-u_{i})\nabla^{2}u_{i}] = \mathrm{d}\{\E[\nabla^{2}u_{i}|u]f(u)\}/\mathrm{d}u+ \littleO(1)$. 
Furthermore, Assumption \ref{Assumption.U.and.K.diff.int}(a) also implies that $\int k^{(2)}(x)\mathrm{d}x=0$, $\int xk^{(2)}(x)\mathrm{d}x=0$, $\int x^{2}k^{(2)}(x)\mathrm{d}x=2$,  and $x^{2}k^{(2)}(x)$ satisfies the hypotheses of Lemma \ref{Lemma:convergence}. Thus, by a second order Taylor expansion and a similar argument to eq. \eqref{Eq:Thm:GEL.KDE.MSE.estimated.beta:E.dk.du},
\begin{align*}
\E[k_{b}^{(2)}(u-u_{i})\nabla u_{i}\nabla^{\transp} u_{i}]
& = \tfrac{1}{b^{2}}\tint \E[\nabla u_{i}\nabla^{\transp}u_{i}|u-bt]f(u-bt)k^{(2)}(t)\mathrm{d}t \\
& = \tfrac{1}{b^{2}}\E[\nabla u_{i}\nabla^{\transp}u_{i}|u]f(u)\tint k^{(2)}(t)\mathrm{d}t 
-\tfrac{1}{b}\mathrm{d}\{\E[\nabla u_{i}\nabla^{\transp}u_{i}|u]f(u)\}/\mathrm{d}u\tint tk^{(2)}(t)\mathrm{d}t \\
&\quad +\tfrac{1}{2}\tint(\mathrm{d}^{2}\{\E[\nabla u_{i}\nabla^{\transp}u_{i}|u-\omega bt]f(u-\omega bt)\}/\mathrm{d}u^{2})t^{2}k^{(2)}(t)\mathrm{d}t  \\
& = \mathrm{d}^{2}\{\E[\nabla u_{i}\nabla^{\transp}u_{i}|u]f(u)\}/\mathrm{d}u^{2} + \littleO(1).
\end{align*}
Since $H\Omega P=0$, from eq. \eqref{Eq.KDE.Diff.delta.2.expansion},  $\E[\hat{\delta}_{2}(u)]=\littleO(n^{-1})$. By Lemma \ref{Lemma.Implied.probabilities.moments} eq. \eqref{Lemma.implied.prob.expectations.prod} and the same argument used in the proof of Theorem \ref{Thm:GEL.KDE.MSE.known.beta}, 
$\E[\hat{\delta}_{3}(u)] = n^{-1}\{
 -c_{\rho}\E[g_{i}^{\transp}Pg_{i}|u] +\E[g_{i}|u]^{\transp}P\zeta_{\lambda} + c_{\rho}(d_{g}-d_{\beta})\}f(u)
+\littleO(n^{-1})$.

\vspace*{0.5\baselineskip}
\noindent\textbf{Variance.} Since $\E[\hat{\delta}_{1}(u)]=\bigO(n^{-1})$, from eq. \eqref{Eq.KDE.Diff.delta.1.expansion}, 
\begin{align*}
\Var(\hat{\delta}_{1}(u)) 
& = n^{-2}\tsum_{i=1}^{n}\tsum_{j=1}^{n}\E[k_{b}^{(1)}(u-u_{i})\nabla^{\transp}u_{i}H\tilde{g}\tilde{g}^{\transp}H^{\transp}\nabla u_{j}k_{b}^{(1)}(u-u_{j})] +\littleO(n^{-1}) \\
& = n^{-1}[\mathrm{d}\{\E[\nabla u_{i}|u]f(u)\}/\mathrm{d}u]^{\transp}\Sigma [\mathrm{d}\{\E[\nabla u_{i}|u]f(u)\}/\mathrm{d}u] +\littleO(n^{-1}).
\end{align*}
Similarly, noting $\E[\hat{\delta}_{2}(u)]=\littleO(n^{-1})$, from Lemma \ref{Lemma.Implied.probabilities.moments}, it is straightforward to verify that $\Var[\hat{\delta}_{2}(u)]=\littleO(n^{-1})$. Furthermore, also using Lemma \ref{Lemma.Implied.probabilities.moments}, as $\E[\hat{\delta}_{3}(u)]=\bigO(n^{-1})$ and $\E[k_{b}(u-u_{i})g_{i}]=\E[g_{i}|u]f(u)$, 
$\Var(\hat{\delta}_{3}(u)) = n^{-1}\E[g_{i}|u]^{\transp}P\E[g_{i}|u]f(u)^{2}+\littleO(n^{-1})$. 
It is straightforward to verify that 
$\Cov[\hat{\delta}_{1},\hat{\delta}_{2}]=\littleO(n^{-1})$, recalling $H\Omega P=0$, 
\begin{equation*}
\Cov[\hat{\delta}_{1}(u)\hat{\delta}_{3}(u)]= 
-n^{-2}\tsum_{i=1}^{n}\tsum_{j=1}^{n}\E[k_{b}^{(1)}(u-u_{i})\nabla^{\transp}u_{i}H\tilde{g}
\tilde{g}^{\transp}Pg_{j}k_{b}(u-u_{j})]
 + \bigO(n^{-2})=\bigO(n^{-2}),
\end{equation*}
\begin{align*}
\Cov[\hat{\delta}_{1}(u),\tilde{f}(u)]
& = n^{-1}\E[k_{b}^{(1)}(u-u_{i})\nabla^{\transp}u_{i}]H\E[g_{j}k_{b}(u-u_{j})] + \littleO(n^{-1})\\
& = n^{-1}[\mathrm{d}\{\E[\nabla u_{i}|u]f(u)\}/\mathrm{d}u]^{\transp}H\E[g_{i}|u]f(u)  + \littleO(n^{-1}),
\end{align*}
$\Cov[\hat{\delta}_{2}(u),\hat{\delta}_{3}(u)]=\littleO(n^{-1})$, $\Cov[\hat{\delta}_{2}(u),\tilde{f}(u)]=\littleO(n^{-1})$, noting again  $H\Omega P=0$, and finally, 
\begin{equation*}
\Cov[\hat{\delta}_{3}(u),\tilde{f}(u)] = -n^{-1}\E[g_{i}|u]^{\transp}P\E[g_{i}|u]f(u)^{2}+\littleO(n^{-1}). 
\end{equation*}
Combining these results gives eqs. 
\eqref{Thm:GEL.KDE.MSE.estimated.beta:delta.rho}--\eqref{Thm:GEL.KDE.MSE.estimated.beta:VAR.f.hat.rho}. 
\hfill$\blacksquare$

\vspace*{0.5\baselineskip}


\noindent\textsc{Proof of Theorem \ref{Thm:GEL.KDFE.MSE.known.beta}.} 
Since $\lim_{x\to-\infty}K(x)=0$ and $\lim_{x\to\infty}K(x)=1$, $2\int K(x)k(x)\mathrm{d}x=1$, and $\int\abs{K(x)k(x)}\mathrm{d}x<\infty$, 
$\E[K((u-u_{i})/b)] =  F(u)+\int k(t)[F(u-bt)-F(u)]\mathrm{d}t$
and
$\E[K((u-u_{i})/b)^{2}] = F(u)+2\int K(t)k(t)[F(u-bt)-F(u)]\mathrm{d}t$. 
$F$ as a c.d.f. is bounded and hence $\E[K((u-u_{i})/b)^{2}]<\infty$, and  $\E[K((u-u_{i})/b)] =  F(u)+\littleO(1)$ and $\E[K((u-u_{i})/b)^{2}] =  F(u)+\littleO(1)$ as $b\to0$ and at all points of continuity of $F$.
Therefore, cf. the proof of Theorem \ref{Thm:GEL.KDE.MSE.known.beta}, 
$\abs{\widetilde{F}_{\rho}(u)-\widetilde{F}(u)}=\littleO[p](1)$. 

Equation \eqref{Thm:GEL.KDFE.MSE.known.beta.Bias} follows by Corollary \ref{Corollary.Implied.probabilities.moments} with $a_{i}=K((u-u_{i})/b)$, $i=1,\ldots,n$. 
Assumptions \ref{Assumption.KDE.default}(a)(i) and $\lim_{\abs{x}\to\infty}\abs{x^{2}k(x)}=0$ imply that $xk(x)$ satisfies conditions of Lemma \ref{Lemma:convergence}. Since $\int xk(x)=0$ and $\E[\abs{\E[g_{i}|u]}]<\infty$, integration by parts and an application of MVT give 
\begin{align}\notag
\E[g_{i}K((u-u_{i})/b)]   
&= \tint_{-\infty}^{\infty} K((u-s)/b)\E[g_{i}|s]\mathrm{d}F(s) 
= [K((u-s)/b)\tint_{-\infty}^{s}\E[g_{i}|t]\mathrm{d}F(t)]_{-\infty}^{\infty} \\
\notag
&\quad + \tint_{-\infty}^{u}\E[g_{i}|t]\mathrm{d}F(t) 
-b\tint_{-\infty}^{\infty} (\E[g_{i}|u-\omega bt]f(u-\omega bt))tk(t)\mathrm{d}t
\\
\label{Eq:Thm:GEL.KDFE.MSE.known.beta:Jh}
& = \tint_{-\infty}^{u}\E[g_{i}|t]\mathrm{d}F(t) + \littleO(b).
\end{align}
Similarly, 
$\E[g_{i}^{\transp}\Omega^{-1}g_{i}K((u-u_{i})/b)] =\int_{-\infty}^{u}\E[g_{i}^{\transp}\Omega^{-1}g_{i}|t]\mathrm{d}F(t)$. Eq. \eqref{Thm:GEL.KDFE.MSE.known.beta.Variance} follows by Corollary \ref{Corollary.Implied.probabilities.moments} and eq. \eqref{Eq:Thm:GEL.KDFE.MSE.known.beta:Jh}. \hfill$\blacksquare$

\vspace*{0.5\baselineskip}
	

Set 
\begin{align}
\label{Eq.KDFE.Diff.delta.1}
\widehat{\Delta}_{1}(u) & = n^{-1}\tsum_{i=1}^{n}[K((u-\hat{u}_{i})/b)-K((u-u_i)/b)];\\
\label{Eq.KDFE.Diff.delta.2}
\widehat{\Delta}_{2}(u) & = n^{-1}\tsum_{i=1}^{n}(n\hat{\pi}_{i}-1)[K((u-\hat{u}_{i})/b)-K((u-u_i)/b)];\\
\label{Eq.KDFE.Diff.delta.3}
\widehat{\Delta}_{3}(u) & = n^{-1}\tsum_{i=1}^{n}(n\hat{\pi}_{i}-1)K((u-u_{i})/b).
\end{align}
Note 
$\widetilde{F}(u) = \widehat{F}(u)+ \widehat{\Delta}_{1}(u)$ and 
$\widehat{F}_{\rho}(u) =  \widehat{F}(u) +\widehat{\Delta}_{2}(u)+\widehat{\Delta}_{3}(u)$.

\vspace*{0.5\baselineskip}

\noindent\textsc{Proof of Theorem \ref{Thm:GEL.KDFE.MSE.estimated.beta.Consistency}}.
Since $k$ is bounded, $K$ is Lipschitz continuous and, by the proof of Theorem \ref{Thm:GEL.KDFE.MSE.known.beta}, $\E[\abs{K((u-u_i)/b)}]<\infty$ for all $u$. Then, as in the proof of Theorem \ref{Thm:GEL.KDE.MSE.estimated.beta.Consistency}, invoking Assumptions \ref{Assumption.GEL.1}--\ref{Assumption.GEL.3} and \ref{Assumption.U.and.K.diff.int.0}(b), from eqs. 
\eqref{Eq.KDFE.Diff.delta.1}--\eqref{Eq.KDFE.Diff.delta.3},
\begin{align*}
\abs{\widehat{\Delta}_{1}(u)}&\leq \tfrac{C}{n^{\alpha/2}b}\norm{n^{1/2}(\hat{\beta}-\beta_{0})}^{\alpha}\tfrac{1}{n}\tsum_{i=1}^{n}d(z_{i})=\littleO[p](1); \\
\abs{\widehat{\Delta}_{2}(u)} &\leq (\max_{1\leq i\leq n}\abs{n\hat{\pi}_{i}-1})\tfrac{C}{n^{\alpha/2}b}\norm{n^{1/2}(\hat{\beta}-\beta_{0})}^{\alpha} \tfrac{1}{n}\tsum_{i=1}^{n}d(z_{i})=\littleO[p](1);\\
\tag*{$\blacksquare$}
\abs{\widehat{\Delta}_{3}(u)} &\leq (\max_{1\leq i\leq n}\abs{n\hat{\pi}_{i}-1}) \tfrac{1}{n}\tsum_{i=1}^{n}\abs{K((u-u_{i})/b)}=\littleO[p](1).
\end{align*}

\vspace*{0.5\baselineskip}


\noindent\textsc{Proof of Theorem \ref{Thm:GEL.KDFE.MSE.estimated.beta}}. \textbf{Preliminaries.} 
From a second order Taylor expansion around $\beta_{0}$,
\begin{align*}
K((u-\hat{u}_{i})/b)
& = K((u-u_{i})/b) -k_{b}(u-u_{i})\nabla^{\transp} u_{i}(\hat{\beta}-\beta_{0}) \\
&\quad +\tfrac{1}{2}(\hat{\beta}-\beta_{0})^{\transp}[k_{b}^{(1)}(u-\bar{u}_{i})\nabla\bar{u}_{i}\nabla^{\transp}\bar{u}_{i}
-k_{b}(u-\bar{u}_{i})\nabla^{2}\bar{u}_{i}](\hat{\beta}-\beta_{0}),
\end{align*}
where $\bar{u}_{i}=u(z_{i},\bar{\beta})$, $i=1,\ldots,n$, with $\bar{\beta}$ on the line segment joining $\hat{\beta}$ and $\beta_{0}$; $\nabla\bar{u}_{i}$ and $\nabla^{2}\bar{u}_{i}$, $i=1,\ldots,n$, are defined analogously. 
By the same argument as in the proof of Theorem \ref{Thm:GEL.KDE.MSE.estimated.beta}, noting that Assumption \ref{Assumption.U.and.K.diff.int.CDF}(a)  implies $k$ is Lipschitz and $nb^{6}\to\infty$ as $n^{\tau/2}b^{2+\tau}\to\infty$, invoking Assumption \ref{Assumption.U.and.K.diff.int}(b), 
\begin{align*}
& \norm{\tfrac{1}{n}\tsum_{i=1}^{n}(k_{b}^{(1)}(u-\bar{u}_{i})\nabla\bar{u}_{i}\nabla^{\transp}\bar{u}_{i}
-k_{b}^{(1)}(u-u_{i})\nabla u_{i}\nabla^{\transp}u_{i})} \\
& \qquad \leq \tfrac{C}{n^{\tau/2}b^{2+\tau}}\norm{n^{1/2}(\hat{\beta}-\beta_{0})}^{\tau}\tfrac{1}{n}\tsum_{i=1}^{n}d_{0}(z_{i})^{\tau}\norm{\nabla u_{i}}^{2} \\
& \qquad\quad+ \tfrac{1}{n}\tsum_{i=1}^{n}\left[\tfrac{1}{n^{1/2}b^{3/2}}\abs{b^{-1/2}k^{(1)}((u-u_{i})/b)}+\tfrac{\littleO[p](1)}{n^{\tau/2}b^{2+\tau}}\right]\left[d_{1}(z_{i})\norm{\nabla u_{i}} +\littleO[p](1)\right]\bigO[p](1)
=\littleO[p](1)
\end{align*}
and 
\begin{align*}
& \norm{\tfrac{1}{n}\tsum_{i=1}^{n}(k_{b}(u-\bar{u}_{i})\nabla^{2}\bar{u}_{i} - k_{b}(u-u_{i})\nabla^{2}u_{i})}
\leq  \tfrac{C}{n^{1/2}b^{2}}\norm{n^{1/2}(\hat{\beta}-\beta_{0})}\tfrac{1}{n}\tsum_{i=1}^{n}d_{0}(z_{i})\norm{\nabla^{2}u_{i}} \\
& \:\:+\tfrac{1}{n^{\alpha/2}b^{1/4}}\norm{n^{1/2}(\hat{\beta}-\beta_{0})}^{\alpha}\tfrac{1}{n}\tsum_{i=1}^{n}\left[\abs{b^{-3/4}k((u-u_{i})/b)}+\tfrac{C}{n^{1/2}b^{7/4}}d_{0}(z_{i})\norm{n^{1/2}(\hat{\beta}-\beta_{0})}\right]d(z_{i})
= \littleO[p](1).
\end{align*}
Therefore, using expansion eq. \eqref{Eq:GEL.stochastic.expansion} and Lemma \ref{Lemma.Implied.probabilities}, 
\begin{align}
\notag
\widehat{\Delta}_{1}(u) &=  n^{-1}\tsum_{i=1}^{n}k_{b}(u-u_{i})\nabla^{\transp} u_{i}H\tilde{g} 
-n^{-1}\tsum_{i=1}^{n}k_{b}(u-u_{i})\nabla^{\transp} u_{i}\begin{bmatrix}-\Sigma & H\end{bmatrix}\tilde{\zeta}  \\
\label{Eq.KDFE.Diff.delta.1.expansion}
& \quad +\tfrac{1}{2}\tilde{g}^{\transp}H^{\transp}n^{-1}\tsum_{i=1}^{n}[k_{b}^{(1)}(u-u_{i})\nabla u_{i}\nabla^{\transp}u_{i}
-k_{b}(u-u_{i})\nabla^{2}u_{i}]H\tilde{g} + \littleO[p](n^{-1}),\\
\label{Eq.KDFE.Diff.delta.2.expansion}
\widehat{\Delta}_{2}(u) & = -n^{-1}\tsum_{i=1}^{n}k_{b}(u-u_{i})\nabla^{\transp} u_{i}H\tilde{g}\tilde{g}^{\transp}Pg_{i} + \bigO[p](n^{-3/2}),\\
\notag
\widehat{\Delta}_{3}(u) & =  n^{-1}\tsum_{i=1}^{n}[
 -g_{i}^{\transp}P\tilde{g}   -\tfrac{\rho_{3}}{2}(g_{i}^{\transp}P\tilde{g})^{2}
+ g_{i}^{\transp}\begin{bmatrix}H^{\transp}&P\end{bmatrix}\tilde{\zeta} 
+\tilde{g}^{\transp}PG_{i}H\tilde{g}
+c_{\rho}\tilde{g}^{\transp}P\tilde{g}]
K((u-u_{i})/b)\\
\label{Eq.KDFE.Diff.delta.3.expansion}
&\mkern580mu +\littleO[p](n^{-1}). 
\end{align}

\vspace*{0.5\baselineskip}
\noindent\textbf{Expectation.} Similarly to the proof of Theorem \ref{Thm:GEL.KDE.MSE.estimated.beta}, from eq. \eqref{Eq.KDFE.Diff.delta.1.expansion},
\begin{align*}
\E[\widehat{\Delta}_{1}(u)] &=  n^{-1}\E[k_{b}(u-u_{i})\nabla^{\transp} u_{i}Hg_{i}]
-n^{-1}\E[k_{b}(u-u_{i})\nabla^{\transp} u_{i}]H\zeta_{\lambda}  \\
& \quad +\tfrac{1}{2}n^{-1}\trace\left(\Sigma\E[k_{b}^{(1)}(u-u_{i})\nabla u_{i}\nabla^{\transp}u_{i}
-k_{b}(u-u_{i})\nabla^{2}u_{i}]\right) + \littleO(n^{-1}).
\end{align*}
Assumption  \ref{Assumption.U.and.K.diff.int.CDF}(a) implies  $k(x)$ satisfies the hypotheses of of Lemma \ref{Lemma:convergence}. Hence 
$\E[k_{b}(u-u_{i})\nabla u_{i}]
= \E[\nabla u_{i}|u]f(u)+\littleO(1)$, 
$\E[k_{b}(u-u_{i})\nabla^{\transp} u_{i}Hg_{i}] = \E[\nabla^{\transp}u_{i}Hg_{i}|u]f(u)+\littleO(1)$, and 
$\E[k_{b}(u-u_{i})\nabla^{2}u_{i}] = \E[\nabla^{2}u_{i}|u]f(u)+\littleO(1)$.
Assumption  \ref{Assumption.U.and.K.diff.int.CDF}(a) also implies  $xk^{(1)}(x)$ satisfies the hypotheses of Lemma \ref{Lemma:convergence}. Hence, by MVT as in eq. \eqref{Eq:Thm:GEL.KDE.MSE.estimated.beta:E.dk.du}, 
$\E[k_{b}^{(1)}(u-u_{i})\nabla u_{i}\nabla^{\transp}u_{i}]
= \mathrm{d}\{\E[\nabla u_{i}\nabla^{\transp}u_{i}|u]f(u)\}/\mathrm{d}u+\littleO(1)$.
Therefore, $\E[\widehat{\Delta}_{1}(u)]=n^{-1}\Delta(u)+\littleO(n^{-1})$ as required.

Likewise, as in the proof of Theorem \ref{Thm:GEL.KDE.MSE.estimated.beta}, from eq. \eqref{Eq.KDFE.Diff.delta.2.expansion}, $\E[\widehat{\Delta}_{2}(u)]=\littleO(n^{-1})$. Finally, by Lemma \ref{Lemma.Implied.probabilities.moments} and proof of Theorem \ref{Thm:GEL.KDFE.MSE.known.beta}, $\E[\widehat{\Delta}_{3}(u)] = n^{-1}\Delta_{\rho}(u) +\littleO(n^{-1})$.


\vspace*{0.5\baselineskip}
\noindent\textbf{Variance.} Using expansions eqs. \eqref{Eq.KDFE.Diff.delta.1.expansion}--\eqref{Eq.KDFE.Diff.delta.3.expansion} for $\widehat{\Delta}_{j}(u)$, $j=1,2,3$, 
$\Cov[\widehat{\Delta}_{1}(u),\widehat{\Delta}_{2}(u)]$, \\
$\Cov[\widehat{\Delta}_{1}(u),\widehat{\Delta}_{3}(u)]$, 
$\Cov[\widehat{\Delta}_{2}(u),\widehat{\Delta}_{3}(u)]$, 
$\Cov[\widetilde{F}(u),\widehat{\Delta}_{2}(u)]$,  and  
$\Var[\widehat{\Delta}_{2}(u)]$ are all $\bigO(n^{-2})$. 
Also, 
\begin{align*}
\Var[\widehat{\Delta}_{1}(u)] &=  n^{-1}\E[k_{b}(u-u_{i})\nabla u_{i}]^{\transp}\Sigma \E[k_{b}(u-u_{j})\nabla u_{j}]  + \bigO(n^{-2}), \\
\Cov[\widetilde{F}(u),\widehat{\Delta}_{1}(u)] &=  n^{-1}\E[k_{b}(u-u_{i})\nabla u_{i}]^{\transp}H\E[g_{j}K((u-u_{j})/b)]  +\littleO(n^{-3/2}),\\
\Var[\widehat{\Delta}_{3}(u)] &= n^{-1}\E[g_{i}K((u-u_{i})/b)]^{\transp}P\E[g_{i}K((u-u_{i})/b)] +\bigO(n^{-2}), \\
\Cov[\widetilde{F}(u),\widehat{\Delta}_{3}(u)] &= -n^{-1}\E[g_{i}K((u-u_{i})/b)]^{\transp}P\E[g_{i}K((u-u_{i})/b)] +\bigO(n^{-2}).
\end{align*}
Eqs. \eqref{Thm:GEL.KDFE.MSE.estimated.beta:Var.F} and \eqref{Thm:GEL.KDFE.MSE.estimated.beta:Var.F.rho} then follow immediately using eq. \eqref{Eq:Thm:GEL.KDFE.MSE.known.beta:Jh} and $\E[k_{b}(u-u_{i})\nabla u_{i}] = \E[\nabla u_{i}|u]f(u)+\littleO(1)$.
If $\mathrm{d}\{\E[\nabla u_{i}|u]f(u)\}/\mathrm{d}u$ is absolutely integrable, then, using Lemma  \ref{Lemma:convergence}, 
$\E[k_{b}(u-u_{i})\nabla u_{i}]  = \E[\nabla u_{i}|u]f(u) - b\int(\mathrm{d}\{\E[\nabla u_{i}|u-\omega bt]f(u-\omega bt)\}/\mathrm{d}u)tk(t)\mathrm{d}t =\E[\nabla u_{i}|u]f(u)+\littleO(b)$. 
\hfill$\blacksquare$



\clearpage
\setcounter{page}{1}
\setcounter{section}{0}

\title{\scshape Supplement B to \textquotedblleft Improved Density and Distribution Function Estimation\textquotedblright: Examples} 
\author{\protect\begin{tabular}{cp{10mm}c}
Vitaliy Oryshchenko\textsuperscript{*}
&  & Richard J. Smith \\ 
Department of Economics  & & c{\it e}mmap, U.C.L and I.F.S.\\
University of Manchester & & Faculty of Economics, University of Cambridge \\
 & & Department of Economics, University of Melbourne \\
 & & ONS Economic Statistics Centre of Excellence
 \protect\end{tabular}}
\date{\today}
\maketitle

\phantomsection
\addcontentsline{toc}{section}{Supplement B: Examples}
\textcolor{white}{ 
\customlabel{Supp:Examples}{B}
}
\vspace*{-\baselineskip} 
\renewcommand{\thepage}{[B.\arabic{page}]}
\renewcommand{\thesection}{B.\arabic{section}}
\renewcommand{\theequation}{B.\arabic{equation}}
\renewcommand{\thefigure}{B.\arabic{figure}}
\renewcommand{\theexample}{B.\arabic{example}}
\renewcommand{\theremark}{B.\arabic{remark}}

\phantomsection
\addcontentsline{toc}{section}{Example \ref{Example:known.beta}: \texorpdfstring{$u$}{u} is not a function of  \texorpdfstring{$\beta$}{beta}}
\begin{example}[$u$ is not a function of $\beta$]\label{Example:known.beta}
When $u=u(z)$ is a function of $z$ but not of $\beta$, $u_{i}$, $i=1,\ldots,n$, is of course observable. Hence the estimators $\tilde{f}$ eq. \eqref{Eq:gelkde.known.beta.unweighted} and $\hat{f}$ eq. \eqref{Eq:gelkde.estimated.beta.unweighted} are identical and the terms $\hat{\delta}_{1}$ and $\hat{\delta}_{2}$ in the proof of Theorem \ref{Thm:GEL.KDE.MSE.estimated.beta} are zero. 
The density estimators $\tilde{f}_{\rho}$ eq. \eqref{Eq:gelkde.known.beta.gel.weighted} and $\hat{f}_{\rho}$ eq. \eqref{Eq:gelkde.estimated.beta.gel.weighted} use different implied probabilities, $\tilde{\pi}_{i}$ versus $\hat{\pi}_{i}$, $i=1,\ldots,n$. Thus, Theorem \ref{Thm:GEL.KDE.MSE.known.beta} with known $\beta_{0}$ is unchanged whereas, in Theorem \ref{Thm:GEL.KDE.MSE.estimated.beta} with estimated $\beta_{0}$, 
$\E[\hat{f}_{\rho}(u)]= \E[\tilde{f}(u)]+n^{-1}\delta_{\rho}(u)+\littleO(n^{-1})$ with $\delta_{\rho}(u)$ defined in eq. \eqref{Thm:GEL.KDE.MSE.estimated.beta:delta.rho}. Eq. \eqref{Thm:GEL.KDE.MSE.estimated.beta:VAR.f.hat.rho} also holds with $\tilde{f}$ replacing $\hat{f}$. 

Classical examples wherein, e.g., a mean, variance, or a third moment of $u$ are either fully or partially known, are included here. For instance, symmetry can be imposed by the moment condition that the third moment around an unknown mean is known to be zero. 

This set-up also allows for situation in which the interest is in the density of $u(z_{1})$, say, but the remaining $d_{z}-1$ variates $z_{2}$ satisfy moment conditions $\E[g(z_{2},\beta_{0})]=0$. Provided $u(z_{1})$ and $g(z_{2},\beta_{0})$ are not independent, (G)EL-based estimators for $f$ will generally enjoy a reduction in variance due to the extra information from the moment condition $\E[g(z_{2},\beta_{0})]=0$.
\end{example}

\phantomsection
\addcontentsline{toc}{section}{Example \ref{Example:reg.on.a.const}: Regression on a constant}
\begin{example}[Regression On A Constant]\label{Example:reg.on.a.const} 
To explain the method behind the proof of Theorem \ref{Thm:GEL.KDE.MSE.estimated.beta} and to provide the background for Example \ref{Example:GEL.with.constant} below, the estimation of the density of the residual $u$ from a regression on a constant is examined, \textit{viz}., 
$y=\beta_{0}+u$, with $\beta_{0}$ estimated by the sample average $\hat{\beta}= \bar{y} = n^{-1}\sum_{i=1}^{n}y_{i} = \beta_{0}+\bar{u}$. 
The estimated residuals are 
$\hat{u}_{i} = y_{i}-\hat{\beta} = u_{i} - \bar{u}$, $i=1,\ldots,n$. If Assumption \ref{Assumption.U.and.K.diff.int}(a) holds, $\hat{f}(u) = \tilde{f}(u) + \hat{\delta}_{1}(u)$, where,   
for some $0\leq\omega\leq1$,
\begin{equation*}
\hat{\delta}_{1}(u)=
 n^{-1}\tsum_{i=1}^{n}k_{b}^{(1)}(u-u_{i})\bar{u}
+\tfrac{1}{2}n^{-1}\tsum_{i=1}^{n}k_{b}^{(2)}(u-u_{i})\bar{u}^{2}
+\tfrac{1}{2}n^{-1}\tsum_{i=1}^{n}[k_{b}^{(2)}(u-u_{i}+\omega\bar{u})-k_{b}^{(2)}(u-u_{i})]\bar{u}^{2}.
\end{equation*}
By H\"{o}lder  continuity of $k^{(2)}$, for some $0<C<\infty$, 
$\abs{k_{b}^{(2)}(u-u_{i}+\omega\bar{u})-k_{b}^{(2)}(u-u_{i})}\leq 
C\abs{n^{1/2}\bar{u}}^{\tau}/n^{\tau/2}b^{3+\tau}\\\to0$ in probability if $n^{\tau/2}b^{3+\tau}\to\infty$, and in mean square if $\E[u^{4}]<\infty$. 
Furthermore, for some $\epsilon>0$, $n^{(1-\epsilon)/2}\bar{u}^{2}$ is essentially bounded w.p.1 as $n\to\infty$. 
To see this, suppose $\E[X_{n}^{2}]<\infty$. Then, for any $\epsilon>0$ and $0<B<\infty$, by Chebyshev inequality, $\sum_{n=1}^{\infty}P(\abs{X_{n}}\geq n^{(1+\epsilon)/2}B)\leq \E[X_{n}^{2}]B^{-2}\sum_{n=1}^{\infty}n^{-(1+\epsilon)}<\infty$. Thus, by the first Borel-Cantelli Lemma, $P(n^{-(1+\epsilon)/2}\abs{X_{n}}\geq B\quad \text{i.o.})=0$, i.e., $n^{-(1+\epsilon)/2}\abs{X_{n}}$ is essentially bounded w.p.1 as $n\to\infty$. 
Since $\E[u^{4}]<\infty$ by assumption, for some $\epsilon>0$, however small, 
$n^{(1-\epsilon)/2}\bar{u}^{2} 
=n^{-(1+\epsilon)/2}(n^{1/2}\bar{u})^{2}$
 is essentially bounded w.p.1 as $n\to\infty$. Next, 
\begin{align*}
 \E[(n^{-1}\tsum_{i=1}^{n}[k_{b}^{(2)}(u-u_{i}+\omega\bar{u})-k_{b}^{(2)}(u-u_{i})])^{2}\bar{u}^{4}]
&\leq  \E[(\max_{1\leq i\leq n}\abs{k_{b}^{(2)}(u-u_{i}+\omega\bar{u})-k_{b}^{(2)}(u-u_{i})})^{2}\bar{u}^{4}] \\
& \leq C^{2}(n^{\tau/2}b^{3+\tau})^{-2}n^{\tau(1+\epsilon)/2}\E[(n^{(1-\epsilon)/2}\abs{\bar{u}}^{2})^{\tau}\bar{u}^{4}]\\ & = \littleO(n^{-2+\tau(1+\epsilon)/2}) = \littleO(n^{-1}b^{3})\quad \text{w.p.1}.
\end{align*}
The first inequality follows from $n^{-1}\sum_{i}a_{i}^{2}\leq \max_{1\leq i\leq n}a_{i}^{2}$, the second by H\"{o}lder continuity of $k^{(2)}$ as above and writing 
$\abs{n^{1/2}\bar{u}}^{2\tau} 
=n^{\tau(1+\epsilon)/2}(\abs{n^{(1-\epsilon)/2}\bar{u}}^{2})^{\tau}$, the third as, by Assumption \ref{Assumption.U.and.K.diff.int}(c), $n^{\tau/2}b^{3+\tau}\to\infty$ and,
 by the extremal H\"{o}lder inequality with exponents $\infty$ and $1$, 
$\E[(n^{(1-\epsilon)/2}\abs{\bar{u}}^{2})^{\tau}\bar{u}^{4}] \leq 
\bigO(n^{-2})$ noting that $n^{(1-\epsilon)/2}\abs{\bar{u}}^{2}$ is essentially bounded w.p.1 as $n\to\infty$ and $\E[\bar{u}^{4}]=\bigO(n^{-2})$ and, finally, as 
$\littleO(n^{-2+\tau(1+\epsilon)/2}) = 
\littleO(n^{-1}b^{3})n^{(\tau-1)/2+9(\tau-1)/[8(3+\tau)]+(4\tau\epsilon-1)/8}$
because  $n^{-3\tau/[2(3+\tau)]}b^{-3}\to0$ by Assumption \ref{Assumption.U.and.K.diff.int}(c), choosing  $\epsilon\leq1/4\tau$ gives the result.

If $f$ is twice differentiable and $f^{(2)}(u)$ and $uf^{(1)}(u)$ are absolutely integrable, applying Lemma \ref{Lemma:convergence}, 
\begin{align*}
\E[\hat{\delta}_{1}(u)]
& = n^{-1}\E[u_{i}k_{b}^{(1)}(u-u_{i})]
+\tfrac{1}{2}\sigma^{2}n^{-1}\E[k_{b}^{(2)}(u-u_{i})]
+\littleO(n^{-1})\\
& = n^{-1}\left(f(u)+uf^{(1)}(u) +\tfrac{1}{2}\sigma^{2}f^{(2)}(u)\right) +\littleO(n^{-1}),
\end{align*}
where $\sigma^{2}=\E[u^{2}]$. 
Since $\Var[\tilde{f}(u)] \sim (nb)^{-1}$, the covariance between $\tilde{f}(u)$ and the remainder term in $\hat{\delta}_{1}(u)$ is of order $\littleO(n^{-1}b)$, and, hence,
\begin{align}
\label{Eq:reg.on.const.Cov.f.tilde.psi.hat}
\mkern-20mu\Cov[\tilde{f}(u),\hat{\delta}_{1}(u)]
& = n^{-1}\E[k_{b}^{(1)}(u-u_{i})]\E[k_{b}(u-u_{j})u_{j}] +\littleO(n^{-1}b)
 = n^{-1}uf^{(1)}(u)f(u) + \littleO(n^{-1}b),\\
\label{Eq:reg.on.const.Var.psi.hat}
\Var[\hat{\delta}_{1}(u)] & = n^{-1}\E[k_{b}^{(1)}(u-u_{i})]^{2}\E[u_{j}^{2}] + \littleO(n^{-1}b^{3})
 = n^{-1}\sigma^{2}f^{(1)}(u)^{2}+ \littleO(n^{-1}b).
\end{align}
Note that $\zeta_{\lambda}=0$,  $\mathrm{d}\{\E[\nabla u_{i}|u]f(u)\}/\mathrm{d}u=-f^{(1)}(u)$, $\mathrm{d}\{\E[\nabla^{\transp}u_{i}Hg_{i}|u]f(u)\}/\mathrm{d}u=f(u)+uf^{(1)}(u)$,
$\mathrm{d}\{\E[\nabla^{2}u_{i}|u]f(u)\}/\mathrm{d} u=0$, and 
$\mathrm{d}^{2}\{\E[\nabla u_{i}\nabla^{\transp}u_{i}|u]f(u)\}/\mathrm{d}u^{2}=f^{(2)}(u)$
 from the unbiasedness of $\hat{\beta}$ and linearity of $u(z,\beta)$; cf. Theorem \ref{Thm:GEL.KDE.MSE.estimated.beta}.

Assuming $f^{(1)}(u)$ is square integrable, and if $\lim_{\abs{u}\to\infty}uf(u)^{2}=0$, $\int uf^{(1)}(u)f(u)\mathrm{d}u=-\frac{1}{2}R(f)$ and, thus, 
\begin{equation*}
\IVar[\hat{f}] = \IVar[\tilde{f}] - n^{-1}(R(f)-\sigma^{2}R(f^{(1)})) + \littleO(n^{-1}).
\end{equation*}
Hence, whenever $R(f)>\sigma^{2}R(f^{(1)})$, $\hat{f}$ achieves a second order reduction in variance relative to $\tilde{f}$.  While this may appear as a `free' reduction in variance, it is not so. Construction of $\hat{f}$ explicitly assumes that $\E[u]$ exists, and the validity of the above result requires the first four moments of $u$ to exist whereas that of $\tilde{f}$ makes no such assumptions.

When the mean $\E[u]$ is known, the (G)EL-reweighted estimator $\tilde{f}_{\rho}$ eq. \eqref{Eq:gelkde.known.beta.gel.weighted} imposing the constraint $\E[u]=0$ will achieve a second order reduction in variance of $n^{-1}\sigma^{-2}u^{2}f(u)^{2}$, i.e., $\IVar[\tilde{f}_{\rho}] = \IVar[\tilde{f}] - n^{-1}\sigma^{-2}\int u^{2}f(u)^{2}\mathrm{d}u+ \littleO(n^{-1})$; see, e.g., \citet[eq. (13), p.56]{chen1997}. 
In particular, for normally distributed $u$, 
$R(\phi_{\sigma})-\sigma^{2}R(\phi_{\sigma}^{(1)})=1/4\sqrt{\pi}\sigma$, which equals 
$\sigma^{-2}\int u^{2}\phi_{\sigma}(u)^{2}\mathrm{d}u$ exactly. 
For the Student $t$ distribution with $\nu>2$ degrees of freedom, 
$R(t_{\nu}) -\sigma^{2}R(t_{\nu}^{(1)})= R(t_{\nu})(2\nu^{2}-3\nu -17)/4(\nu^{2}-4)$,
which is positive for $\nu>4$, the condition for the first four moments of $u$ to exist, whereas 
$\sigma^{-2}\int u^{2}t_{\nu}(u)^{2}\mathrm{d}u = R(t_{\nu})(\nu-2)/(2\nu-1)$ which is always larger than $R(t_{\nu})-\sigma^{2}R(t_{\nu}^{(1)})$. This difference may be interpreted as the cost of having to estimate the mean of $u$. 

The same or similar terms appear in the expansions for the variance of $\hat{f}$ in other contexts (the $\bigO(n^{-1})$ bias terms tend to be ignored as their contribution to MISE is $\littleO(n^{-1})$);  
cf. \citet[eq.(3.5)]{mushal2010}. As the next example demonstrates, these same effects appear in a large class of parametric moment condition models. 
\end{example}


\phantomsection
\addcontentsline{toc}{section}{Example \ref{Example:GEL.with.constant}: GEL with a constant and zero mean restriction}
\begin{example}[GEL With A Constant And Zero Mean Restriction]\label{Example:GEL.with.constant}
Consider GEL estimation based on moment indicator functions of the form $g(z,\beta) = u(z,\beta)\alpha(w)$ where $u(z,\beta)$ is scalar, $\beta$ a $d_{\beta}$-vector of parameters, and $\alpha(w)$ a $d_{g}$-vector of functions of $w$. Suppose that $u(z,\beta_{0})$ is independent of $w$,
Assumption \ref{Assumption.reformulation.bijection} holds, 
and the moment condition $\E[g(z,\beta_{0})]=0$ includes the restriction $\E[u(z,\beta_{0})]=0$. 
Furthermore, it is assumed that $u(z,\beta)$ contains a constant; the inclusion of an explicit constant is not essential as the results here continue to hold if $\E[\partial u(z,\beta_{0})/\partial\beta^{\transp}|w]\gamma=c$ for some non-zero vector $\gamma$ and scalar $c$, in which case $\E[\alpha(w)]=G\gamma/c$.
Without loss of generality let $\alpha_{1}(w)=1$ and $\partial u(z,\beta_{0})/\partial\beta_{1} =-1$.
 
Since $u$ and $w$ are independent, $\E[g_{i}|u]=u\E[\alpha(w)]$,  $\Omega=\sigma^{2}\E[\alpha(w)\alpha(w)^{\transp}]$, where $\sigma^{2}=\E[u^{2}|w]=\E[u^{2}]$. Then, because the first column of $G$ is $-\E[\alpha(w)]$, as $PG=0$, $\E[g_{i}|u]^{\transp}P\E[g_{i}|u]=0$. That is, there is no second order reduction in variance due to re-weighting. 

Since the first column (and row) of $\Omega$ is $\sigma^{2}\E[\alpha(w)]$, 
$\Omega^{-1}\E[g_{i}|u] = u\sigma^{-2}\bv_{1}$, where $\bv_{j}$ is the $j$th unit $d_{g}$-vector, $j=1,\ldots,d_{g}$.
For an $n\times m$ matrix $A$, let $A_{(s:t)}$, $1\leq s\leq t\leq m$, denote the $n\times (t-s+1)$ submatrix comprised of columns $j=s,\ldots,t$ of $A$. Noting that $\Omega^{-1}\E[\alpha(w)]=\bv_{1}/\sigma^{2}$ and $\E[\alpha(w)]^{\transp}\Omega^{-1}\E[\alpha(w)]=1/\sigma^{2}$, partition $\Sigma$ and $H$ as
\begin{equation*}
\Sigma 
=\sigma^{2}\left[\begin{smallmatrix}1+\bv_{1}^{\transp}G_{(2:p)}Q^{-1}G_{(2:p)}^{\transp}\bv_{1}&\bv_{1}^{\transp}G_{(2:p)}Q^{-1}\\Q^{-1}G_{(2:p)}^{\transp}\bv_{1}&Q^{-1}\end{smallmatrix}\right], \qquad
H =-\left[\begin{smallmatrix}
\bv_{1}^{\transp}
-\bv_{1}^{\transp}G_{(2:p)}Q^{-1}G_{(2:p)}^{\transp}(\sigma^{2}\Omega^{-1}-\bv_{1}\bv_{1}^{\transp}) \\
 -Q^{-1}G_{(2:p)}^{\transp}(\sigma^{2}\Omega^{-1}-\bv_{1}\bv_{1}^{\transp})\end{smallmatrix}\right],
\end{equation*}
where $Q =  G_{(2:p)}^{\transp}(\sigma^{2}\Omega^{-1}-\bv_{1}\bv_{1}^{\transp})G_{(2:p)}$ and 
$\bv_{1}^{\transp}G_{(2:p)}= (\E[\partial u(z,\beta_{0})/\partial\beta_{2}],\ldots,\E[\partial u(z,\beta_{0})/\partial\beta_{p}])$ is the first row of $G_{(2:p)}$.
Thus, $H\E[g_{i}|u] =-u\bv_{1}$. 

As the first element of $\mathrm{d}\{\E[\nabla u_{i}|u]f(u)\}/\mathrm{d}u$  is $-f^{(1)}(u)$,  $[\mathrm{d}\{\E[\nabla u_{i}|u]f(u)\}/\mathrm{d}u]^{\transp}H\E[g_{i}|u]f(u) = uf^{(1)}(u)f(u)$, same as eq. \eqref{Eq:reg.on.const.Cov.f.tilde.psi.hat}. 
Partition $\mathrm{d}\{\E[\nabla u_{i}|u]f(u)\}/\mathrm{d}u$ as $\mathrm{d}\{\E[\nabla u_{i}|u]f(u)\}/\mathrm{d}u=\left(-f^{(1)}(u),\right. \\ \left.[\mathrm{d}\{\E[\nabla u_{i}|u]f(u)\}/\mathrm{d}u]_{(2:p)}\right)^{\transp}$. Hence,
\begin{align}\notag
& \mkern-10mu [\mathrm{d}\{\E[\nabla u_{i}|u]f(u)\}/\mathrm{d}u]^{\transp}\Sigma [\mathrm{d}\{\E[\nabla u_{i}|u]f(u)\}/\mathrm{d}u]
 =\sigma^{2}f^{(1)}(u)^{2}
+\sigma^{2}f^{(1)}(u)^{2}\bv_{1}^{\transp}G_{(2:p)}Q^{-1}G_{(2:p)}^{\transp}\bv_{1}
\\ \notag & \mkern230mu
-2\sigma^{2}f^{(1)}(u)\bv_{1}^{\transp}G_{(2:p)}Q^{-1}[\mathrm{d}\{\E[\nabla u_{i}|u]f(u)\}/\mathrm{d}u]_{(2:p)}
\\ \label{Eq:J1.Sigma.J1.generic.GEL.with.constant} & \mkern230mu
+ \sigma^{2}[\mathrm{d}\{\E[\nabla u_{i}|u]f(u)\}/\mathrm{d}u]_{(2:p)}^{\transp}Q^{-1}[\mathrm{d}\{\E[\nabla u_{i}|u]f(u)\}/\mathrm{d}u]_{(2:p)}
\end{align}
The first term in \eqref{Eq:J1.Sigma.J1.generic.GEL.with.constant} is the same as the main term in \eqref{Eq:reg.on.const.Var.psi.hat}. The remaining terms represent the additional increase in the variance of $\hat{f}(u)$ due to the estimation error in $\beta_{2},\ldots,\beta_{p}$. 

The independence of $u$ and $w$ is crucial to the above argument implying $\E[g_{i}|u]=u\E[\alpha(w)]$, and $P$ annihilates $\E[\alpha(w)]$. The next example illustrates that these relationships need not hold in the dependent case.
\end{example}

\phantomsection
\addcontentsline{toc}{section}{Example \ref{Sec.Example:normal.over.GG}: Linear regression model with 
\texorpdfstring{$\E[u|x]=0$}{E[u|x]=0} but dependent \texorpdfstring{$u$}{u} and \texorpdfstring{$x$}{x}}
\begin{example}[Linear Regression Model With 
\texorpdfstring{$\E[u|x]=0$}{E[u|x]=0} But Dependent \texorpdfstring{$u$}{u} And \texorpdfstring{$x$}{x}]\label{Sec.Example:normal.over.GG}
For simplicity, consider the linear regression model 
\begin{equation}\label{Eq:linear.model}
y = \delta_{0} + \gamma_{0} x + u, 
\end{equation}
where $\E[u|x]=0$. Here $\beta = (\delta,\gamma)^{\transp}$ and $z=(y,x)^{\transp}$. 

Estimation of $\beta_{0}$ may be based on the unconditional moment restriction $\E[g(z,\beta_{0})]=0$ where
\begin{equation}\label{Eq:Example.LM.GaussOverGG.Moment.Indicator.Fn}
g(z,\beta) = u(z,\beta)(1,\:x,\:x^{2},\:\ldots,\:x^{q-1})^{\transp}, \quad q\geq2
\end{equation}

Suppose now that $u$ and $x$ are distributed with joint density
\begin{equation}
\label{Eq:normal.over.GG.joint.pdf}
f_{U,X}(u,x) = \frac{2(\nu/2)^{\nu/2}}{\sqrt{2\pi}\omega\Gamma(\nu/2)} x^{\nu}\mathrm{e}^{-x^{2}(\nu+u^{2}/\omega^{2})/2}, 
\quad x\geq0, \: -\infty<u<\infty, \: \nu>0, \:\omega>0.
\end{equation}
The marginal distributions of $u$ and $x$ are the non-standardized Student $t$ distribution with $\nu$ degrees of freedom and scale parameter $\omega$, and the generalized gamma \citep{stacy1962} with parameters $p=2$, $d=\nu$, and $a=(2/\nu)^{1/2}$. \footnote{$x$ is distributed as $(w/\nu)^{1/2}$ where $w\sim\chi_{\nu}^{2}$ and, if $\nu=1$, as a standard half-normal random variable. The joint density  eq. \eqref{Eq:normal.over.GG.joint.pdf} is that of $u=z/x$ and $x$, where $z\sim N(0,\omega^{2})$, independent of $x$.}
The moments of $x$ are $m_{k}=\E[x^{k}] =  (2/\nu)^{k/2}\Gamma((\nu+k)/2)/\Gamma(\nu/2)$, $k>-\nu$, and satisfy the recursion $m_{k+2}= (1+k/\nu)m_{k}$. 
 The odd moments of $u$ of order $k<\nu$ are zero, while the even moments are $\E[u^{2k}] = \omega^{2k}\pi^{-1/2}\nu^{k}\Gamma(\nu/2-k)\Gamma(k+1/2)/\Gamma(\nu/2)$, $k<\nu/2$.

The conditional density of $u$ given $x$ is $f_{U|X}(u,x)=\phi_{\omega/x}(u)$ and, hence, $\E[u|x] = 0$, but $u$ and $x$ are not independent. If $\nu>2$, $\E[u^{2}|x]=\omega^{2}/x^{2}$.
The conditional moments of $x$ given $u$ are 
$m_{k|u}(u)  =\E[x^{k}|u]  = m_{k+1}/m_{1}(1+(u/\omega)^{2}/\nu)^{k/2}$, $k>-\nu-1$. The transformation in Assumption \ref{Assumption.reformulation.bijection} has $v(z,\beta)=x$ and, hence, $\E[g_{i}|u] = u(1,\: m_{1|u}(u),\: m_{2|u}(u),\: \ldots,\: m_{q-1|u}(u))^{\transp}$.

To describe the quantities involved, let $\M[q]{s}{}{}=\{m_{i+j-2-s}\}_{i,j=1}^{q}$ be a $q\times q$ matrix composed of the $(i+j-2-s)$th moments of $x$. Note that, if $q>2$, then $\M[q]{0}{(s)}{\transp} \M[q]{2}{}{-1}\M[q]{0}{(t)}{} =m_{s+t}$ for $s,t=1,2,\ldots$, $(s\wedge t)\leq q-2$, and  $\M[q]{2}{}{-1}\M[q]{0}{(t)}{} = \bv_{t+2}$ for $1\leq t\leq q-2$. 
The relevant (G)EL matrices are $\Omega = \omega^{2}\M[q]{2}{}{}$, $G = -\M[q]{0}{(1:2)}{}$ and, if $q\geq4$, 
\begin{equation*}
\Sigma = \tfrac{\omega^{2}\nu}{\nu(\nu+2)-(\nu+1)^{2}m_{1}^{2}}
\left[\begin{smallmatrix} \nu+2& -(\nu+1)m_{1}  \\ -(\nu+1)m_{1}  & \nu \end{smallmatrix}\right], \qquad
H  = -\tfrac{1}{\omega^{2}}\Sigma \left[\begin{smallmatrix}\bv_{q,3}^{\transp} \\ \bv_{q,4}^{\transp}  \end{smallmatrix} \right],
\end{equation*}
and
\begin{equation*}
P = \tfrac{1}{\omega^{2}}\M[q]{2}{}{-1}-\tfrac{1}{\omega^{2}}\tfrac{\nu}{\nu(\nu+2)-(\nu+1)^{2}m_{1}^{2}}
\left[  (\nu+2)\bv_{3}\bv_{3}^{\transp}
 -(\nu+1)m_{1}\left( \bv_{3}\bv_{4}^{\transp}  +\bv_{4}\bv_{3}^{\transp} \right)
 + \nu \bv_{4}\bv_{4}^{\transp} \right].
\end{equation*}

\begin{remark}
For the exactly identified case, $q=2$, $G$ is square and invertible. Hence, $\Sigma = G^{-1}\Omega G^{\transp-1}$, $H = G^{-1}$, and $P=0$. Closed form expressions for $\Sigma$, $H$, and $P$ when $q=3$ can be obtained in a straightforward fashion. That $\Sigma$ remains unaltered as $q$ increases above $4$ is of course due to the special form of the conditional variance of $u$. 
Figure \ref{Fig:LMRelEffSigmaQ234} displays the relative efficiency of $\hat{\beta}$ based on the first $q$  compared with the first $q^{\prime}$ moment conditions, $[\det(\Sigma_{q})/\det(\Sigma_{q^{\prime}})]^{1/p}$, for various values of $\nu$. 
\end{remark}

\begin{figure}[htbp]\centering
\includegraphics[width=0.5\linewidth]{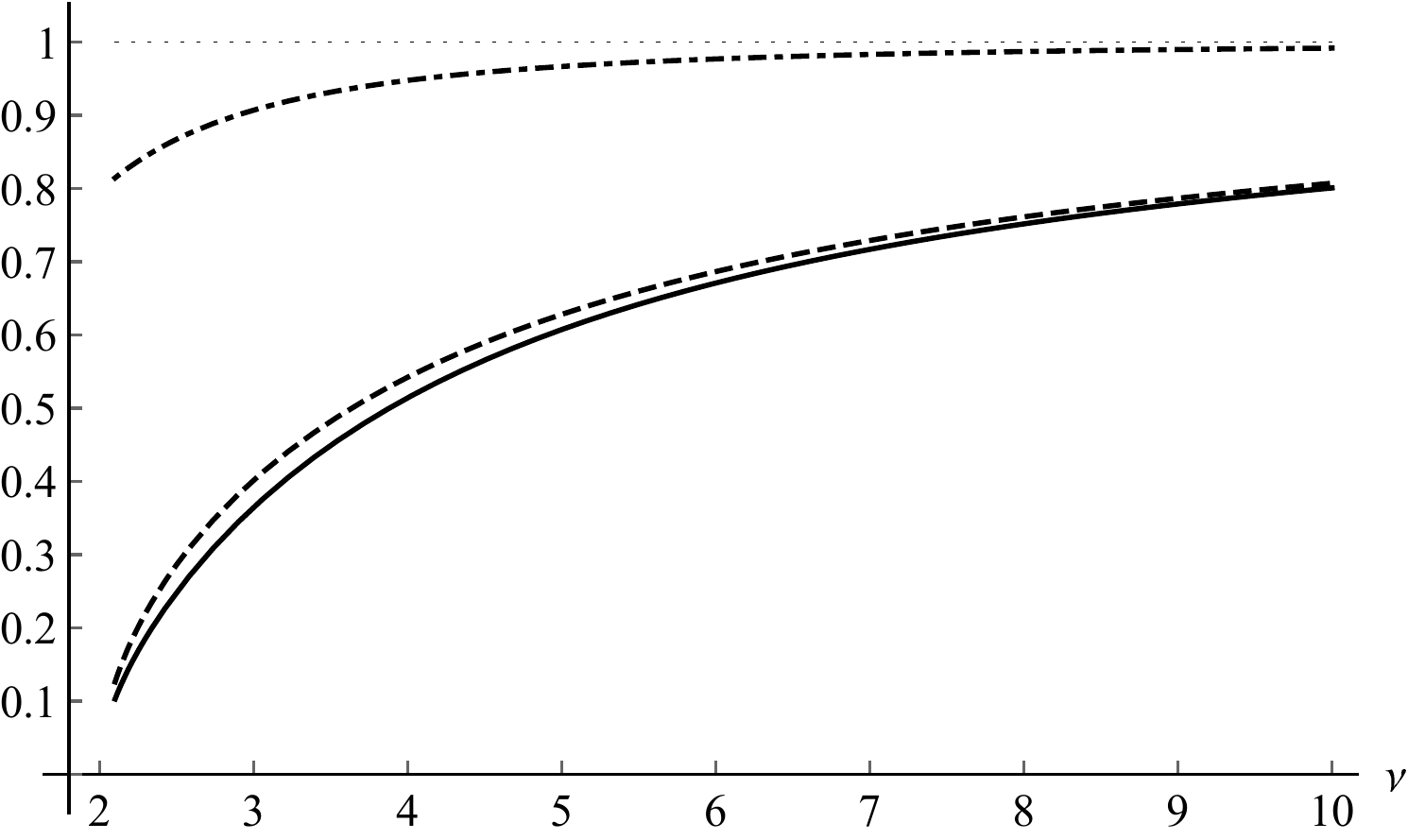}
\caption{Relative efficiency of $\hat{\beta}$ based on 
$q\geq 4$ vs. $q=2$ (solid line), 
$q=3$ vs. $q=2$ (dashed line), 
and $q\geq 4$ vs. $q=3$ (dash-dotted line) moment conditions as a function of $\nu$.}
\label{Fig:LMRelEffSigmaQ234}
\end{figure}

If $q\geq4$, only the moment indicators $x^{j-1}u(z,\beta)$, $j=3,4$, are used to estimate $\beta_{0}$. Information in the remaining moment conditions, however, can be usefully exploited to improve the efficiency of the density estimators $\hat{f}$ and $\hat{f}_{\rho}$. The quantities entering the integrated variance eqs. \eqref{Thm:GEL.KDE.MSE.estimated.beta:VAR.f.hat} and \eqref{Thm:GEL.KDE.MSE.estimated.beta:VAR.f.hat.rho} can be computed as 
$\trace\left(\Sigma \tint [\mathrm{d}\{\E[\nabla u_{i}|u]f(u)\}/\mathrm{d}u][\mathrm{d}\{\E[\nabla u_{i}|u]f(u)\}/\mathrm{d}u]^{\transp}\mathrm{d}u\right)$, 
$\trace\left(H \int\E[g_{i}|u]\right.\\\left.\times [\mathrm{d}\{\E[\nabla u_{i}|u]f(u)\}/\mathrm{d}u]^{\transp}f(u)\mathrm{d}u \right)$, and 
$\trace\left(P \int\E[g_{i}|u]\E[g_{i}|u]^{\transp}f(u)^{2}\mathrm{d}u \right)$, where
\begin{equation*}
\int \tfrac{\mathrm{d}\{\E[\nabla u_{i}|u]f(u)\}}{\mathrm{d}u}\left(\tfrac{\mathrm{d}\{\E[\nabla u_{i}|u]f(u)\}}{\mathrm{d}u}\right)^{\transp} \mathrm{d}u
= \tfrac{\Gamma((\nu+3)/2)}{\omega^{3}\pi^{1/2}\nu^{3/2}\Gamma(\nu/2)}\left[\begin{smallmatrix}
\tfrac{\nu\Gamma(\nu+3/2)\Gamma((\nu+3)/2)}{\Gamma(\nu/2+1)\Gamma(\nu+3)} & 
\tfrac{\nu^{1/2}\Gamma(\nu+3)}{2^{1/2}\Gamma(\nu+7/2)}\\
\tfrac{\nu^{1/2}\Gamma(\nu+3)}{2^{1/2}\Gamma(\nu+7/2)}&
\tfrac{\Gamma(\nu+5/2)\Gamma(\nu/2+2)}{2\Gamma(\nu+2)\Gamma((\nu+5)/2)}	\end{smallmatrix}\right],
\end{equation*}
the $q\times 2$ matrix $\tint\E[g_{i}|u][\mathrm{d}\{\E[\nabla u_{i}|u]f(u)\}/\mathrm{d}u]^{\transp}f(u)\mathrm{d}u$ has rows 
\begin{equation*}
\tfrac{1}{(2\pi)^{1/2}\omega}\left[
\tfrac{(2/\nu)^{i/2}\Gamma((\nu+3)/2)\Gamma(\nu+i/2)\Gamma((\nu+i)/2))}{\Gamma(\nu/2)^{2}\Gamma(\nu+(i+3)/2)}
, \quad 
\tfrac{(2/\nu)^{(i-1)/2}(\nu+2)\Gamma((\nu+1)/2)\Gamma(\nu+(i+1)/2)}{2\Gamma(\nu/2)\Gamma(\nu+i/2+2)}
\right], \quad i=1,\ldots,q, 
\end{equation*}
and the $q\times q$ matrix $\int\E[g_{i}|u]\E[g_{i}|u]^{\transp}f(u)^{2}\mathrm{d}u$  with $(i,j)$th element 
\begin{equation*}
\omega m_{i}m_{j}\tfrac{\nu^{3/2} \Gamma(\nu+(i+j-3)/2)}{4\pi^{1/2}\Gamma(\nu+(i+j)/2)}, \quad i,j=1,\ldots,q.
\end{equation*}

\begin{figure}[htbp]\centering\small
\begin{tabular}{@{}c@{}@{}c@{}}
\multicolumn{2}{c}{}\\
(A) $\omega\tint [\mathrm{d}\{\E[\nabla u_{i}|u]f(u)\}/\mathrm{d}u]^{\transp}\Sigma [\mathrm{d}\{\E[\nabla u_{i}|u]f(u)\}/\mathrm{d}u]\mathrm{d}u$ & (A: zoom) \\
\includegraphics[width=0.5\linewidth]{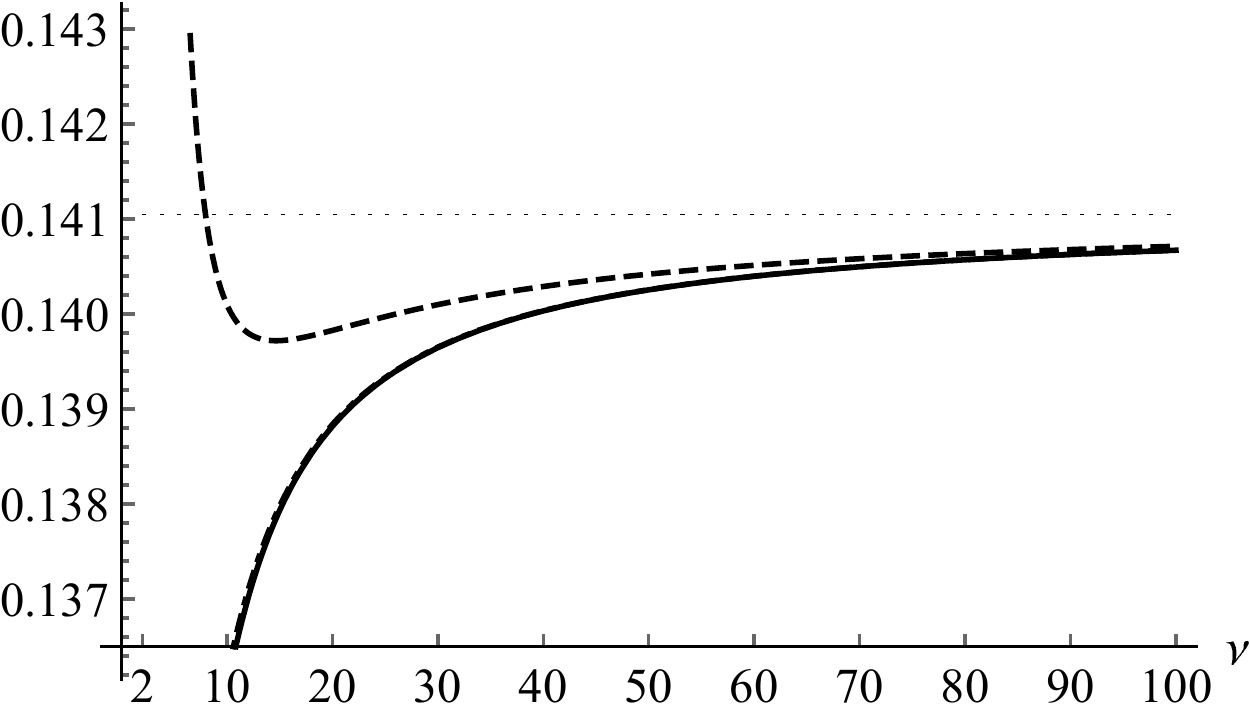} &
\includegraphics[width=0.5\linewidth]{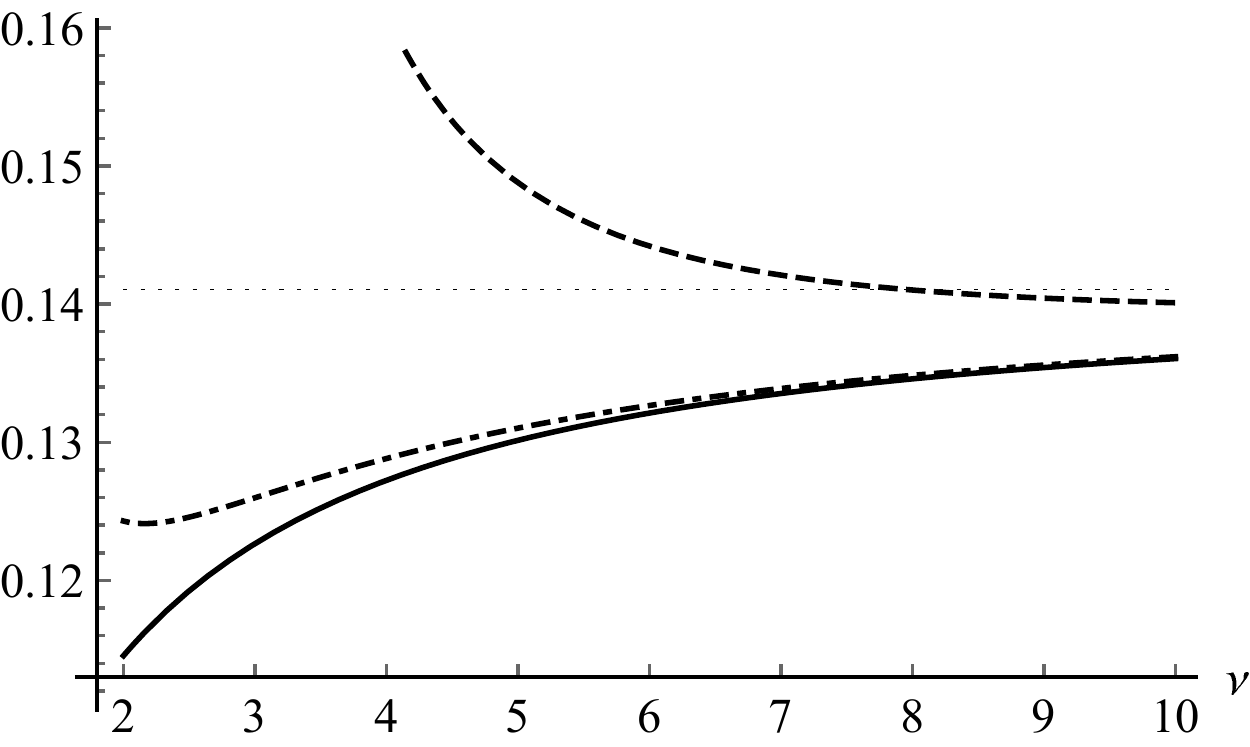} \\[4pt]
(B) $\omega\tint [\mathrm{d}\{\E[\nabla u_{i}|u]f(u)\}/\mathrm{d}u]^{\transp}H\E[g_{i}|u]f(u)\mathrm{d}u$ & (B: zoom) \\
\includegraphics[width=0.5\linewidth]{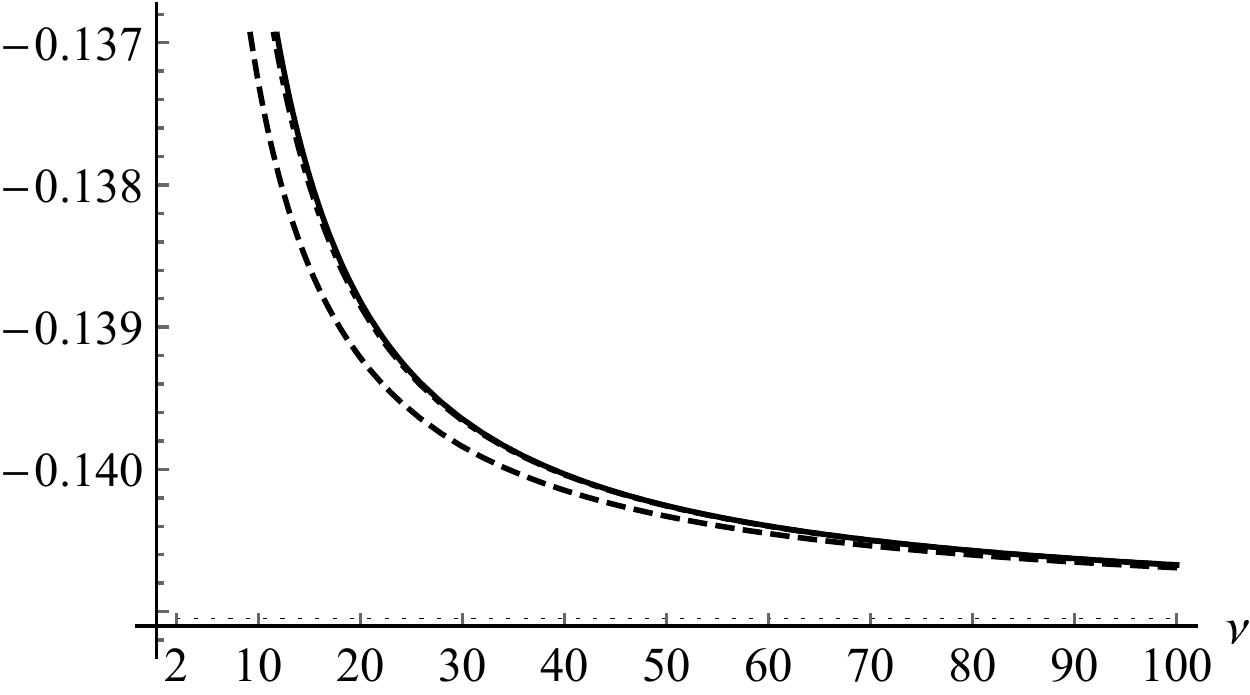} &
\includegraphics[width=0.5\linewidth]{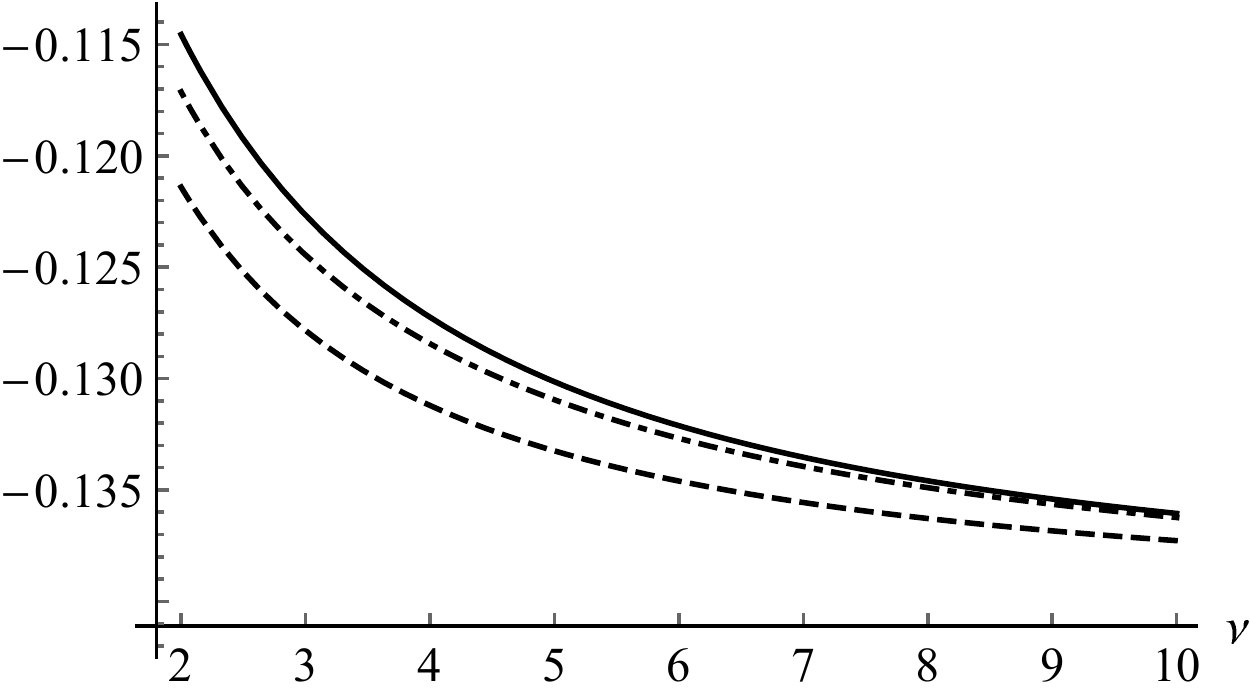} \\[2pt]
\multicolumn{2}{l}{Legend to panels A and B: dashed line: $q=2$; dash-dotted line: $q=3$; solid line:  $q\geq4$.}\\[10pt]
(C) $\omega\int\E[g_{i}|u]^{\transp}P\E[g_{i}|u]f(u)^{2}\mathrm{d}u$ & (D) Overall effect: $(A)+2(B)-(C)$ \\
\includegraphics[width=0.5\linewidth]{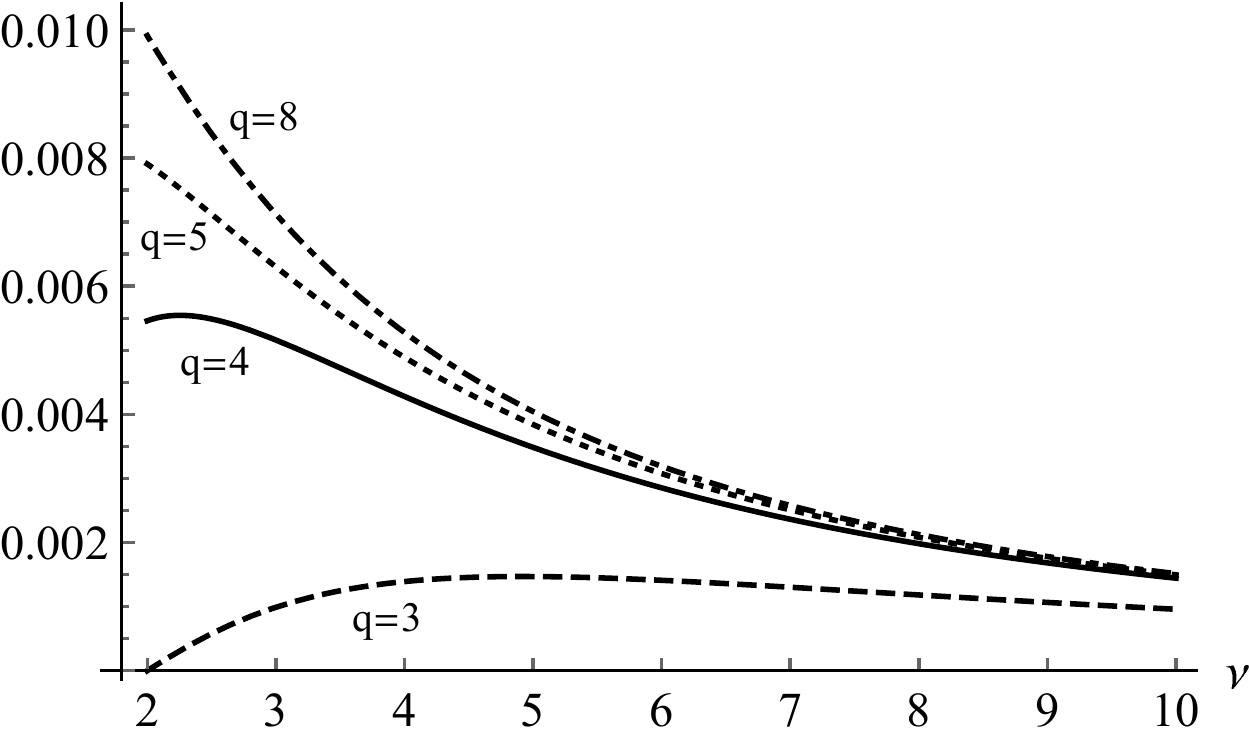} & 
\includegraphics[width=0.5\linewidth]{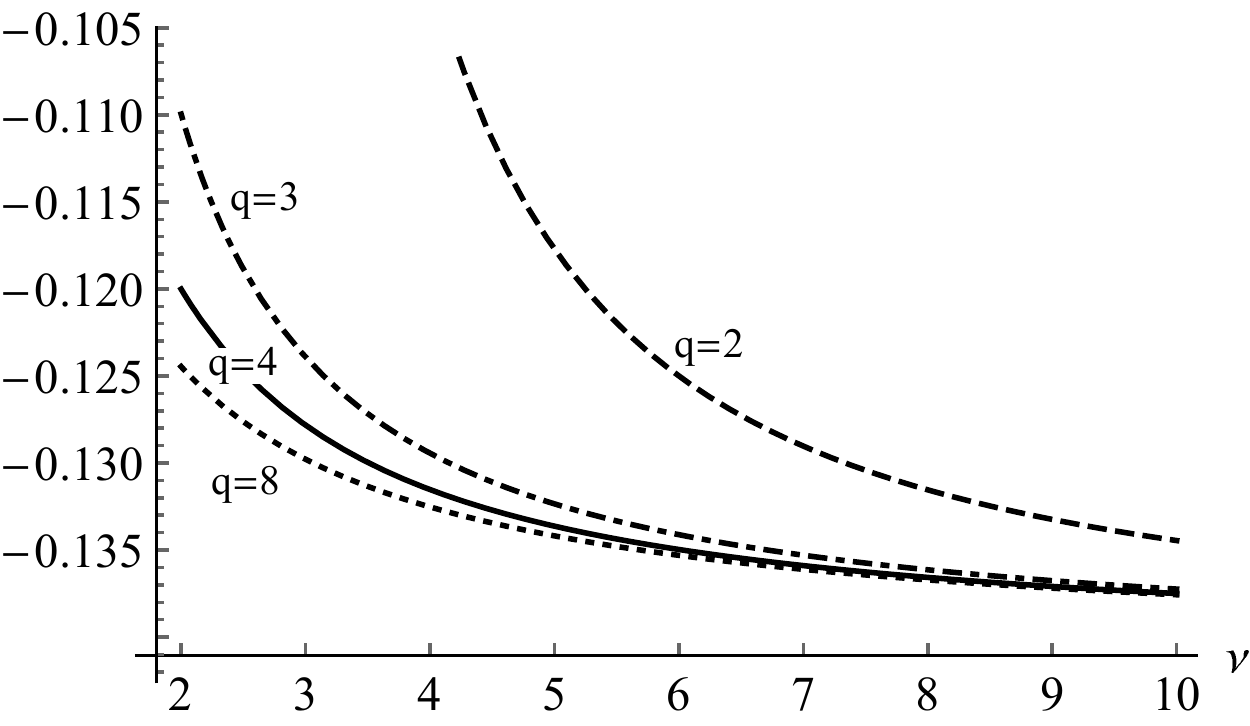} 
\end{tabular}
\caption{Quantities entering the integrated variance of $\hat{f}$ and $\hat{f}_{\rho}$ in Example \ref{Sec.Example:normal.over.GG}  ($\times n\omega$) }
\label{Fig:LMtrJ1J1Sigma.ExLMtrJ1hH}
\end{figure}

\begin{remark}
Figure \ref{Fig:LMtrJ1J1Sigma.ExLMtrJ1hH} shows the values of the above quantities and the overall effect on the integrated variance for selected values of $q$ and $\nu>2$; note that the validity of asymptotic expansions requires $\nu>4$, but variance is defined for $\nu>2$. 
While the main reduction in variance is still due to the zero mean restriction as in Example \ref{Example:GEL.with.constant} (Panels A and B), there are small additional gains due to re-weighting (Panel C). The latter do increase as more moment conditions are added. 
\end{remark}
\end{example}

\end{document}